%% file: Arxiv.tex
\documentclass[11pt]{iopart}

\usepackage[top=1in, bottom=1in, left=1.25in, right=1.25in]{geometry}

\usepackage{cite}
\usepackage{graphicx}
\usepackage{dcolumn}

\usepackage{calrsfs}
\usepackage{relsize}

\usepackage{siunitx}
\usepackage{ulem}
\usepackage{bm}

\usepackage[colorlinks=true,citecolor=black,linkcolor=black,urlcolor=black]{hyperref}
\usepackage[colorinlistoftodos]{todonotes}

\newcommand{\ud}{\mathrm{d}}
\newcommand{\ds}{\displaystyle}
\newcommand{\vecB}{\ensuremath{\vec{B}}}

\usepackage{iopams}  

\newcommand{\iint}{\ensuremath{\int\int}}
\newcommand{\underset}[2]{\begin{tabular}[t]{@{}c@{}}\ensuremath{#2}\\[-0.3em]\scriptsize\ensuremath{#1}\end{tabular}}
    
\begin{document}

\title[]{The irony of the magnet system for Kibble balances -- a review}

\author{Shisong Li$^1$, Stephan Schlamminger$^2$}

\address{1. Department of Electrical Engineering, Tsinghua University, Beijing 100084, China\\
2. National Institute of Standards and Technology (NIST), Gaithersburg, 20899 MD, United States}
\ead{shisong.li@outlook.com; stephan.schlamminger@nist.gov}
\vspace{10pt}

\begin{abstract}
The magnet system is an essential component of the Kibble balance, a device that is used to realize the unit of mass. It is the source of the magnetic flux, and its importance is captured in the geometric factor $Bl$. Ironically, the $Bl$ factor cancels out and does not appear in the final Kibble equation. Nevertheless, care must be taken to design and build the magnet system because the cancellation is perfect only if the $Bl$ is the same in both modes: the weighing and velocity mode. This review provides the knowledge necessary to build a magnetic circuit for the Kibble balance. In addition, this article discusses the design considerations,  parameter optimizations, practical adjustments to the finished product, and an assessment of systematic uncertainties associated with the magnet system.
\end{abstract}

%

\submitto{\MET}

\maketitle



    \input{01_Introduction}

	\input{02_BasicReview}

	\input{03_DifferentMagSystems} 
	\input{04_DesignMagnets}
	\input{05_TestMeasurement}
	\input{06_MagneticEffects}
	\input{07_Conclusion}

	\clearpage
	\input{08_Appendix}

	\input{09_References}


\end{document}

%% file: 01_Introduction.tex
\section{Introduction}
	\label{sec1}
	
Today, the Kibble balance~\cite{Kibble1976} is a precision instrument that is used to realize the unit of mass, and it can weigh masses ranging from grams to kilograms. It is one of two methods for the primary realization of the mass unit, the other being the X-ray crystal density method (XRCD) \cite{fujii2016realization}. 

Previously, the Kibble balance was called the watt balance. The community agreed to the name change to honor the late Dr. Bryan Kibble, who invented this measurement technique. Before 2019, the balance was used to determine the Planck constant, $h$, utilizing a mass that was traceable to the international prototype kilogram (IPK) as an input quantity. According to  \textit{Nature}~\cite{jones2012tough},  in 2012, the watt balance was one of the six most difficult experiments.    

By the end of July 2017, the different measurements of $h$ had converged sufficiently to initiate the revision of the international system of units, the SI (abbreviation for the French expression {\bf S}yst\`eme {\bf I}nternational d'Unit\'es).  Based on data available at that time, final values were calculated for the Planck constant $h$, the Avogadro constant $N_\mathrm{A}$, the elementary charge $e$, and the Boltzmann constant $k$~\cite{codata17}. The assigned numerical values to these four constants define four of the seven base units in the SI~\cite{cgpm2018}. These are the kilogram, the ampere, the mole, and the kelvin. On May 20th, 2019, the revised SI came into effect. Since then, the Kibble balance and XRCD  have replaced the international prototype kilogram as the starting point of worldwide mass dissemination. Kibble balance experiments are carried out at many National Metrology Institutes (NMIs) and the Bureau International des Poids et Measures (BIPM), e.g. \cite{NRC,NIST,LNE,METAS,BIPM,NIM,MSL,KRISS,UME,NPL3}.
	
The principle of the Kibble balance is based on the measurement of the integral of the magnetic flux density $B$ along the coil wire $l$ or the gradient of the coil flux linkage $\Phi$ over the vertical direction $z$, the so-called geometrical factor, given by (see \ref{sec:AppendixA})
\begin{equation}
Bl=\frac{\partial\Phi}{\partial z}=\int (\mathrm{d}\textbf{l}\times \textbf{B})_\mathrm{z}
\end{equation}
in two separated phases. In the weighing phase, the coil is excited by current $I$. The electromagnetic force is adjusted such that it is equal and opposite to  the weight of a test mass, 
\begin{equation}
(Bl)_\mathrm{w}=\frac{mg}{I},
\label{eq:BLw}
\end{equation}
where $m$ and $g$ denote the test mass and local gravitational acceleration, respectively. 
In the velocity phase, the current is removed, and the open coil is connected to a voltmeter with a high input impedance. The $Bl$ factor is measured by moving the coil along the vertical direction with a velocity $v$. The quotient of the induced voltage $U$  to the velocity $v$ is equal to  $Bl$ as
\begin{equation}
(Bl)_\mathrm{v}=\frac{U}{v}.
\label{eq:BLv}
\end{equation}
Ideally and theoretically that is the case, $(Bl)_\mathrm{w}$ and $(Bl)_\mathrm{v}$ are the same, and, hence, the right side of equation~(\ref{eq:BLw}) is equal to the right side of equation~(\ref{eq:BLv}).  Then, after crosswise multiplication, the equation of virtual power, also known as the Kibble equation,
\begin{equation}
mgv=UI,
\label{eq:balance}
\end{equation}
is obtained. The Kibble equation can only be obtained if $(Bl)_\mathrm{w}=(Bl)_\mathrm{v}$. 
At this point, it is necessary to reflect on the relative uncertainties that are required. The best Kibble balance can measure a \SI{0.5}{\kilo\gram} mass with a relative uncertainty just below \SI{1e-8}{}~\cite{NRC}. So the question is, can the ratio $(Bl)_\mathrm{w}/(Bl)_\mathrm{v}$ be trusted to be one within $\pm\SI{1e-8}{}$? We will scrutinize this assumption in the sections below.

The current in the weighing phase is measured as a voltage drop $V$ on a standard resistor $R$, i.e. $I=V/R$. Solving for mass yields,
\begin{equation}
m=\frac{UV}{gvR}.
\label{eq:mass}
\end{equation}
The two different electrical measurements, voltage and resistance, can be traced back to quantum effects. The Josephson effect~\cite{Josephson1962} allows to realize a voltage
\begin{equation}
U = \frac{n_\mathrm{U} f_\mathrm{U}}{ K_\mathrm{J}},
\label{eq:KJ}
\end{equation}
with the Josephson constant $K_\mathrm{J} = {2e}/{h}$. Here $n_\mathrm{U}$ is the number of Josephson Junctions used (typically between $10^4$ and $10^5$) and $f_\mathrm{U}$ is the microwave frequency that is used to irradiate the Josephson junctions (several tens of \SI{}{\giga \hertz}). For more information, the reader may consult a recent review on Josephson voltage standards, for example~\cite{R_fenacht_2018}.

The standard resistor is compared against, read: ``is a fraction $\eta$ of'', the quantum Hall value~\cite{Hamilton2000,klitzing1980new},
\begin{equation}
R = \eta R_\mathrm{K},
\label{eq:RK}
\end{equation}
with the von Klitzing constant $R_\mathrm{K} = {h}/{e^2}$. Recent reviews of the quantum Hall effect can be found in ~\cite{jeckelmann2001quantum,Klitzing2017}.

Using $n_\mathrm{V}$ and $f_\mathrm{V}$ for the corresponding values in the $V$ measurement, equation~(\ref{eq:mass}) can be written as
\begin{equation}
m=\frac{n_\mathrm{U} n_\mathrm{V}}{4\eta} \frac{f_\mathrm{U} f_\mathrm{V}}{gv} h.
\label{eq:mass-h}
\end{equation}

The quantum aspects pertain to the electrical measurement chain employed in the Kibble balance. These aspects are crucial in bridging the gulf between classical and quantum mechanics~\cite{haddad2016bridging}, and are the prime connection that enables the  Kibble balance to realize the unit of mass.  Despite the fact that quantum mechanics plays a critical role in the Kibble balance, this article focuses on classical physics: the electromechanical transducer. Hence, equation~(\ref{eq:mass}) suffices to understand our considerations.

The Kibble balance would not work without a magnet system. Its importance is visible in the individual measurements made in weighing, $(Bl)_\mathrm{w}$, and velocity mode, $(Bl)_\mathrm{v}$. In the final equation, i.e., the Kibble equation (\ref{eq:mass}), however,  the geometric factor drops out, and it seems the result is independent of $Bl$. So, why spend energy and effort designing a perfect magnet system? Because, as we shall see in the next paragraph, the $Bl$ does matter.

According to equation (\ref{eq:mass}), the obtained value for the mass, $m$, is given by five measurements. To find the lowest possible uncertainty for a Kibble balance experiment, a simple uncertainty propagation is performed. The relative uncertainty in the mass, neglecting correlations, is given by
\begin{equation}
\left(\frac{\sigma_\mathrm{m}}{m}\right)^2=\left(\frac{\sigma_\mathrm{v}}{v}\right)^2+\left(\frac{\sigma_\mathrm{g}}{g}\right)^2+\left(\frac{\sigma_\mathrm{R}}{R}\right)^2+\left(\frac{\sigma_\mathrm{U}}{U}\right)^2+\left(\frac{\sigma_\mathrm{V}}{V}\right)^2,
\label{eq:uncertainty1}
\end{equation}
where $\sigma_\mathrm{X}$ denotes the absolute uncertainty in the measurement of quantity $X$. It is reasonable to assume the same uncertainties for the two voltage measurements, $\sigma_\mathrm{U} = \sigma_\mathrm{V}$. Following the derivation in \cite{schlamminger2013design}, we replace $U$ with $Bl v$ and $V$ with $mgR/(Bl)$ and obtain
\begin{eqnarray}
\left(\frac{\sigma_\mathrm{m}}{m}\right)^2&=&\left(\frac{\sigma_\mathrm{v}}{v}\right)^2+\left(\frac{\sigma_\mathrm{g}}{g}\right)^2+\left(\frac{\sigma_\mathrm{R}}{R}\right)^2\nonumber\\
&&+\left(\frac{1}{Bl}\right)^2\left(\frac{\sigma_\mathrm{U}}{v}\right)^2+(Bl)^2\left(\frac{\sigma_\mathrm{U}}{mgR}\right)^2.
\label{eq:uncertainty}
\end{eqnarray}
The first three terms in the sum on the right are independent of $Bl$. The fourth term is inversely proportional to $(Bl)^2$ and the last term proportional to $(Bl)^2$. 

To minimize the relative uncertainty in the mass, one has to maximize $Bl$ according to the fourth term and minimize $Bl$ per the fifth term. This dilemma provides the designer with an opportunity. There must be an optimal $Bl$ that minimizes the relative uncertainty of the mass. It is given by,
\begin{equation}
(Bl)_\mathrm{op}=\sqrt{\frac{mgR}{v}}.
\label{eq:Bl_op}
\end{equation}

\begin{figure}
	\centering
	\includegraphics[width=0.7\textwidth]{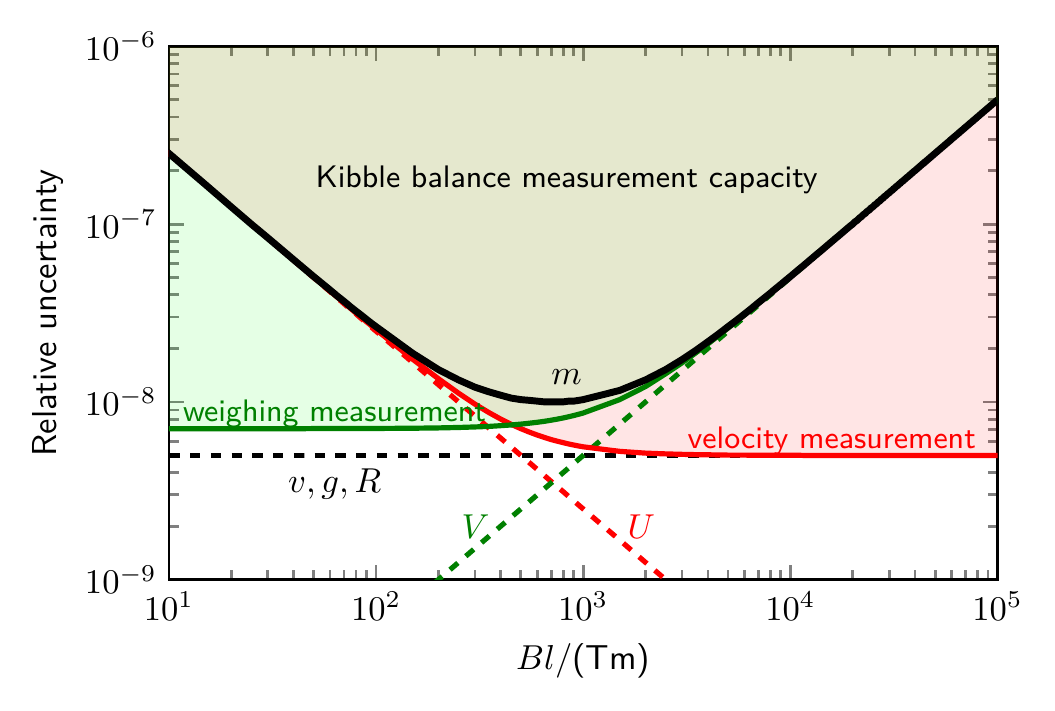}
	\caption{The relative  measurement uncertainty for mass as a function of the chosen $Bl$, according to equation~(\ref{eq:uncertainty}). In this example, $\sigma_\mathrm{v}/v=\sigma_\mathrm{g}/g=\sigma_\mathrm{R}/R=5\times10^{-9}$, $\sigma_\mathrm{U}=5$\,nV, $v=2$\,mm/s, $mgR=1000\,\Omega$N. The smallest relative uncertainty of $1\times10^{-8}$  is obtained at $(Bl)_\mathrm{op}\approx700$\,Tm \cite{schlamminger2013design}.}
	\label{fig:uncertainty_Bl}
\end{figure}

Figure~\ref{fig:uncertainty_Bl} shows the relative uncertainty of the mass as a function of $Bl$. The graph is obtained for typical parameters of a Kibble balance, $m=1$\,kg, $R=100\,\Omega$, and $v=2\,$mm/s. The uncertainties are assumed to be $\sigma_\mathrm{v}/v=\sigma_\mathrm{g}/g=\sigma_\mathrm{R}/R=5\times10^{-9}$, $\sigma_\mathrm{U}=5$\,nV. For these parameters, the minimum is achieved at $(Bl)_\mathrm{op}\approx\SI{700}{\tesla\meter}$. Note that $(Bl)_\mathrm{op}$ only depends on the parameters and not their uncertainties. 
Furthermore, $(Bl)_\mathrm{op}$ is proportional to the square root of the product of weight and resistance, $\sqrt{mgR}$. 
One can achieve the same performance by scaling both quantities inversely to each other. Using the same magnet system with $m=\SI{100}{\gram}$ and $R=\SI{1}{\kilo\ohm}$ will produce the same relative uncertainty as if it were used with $m=\SI{1}{\kilo\gram}$ and $R=\SI{100}{\ohm}$. This scaling is only true for the magnet system. The weighing system may have a different requirement for smaller and larger masses.

Since $Bl$ is a product, the magnet designer can fix one factor and adjust the other to achieve the desired value. But what is the best strategy? Should the designer increase  $B$ or $l$ to obtain the highest possible value? Or is a trade-off the best strategy? A quantity to consider for this decision is the resistive loss in the coil in the weighing mode $P=R_\mathrm{c} I^2$, with  $R_\mathrm{c}$ denoting the coil resistance. With the mass of test mass $m$, the  optimal $Bl$ and $\rho$, the wire resistance per unit length, the electrical loss can be written in two ways
\begin{equation}
P= \frac{\rho m^2g^2}{(Bl)_\mathrm{op}^2} \times  l= \frac{\rho m^2g^2}{(Bl)_\mathrm{op}}\times \frac{1}{B}.
\label{eq:coil_heating}
\end{equation}
Both formulas have a factor that only depends on the product $(Bl)_\mathrm{op}$, $\rho$, and $m$. But the second-factor changes with the free parameter. The equations show that a large flux density, and hence a small $l$, is desired because this scenario leads to a decrease in the power dissipation in the wire. The opposite is true for increasing the length of the wire. It would increase the electrical power dissipated in the coil. Consequently, it's best to make $B$ as large as possible and therefore $l$ as small as possible.

We described how to choose a value for $Bl$ and how to divide it up to $B$ and $l$, but still unclear is the magnetic flux source.  While this topic is discussed in greater detail in section~\ref{sec2}, a brief overview is given here. 

Through the history of the Kibble balance, the source of the magnetic field evolved.
The predecessor of the Kibble balance, the Ampere balance \cite{ab1,ab2}, used air-cored coils to produce the magnetic flux. These coils were wound with copper wire, and not superconductor wire.  We refer to such coils as conventional coils, as opposed to super-conducting coils. The Ampere balance was used to realize the unit of electrical current from 1948 to 1990.  Because of the similarity to the Kibble balance, the magnet system of the Ampere balance can be analyzed within the same framework, see section  \ref{sec:ConvCoil}. 
The conventional coils cannot produce a strong magnetic field. The magnetic flux density at the measurement position was only a few mT, which limited the weighing capacity to a few grams. Although the wire length $l$ or excitation current can be increased to reach a higher force. In both cases, the self-heating increases which yields to larger uncertainty components caused by adverse effect of the heating, consistent with equation~(\ref{eq:coil_heating}).

Later, different magnetic systems, such as permanent magnet systems \cite{npl1,NPL,metas1}, superconducting coils \cite{nist3}, yoke-based electromagnet \cite{zhang2015coils}  were designed to increase the magnetic field $B$ and, at the same time, reduce the ohmic dissipation. During this time, researchers developed concepts for improving the field profile, e.g. \cite{BIPMmag2006,LNEmag,NISTmag,METAS}. After more than a decade of iterations, the designs finally converged on the yoke-based permanent magnet system \cite{steiner2012history, Stephan16}. The success of the yoke-based permanent magnet system relies on two advantages over other designs. First, the yoke can provide a well-defined boundary condition in both radial and azimuthal directions. In this case, the magnetic field design is greatly simplified from a complex three-dimensional problem to a one-dimensional optimization. Second, the permanent magnet system does not contain components that must be powered. Instead, the rare-earth magnetic material is magnetized during production and remains magnetized for the lifespan of the experiment. Hence, such a system provides the magnetic flux with reduced complexity (no power supplies needed) and maintenance cost. The different types of magnet systems are reviewed in more detail in section \ref{sec2}.

Yoke-based permanent magnets supply an intense, uniform, and stable magnetic field for Kibble balance measurement. Hence, these days they are the workhorse for Kibble balances. Various designs for magnet systems exist, and each magnet system has different design parameters that can be optimized. Section~\ref{sec3} discusses the most important considerations for the magnet designer. Among the topics discussed are the selection of the dimensions and the materials for the magnet system. The reader can find tips regarding the assembly, manufacturing, and final evaluation of the magnet system. 

Sometimes, the field profile of the assembled magnet is not satisfactory, and adjustments must be made. Section~\ref{sec4} discusses how to shim the magnet to achieve the desired profile. It further details different techniques to characterize the magnet system.

Even a perfect magnet will have contributions to the uncertainty budget of the Kibble balance. There will always be imperfections. The magnetic field will never be perfectly symmetric. Furthermore, magnetic materials are inherently nonlinear, and nonlinear effects can bias the measurement. The text in section \ref{sec5} discusses the major magnetic systematic concerns for different measurement schemes.

%% file: 02_BasicReview.tex
\section{Brief  review of the physics of magnetic fields}
	\label{secnew2}

This section aims to make the reader familiar with the terminology used for the design of static magnetic fields. We introduce the symbols and show basic formulas typically used in textbooks about the subject. A reader that is familiar with the topic may skip this section.

\subsection{The magnetic flux density}

The most used quantity in the context of the article is the magnetic flux density $B$. It is a vectorial quantity $\vecB$. If only the scalar is printed, the magnitude of the vector is indicated $B=|\vec{B}|$. The magnetic flux density is a source-free vector field. That means the field lines have neither beginning nor end. They are closed. Maxwell's second equation describes this fact. It says that the divergence of the magnetic flux density is 0,  $\vec{\nabla} \cdot \vecB=0$. 

In Cartesian coordinates, \vecB\ has the components $B_\mathrm{x}, B_\mathrm{y}$, and $B_\mathrm{z}$ along the three coordinate vectors $\vec{e}_\mathrm{x},\vec{e}_\mathrm{y},$ and  $\vec{e}_\mathrm{z}$. However since most coils are wound on a circular former it is often more convenient to use cylindrical coordinates, $\vecB=B_\mathrm{r} \vec{e}_\mathrm{r}+B_\varphi \vec{e}_\varphi +B_\mathrm{z} \vec{e}_\mathrm{z}$, where $r$, $\varphi$ and $z$ are the cylindrical coordinates, and $\vec{e}_\mathrm{r}$, $\vec{e}_\varphi$, $\vec{e}_\mathrm{z}$ are the corresponding unit vectors. 

In cylindrical coordinates, the divergence of \vecB\ is given by
\begin{equation}
\vec{\nabla}\cdot \vecB = \frac{1}{r}\frac{\partial (rB_\mathrm{r})}{\partial r}+\frac{1}{r}\frac{\partial B_\varphi}{\partial \varphi} +\frac{\partial B_\mathrm{z}}{\partial z} =0.
\end{equation}

This result leads to an important corollary. If the flux density has cylindrical symmetry, that is, it is independent of the azimuth $\varphi$, then
\begin{equation}
\frac{B_\mathrm{r}}{r} + \frac{\partial B_\mathrm{r}}{\partial r}= -\frac{\partial B_\mathrm{z}}{\partial z}.
\end{equation}
If, furthermore, the radial component of the field is inverse proportional to $r$, that is, $B_\mathrm{r} =B_\mathrm{c} r_\mathrm{c}/r$, then $\partial B_\mathrm{z}/\partial z=0$. Here, $B_\mathrm{c}$ is the radial field at the mean radius of the coil $r_\mathrm{c}$, $B_\mathrm{r}(r_\mathrm{c})=B_\mathrm{c}$. For such a field, the vertical component of $B$ does not change with $z$.

\subsection{The magnetic field}

The magnetic field is abbreviated with $\vec{H}$. In a vacuum,  $\vec{H}$ is except a factor, named the vacuum permeability, the same as $\vecB$. It is $\vec{H}=\mu_0^{-1}\vec{B}$. Inside matter, however, the two vectorial quantities differ. The magnetization of the material decreases the magnetic field, $\vec{H}=\mu_0^{-1}\vec{B}-\vec{M}$. Inside a permanent magnet,  for example Samarium-Cobalt, the $\vec{B}$ and $\vec{H}$ are in opposite directions. The $H$ field is an important quantity to analyze magnetic circuits. Integrating $\vec{H}$ along a closed path yields
\begin{equation}
\oint \vec{H}\cdot\mbox{d}\vec{l} = I_\mathrm{t},
\label{eq:sec2:oh}
\end{equation}
where $I_\mathrm{t}$ is the total current flowing through the surface enclosed by the path. Equation~(\ref{eq:sec2:oh}) makes it easy to remember that the unit of $H$ is A/m. A typical case here is the analysis of a magnetic circuit without current in the coil. In this case, the enclosed integral evaluates to zero.

\begin{figure}
	\centering
	\includegraphics[width=0.5\textwidth]{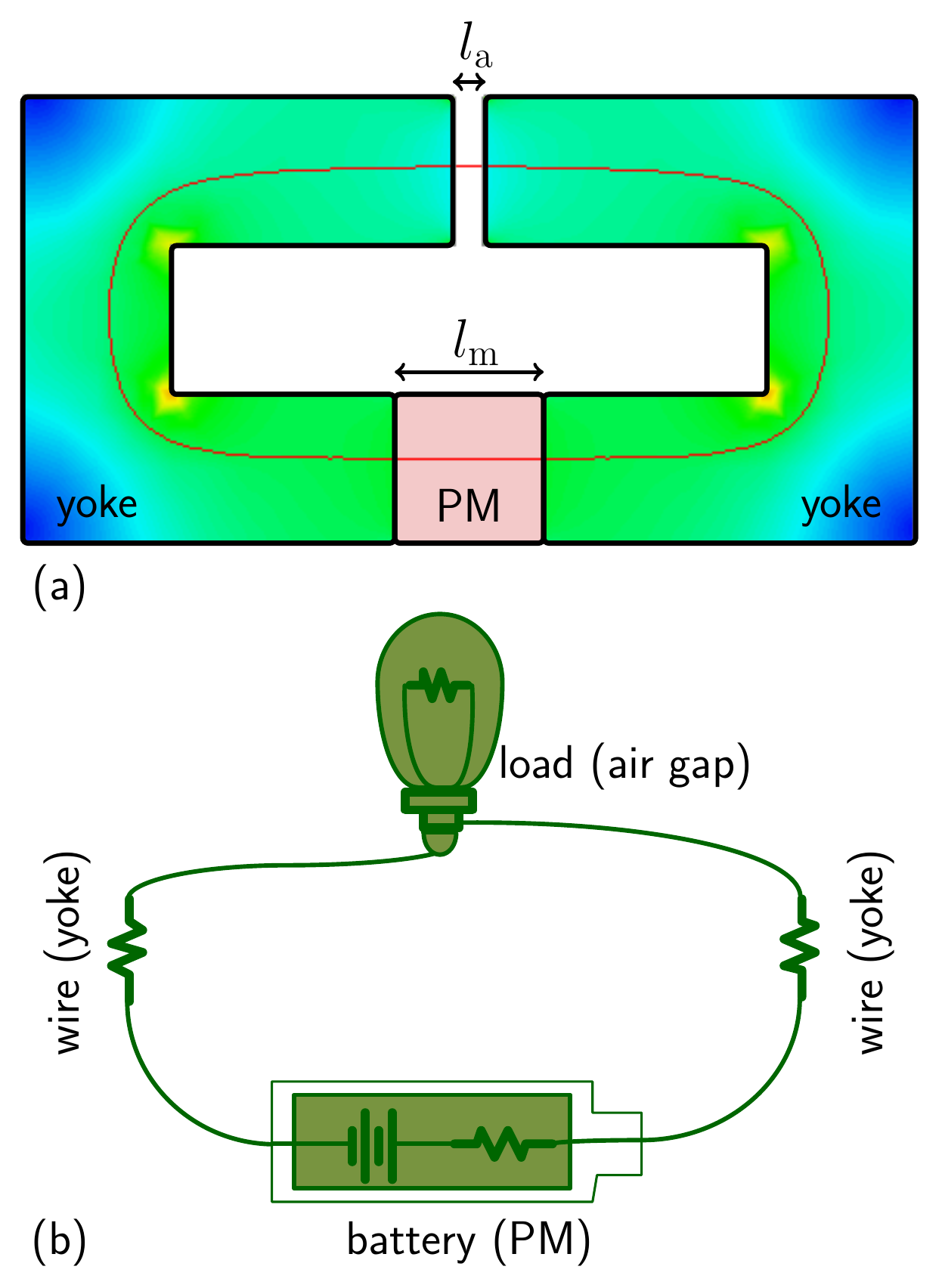}
	\caption{(a) is a simple magnetic circuit with a permanent magnet (PM) and an air gap. The red line denotes a major magnetic flux line through the circuit. (b) shows an equivalent electrical circuit for understanding the magnetic circuit, where the permanent magnet, yoke, and air gap are turned into the battery (with an internal resistance), wire resistance, and load reluctance, respectively. }
	\label{fig:sec2:simplecircuit}
\end{figure}

\subsection{The magnetic circuit}
\label{magcircuit}

A very simple magnetic circuit is shown in figure~\ref{fig:sec2:simplecircuit}(a). It shows a permanent magnet of width $l_\mathrm{m}$, an iron yoke, and an air gap of width $l_\mathrm{a}$. The red line is the closed contour over which the integral in equation~(\ref{eq:sec2:oh}) is calculated. There is no current enclosed inside the contour and no external magnetic flux is considered, hence, the integral must evaluate to zero. We can split the path along the contour into three regions, the permanent magnet, the iron yoke (length $l_\mathrm{y}$), and the air gap. In the air gap $\vec{H}=\mu_0^{-1}\vec{B}$. In the iron,  $\vec{H}=(\mu_0\mu_\mathrm{r})^{-1}\vec{B}$ ($\mu_\mathrm{r}$ is the yoke relative permeability) and finally, in the magnet, we have $H_\mathrm{m}$. The complete integral is
\begin{equation}
H_\mathrm{m} l_\mathrm{m} + \frac{B_\mathrm{y}}{\mu_0\mu_\mathrm{r}} l_\mathrm{y}  + \frac{B_\mathrm{a}}{\mu_0} l_\mathrm{a} =0.
\label{eq:sec2:Hm}
\end{equation}
If the cross-sectional areas in the air gap and the yoke and the magnet are the same, a single symbol $S$ is enough to denote this area. In this case,  $B$ is identical in the magnet, the yoke, and the air gap, because the flux $ \Phi=BS$ is conserved. Note, for simplicity we ignore fringe fields here. It is $B=B_\mathrm{y}=B_\mathrm{a}=B_\mathrm{m}$. Equation~(\ref{eq:sec2:Hm}) can now be easily solved for $B$, it is
\begin{equation}
B = \frac{-H_\mathrm{m} l_\mathrm{m}}{\displaystyle \frac{\ds l_\mathrm{y}}{\mu_0\mu_\mathrm{r} }+\frac{\ds l_\mathrm{a}}{\mu_0}}.
\label{eq:sec2:B}
\end{equation}

\begin{figure}
	\centering
	\includegraphics[width=0.7\textwidth]{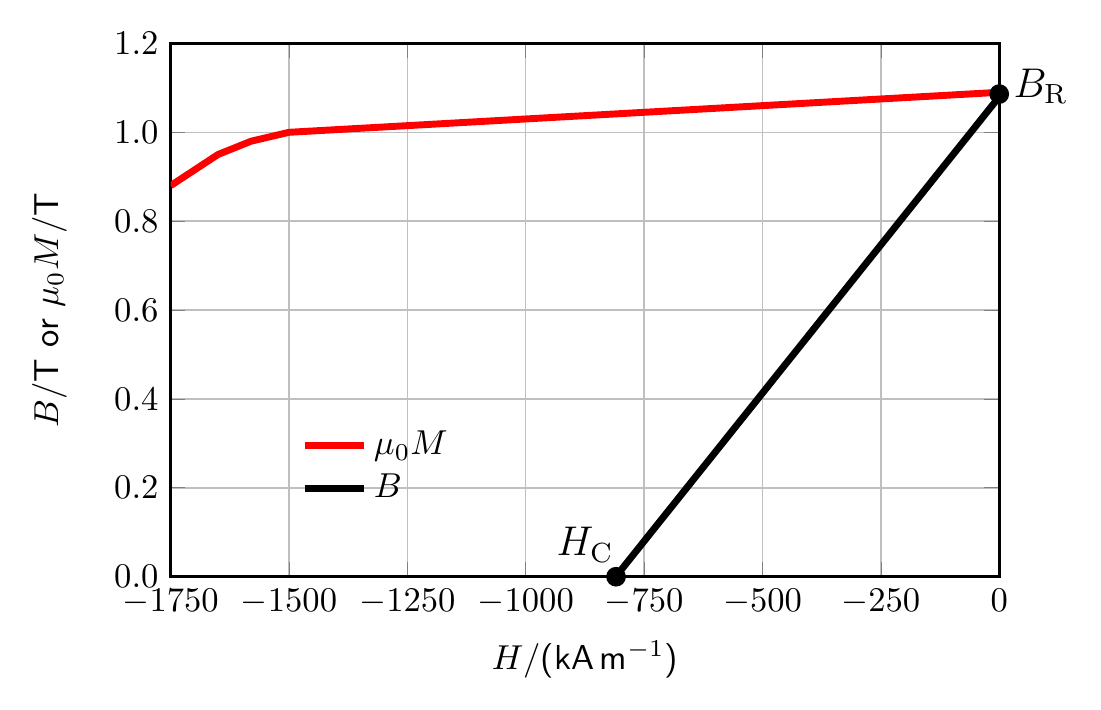}
	\caption{Measured demagnetization curve, i.e., the part of the magnetic hysteresis that is in the second quadrant, for a SmCo sample \cite{NISTmag}. The sample was measured at \SI{26}{\celsius}.
	$B=\mu_0 H +\mu_0 M$. Since the magnetization changes by less then 10\% for $H$ ranging from \SI{-800}{\kilo\ampere \per \meter} to \SI{00}{\ampere \per \meter}, the magnetic flux density is almost a linear function of $H$ in this range.}
	\label{fig:sec2:SmCoHyst}
\end{figure}

The next question we would like to investigate is: What is the magnetomotive force of permanent magnet material? Can it be doubled by doubling the length of the magnet? As we shall see, it is not that simple. As the black curve in figure \ref{fig:sec2:SmCoHyst} shows, commonly used modern magnet materials, e.g., Samarium Cobalt (SmCo) and Neodymium-Iron-Boron (NdFeB), show a linear behavior in the second quadrant (positive $B$, negative $H$). The magnetic flux is given by 
\begin{equation}
   B=\mu_\mathrm{m}\mu_0 H +B_\mathrm{R} ~~\mbox{or}~~B=\mu_\mathrm{m}\mu_0(H-H_\mathrm{C}),
\end{equation}
where $\mu_\mathrm{m}$ is the relative permeability of the permanent magnet, $B_\mathrm{R}$ its remanenence, which is the magnetic flux  in the absence of $H$, and $H_\mathrm{C}$ the coercivity. Note that $B_\mathrm{R}=-\mu_\mathrm{m}\mu_0 H_\mathrm{C}$. Applying 
\begin{equation}
H_\mathrm{m} = \frac{B}{\mu_\mathrm{m}\mu_0} -\frac{B_\mathrm{R}}{\mu_\mathrm{m}\mu_0} =\frac{B}{\mu_\mathrm{m}\mu_0}+H_\mathrm{C}
\end{equation}
into equation~(\ref{eq:sec2:Hm}) and using $\Phi=SB$ yields,
\begin{equation}
\Phi\left(\frac{l_\mathrm{m}}{S\mu_\mathrm{m}\mu_0}+\frac{l_\mathrm{y}}{S\mu_0\mu_\mathrm{r}}  + \frac{l_\mathrm{a}}{S\mu_0}\right) = \frac{B_\mathrm{R}}{\mu_\mathrm{m}\mu_0}l_\mathrm{m}=-H_\mathrm{C}l_\mathrm{m}.
\label{eq:sec2:Ohmslaw}
\end{equation}

Besides using Ampere's law, Ohm's law in magnetism can be used to understand the magnetic circuit and derive (\ref{eq:sec2:Ohmslaw}). Similar to Ohm's law, $I=U/R$, the magnetic version is as
\begin{equation}
\Phi =  \frac{\mathcal{F}} {\mathcal{R}},
\label{eq:sec2:Flux}
\end{equation}
where $\mathcal{R}$ is the magnetic reluctance of the circuit, $\mathcal{F}$ the magnetomotive force (MMF). The magnetomotive force corresponds to the voltage (electromotive force, EMF), the flux to the current, and the reluctance to the resistance in the original law by Ohm. The MMF is supplied by the permanent magnet and is given by $\mathcal{F}=-H_Cl_m$. While the reluctances of three components form a serial circuit and add $\mathcal{R} =  R_\mathrm{a}+R_\mathrm{y} + R_\mathrm{m}$. The individual reluctances of  the air gap, the yoke and the permanent magnet, are
\begin{equation}
R_\mathrm{a} = \frac{l_\mathrm{a}}{S \mu_0},\;R_\mathrm{y} = \frac{l_\mathrm{y}}{S \mu_0\mu_\mathrm{r}},\;R_\mathrm{m} = \frac{l_\mathrm{m}}{S \mu_0\mu_\mathrm{m}},
\label{eq:sec2:R}
\end{equation}
respectively.
Equation (\ref{eq:sec2:Ohmslaw}) is obtained by replacing $\mathcal{R}$ by the sum, of the components in  equation (\ref{eq:sec2:R}) and $\mathcal{F}$ by $-H_Cl_m$. 

The following points are helpful when using Ohm's law to design or analyze a magnet system:

\begin{enumerate}
    \item The reluctance of the yoke is very low because the relative permeability of iron, $\mu_\mathrm{r}$ is very high, of order \SI{1e3} or even larger. So, $R_\mathrm{y}<<R_\mathrm{a}$ and $R_\mathrm{y}<<R_\mathrm{m}$, and hence $\mathcal{R}\approx R_\mathrm{a}+R_\mathrm{m}$. The yoke in the magnetic circuit plays a similar role to the wire in the electric circuit. It guides the flux with very low reluctance, just as the wire in an ideal electrical circuit is thought to have negligible resistance.

    \item  The permanent magnet corresponds to the voltage source (battery) as shown in figure \ref{fig:sec2:simplecircuit}(b). The MMF supplied is $\mathcal{F}=-H_\mathrm{C}l_\mathrm{m}$. However, the magnet comes with its own reluctance $R_\mathrm{m}$, similar to the internal resistance in a voltage source.
    
    \item With the last two items, the effect of doubling the length of the active magnetic material on the flux can be analyzed. the flux is
    \begin{equation}
    \Phi'=\frac{2\mathcal{F}}{2R_\mathrm{m}+R_\mathrm{a}}<2\frac{\mathcal{F}}{R_\mathrm{m}+R_\mathrm{a}}= 2\Phi.
    \end{equation}
    A longer permanent magnet increases the MMF, but also the total reluctance. Hence the flux increase is smaller than proportional to the length. In the limit $R_\mathrm{a}<<R_\mathrm{m}$, the magnetic flux remains constant and does not change with increasing $l_\mathrm{m}$.    
        
    \item The load in the electric circuit corresponds to the air gap in the magnetic circuit. Typically, it has the largest reluctance in the circuit, given by $R_\mathrm{a}$. The ratio of the reluctances is proportional to the length ratios (assuming identical area),
    \begin{equation}
    \frac{R_\mathrm{m}}{R_\mathrm{a}}\approx\frac{l_\mathrm{m}}{l_\mathrm{a}},
    \end{equation}
    because the relative permeabilities of rare earth magnets are close to one.  While for some electrical circuits, the internal resistance of the source can be neglected, the same is not true for magnet systems.
\end{enumerate}

\subsection{The magnetic reluctance}

Above, we have already used  the equation for the reluctance of an element with a  cross-sectional area $S_\mathrm{x}$, flux path length (thickness) of $l_\mathrm{x}$, and a relative magnetic permeability $\mu_\mathrm{x}$ is
\begin{equation}
R_\mathrm{x} =\frac{l_\mathrm{x}}{\mu_0\mu_\mathrm{x} S_\mathrm{x}}.
\label{eq:ra}
\end{equation}
Equation (\ref{eq:ra}) has a similar form as the resistance of a conductive block with the same geometrical parameters ($l_\mathrm{x}, S_\mathrm{x}$), i.e. $R=\rho \,{l_\mathrm{x}}/{S_\mathrm{x}}$. To obtaine the equation for the magnetic reluctance one has to replace the resistivity $\rho$ by the permeability $1/(\mu_\mathrm{x}\mu_0)$. 

Designing a magnet for a Kibble balance, very often one has to work with cylindrical gaps, inner radius $r_\mathrm{i}$, outer radius $r_\mathrm{o}$ and height $h_\mathrm{a}$. For such a gap, the reluctance is
\begin{equation}
R_\mathrm{a} =\frac{1}{\mu_0\mu_\mathrm{a} h_\mathrm{a} 2\pi} \ln{\frac{r_\mathrm{o}}{r_\mathrm{i}}}\approx \frac{\delta}{\mu_0\mu_\mathrm{a} h_\mathrm{a} 2\pi r_\mathrm{i}},
\end{equation}
where the approximation is a Taylor expansion of the natural logarithm for $r_\mathrm{o}=r_\mathrm{i}+\delta$ and $\delta<<r_\mathrm{i}$.

\subsection{The magnetic force required to split the magnet}
In some magnet systems, the coil is surrounded by the yoke. In other words, the air gap is inside the magnet with a few access holes that allow the coil suspension and laser beams to penetrate. The advantage of such an internal air gap is that the coil is shielded from fluctuating environmental magnetic fields. The disadvantage is that the magnet needs to be taken apart to insert the coil. This process is known as ``splitting the magnet.''

\begin{figure}
	\centering
	\includegraphics[width=0.7\textwidth]{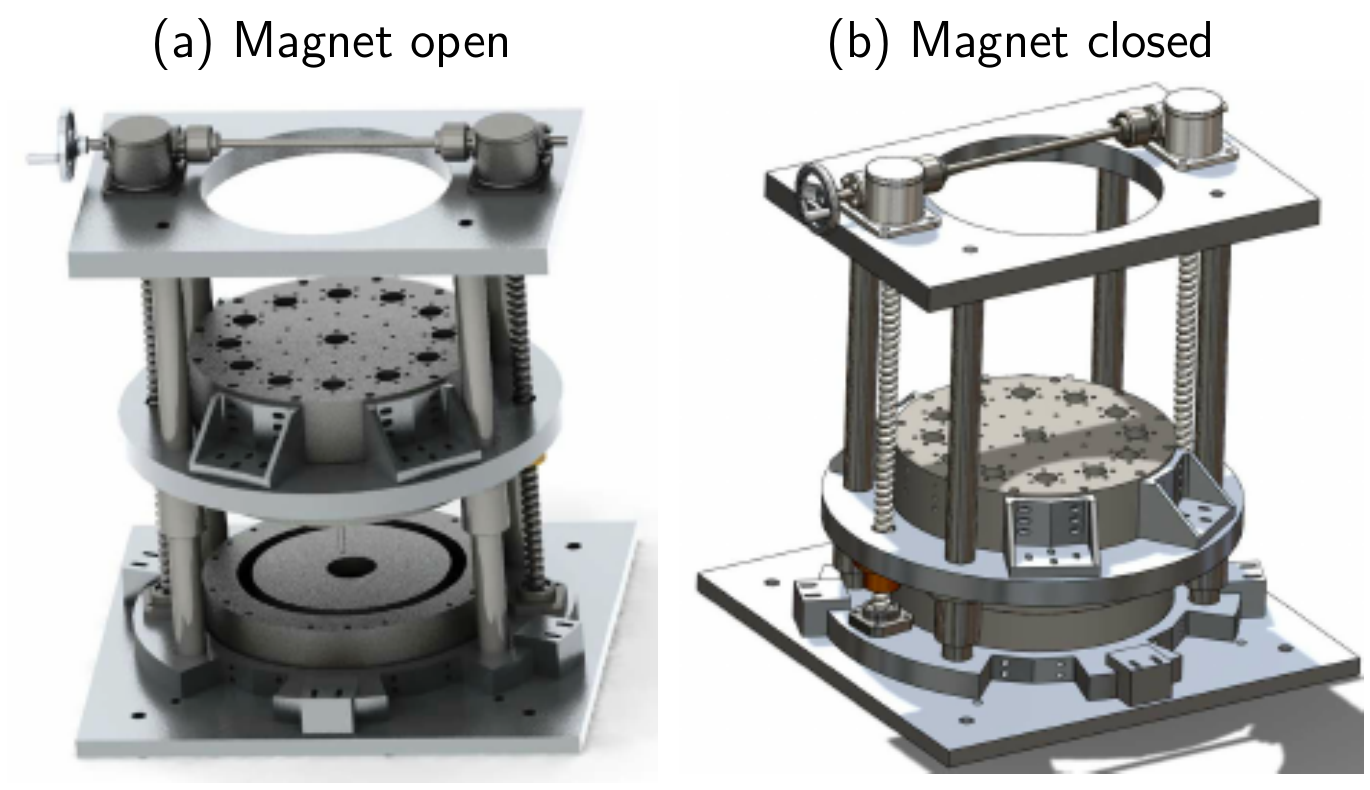}
	\caption{An example of a magnet splitter. (a) and (b) show the open and close status, respectively. Reproduced from \cite{NISTmag}.}
	\label{fig:sec2:splitter}
\end{figure}

An important parameter to design a magnet splitter is the size of the force  that is required to split the magnet. Figure~\ref{fig:sec2:splitter} shows a rendering of such a device that is needed to open the magnet.

The Maxwell stress tensor provides a simple method to calculate the force that acts on an object in a given space~\cite{Griffiths_1990}. The force is given by the surface integral,
\begin{equation}
\mathbf{F} = \oint \mathbf{T}\cdot \mathrm{d}\mathbf{S}.
\label{eq:Fsurf}
\end{equation}
The nine components of the  Maxwell stress tensor $\mathbf{T}$ are given by
\begin{equation}
T_{ij} = \frac{1}{\mu_0} \left(B_i B_j - \frac{1}{2} \delta_{ij} |B|^2\right),
\end{equation}
where $i$ and $j$ indicate the three directions of the Cartesian coordinates, $x$, $y$, and $z$, or a permutation depending on how the problem is set up. The symbol $\delta_{ij}$ denotes the Kronecker delta, which is  $\delta_{ij}=1$ for $i=j$ and   $\delta_{i,j}=0$  for $i\neq j$.

An example of a force calculation using eq.~(\ref{eq:Fsurf}) is given  in section~\ref{sec5:force}.

%% file: 03_DifferentMagSystems.tex
\section{Evolution of different magnet systems}
\label{sec2}

Before discussing the short historical evolution of magnet systems in Kibble balance, we would like to put forward three generally accepted properties that these magnet systems should have.

\begin{enumerate}
    \item  The magnet system shall provide a large and uniform magnetic flux density throughout the coil at the weighing position and in the volume that the coil traverses in velocity mode.
    \item The total magnetic flux penetrating the coil and its gradient shall be independent of external (environmental) and internal factors, most importantly, the coil current.
    \item Manufacturing, operation, and maintenance shall be simple and if possible, economical.
\end{enumerate}
	
Today, when Kibble's idea is almost half a century old, the thinking on the magnet system has clarified enough that these three points may sound trivial. Historically, however, that has not always been the case. As is shown below, researchers were reluctant to introduce iron to the magnet system out of worry that the nonlinear effects may compromise Kibble's idea. For the remainder of the text, we will use the three points above to evaluate various types of magnet systems.

\subsection{Conventional coil system}
\label{sec:ConvCoil}

Long before the Kibble balance, a different type of electrical balance was used in metrology to define the unit of current, the ampere. In the international system of units that was valid until 20th May 2019, the ampere was defined as the constant current that would produce a force of \SI{2e-7}{\newton \per \meter} between two straight parallel conductors placed one meter apart. In the formal definition, these conductors have negligible cross-section and extend to infinity. This definition links the only electrical unit in the SI to the mechanical unit via the force between two current-carrying wires. The practical realization of the unit of current was carried out with an Ampere balance, sometimes also referred to as current balance or magnetometer ~\cite{steiner2012history}.

In the Ampere balance, the force between a fixed and a movable coil connected to a balance was measured~\cite{ab1,ab2}. The electromagnetic force between the two coils can be written as
	\begin{equation}
	F=\frac{\partial M}{\partial z}I_\mathrm{F}I=(Bl)_\mathrm{w}I,
	\end{equation} 
where $I_\mathrm{F}$ and $I$ are the currents through the fixed and movable coils, respectively. Here, $M$  is the mutual inductance of two coils and  $\partial M/\partial z$ the gradient of $M$ along the vertical direction $z$. Note, $(\partial M/\partial z)I_\mathrm{F}$ is identical to the geometric factor $Bl$.
	
\begin{figure}
		\centering
		\includegraphics[width=0.7\textwidth]{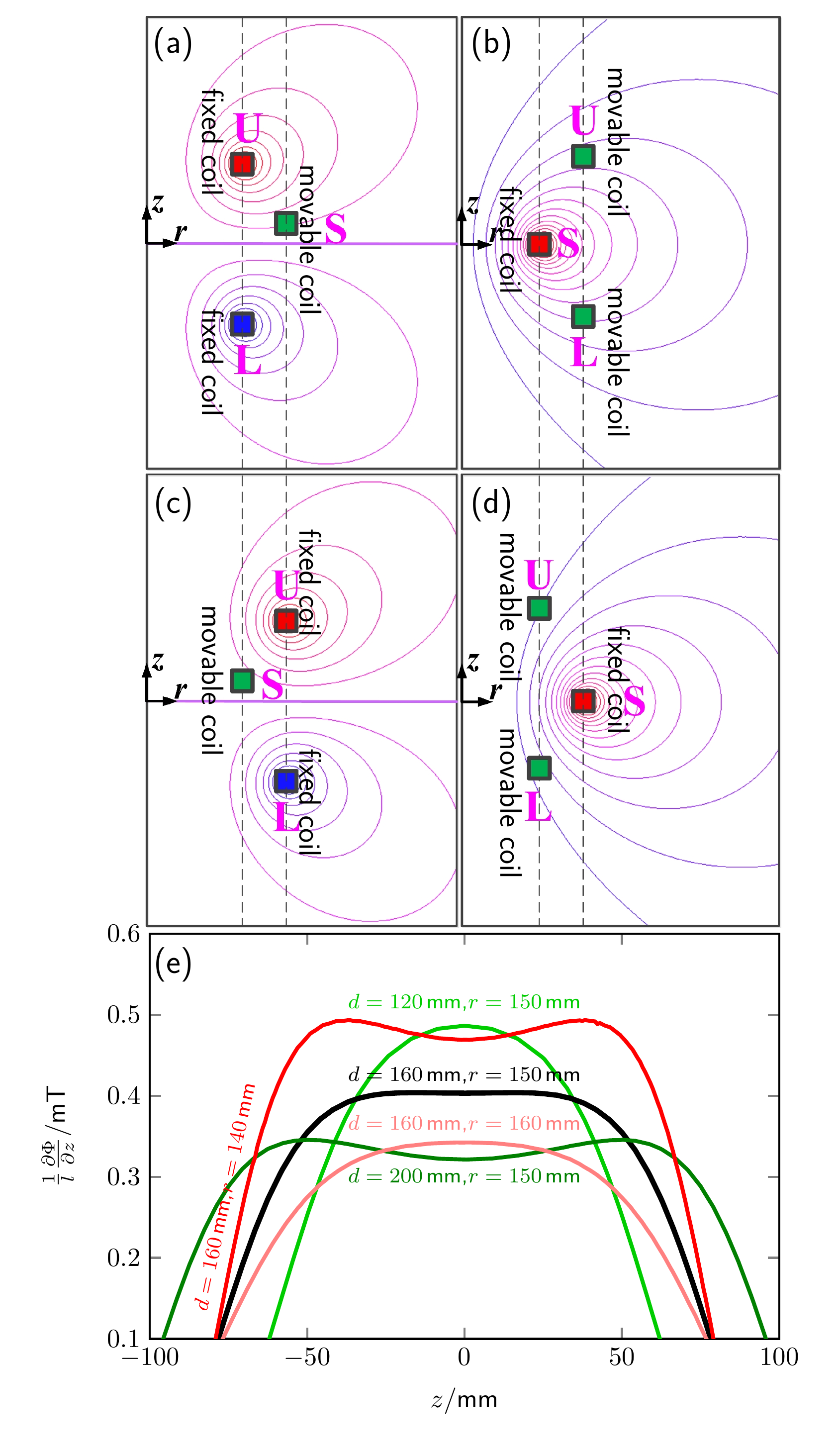}
		\caption{Panels (a)-(d) show four different coils that can be used in the Ampere balances. The inner coil radius and outer coil radius are respectively $r_0$ and $r$. The distance between the coils that form a pair is $d$. Panel (e) shows the magnetic field produced by the fixed coil arrangement with ampere-turns of each fixed segment $NI_\mathrm{F}=200\,$A. In this case, $r_0=100$\,mm. Note the small magnitude of the produced flux density.}
		\label{fig:coil_system}
	\end{figure}

The four panels in figure~\ref{fig:coil_system} show typical coil configurations used in Ampere balances. Each configuration requires three coils. The difference is whether one coil or two coils are stationary and, correspondingly, two coils or one coil are moving. The coils in the pair whether they are moving or not,  have identical parameters (diameter and number of turns) but are connected in serial opposition. In the left column of figure \ref{fig:coil_system} the fixed coil assembly is the coil pair, and in the right column, it is the single coil.
The second choice is which coil assembly has a smaller radius. In the top row of figure~\ref{fig:coil_system} the fixed coil assembly is on the inside (smaller radius), whereas in the second row, it is on the outside (larger radius). Interestingly, as long as the inner radii, outer radii, and coil separation do not change, all four configurations produce the same $Bl$, shown in the last row of figure~\ref{fig:coil_system}. The fact that the four configurations produce the same $Bl$ can be seen by writing the mutual inductance as a sum of the inductances between the single-coil (S) and two other individual coils (upper U and lower L), i.e. 
$M(z)=M_\mathrm{SU}(z)-M_\mathrm{SL}(z)$. 
If the mutual induction of one inner and one outer coil as a function of vertical separation is given by $M_\mathrm{1}(z)$, then $M_\mathrm{SU}(z)=M_{1}(d/2-z)$ and  $M_\mathrm{SL}(z)=-M_{1}(d/2+z)$. Hence, 
$M(z)=M_{1}(d/2-z) + M_{1}(d/2+z)$, and most importantly  $M(z)=M(-z)$. 

One merit of these coil systems is that the field gradient is zero at the symmetry plane, $z=0$, since $\left.\partial ^2M/\partial z^2\right|_\mathrm{z}=-\left.\partial ^2M/\partial z^2\right|_\mathrm{-z}$, it follows that   $\left.\partial ^2M/\partial z^2\right|_\mathrm{z=0}=0$.  Typically, $z=0$ is chosen as the weighing point. Then, the magnetic force $F$ is independent of small variations of the vertical position of the movable coil. The second benefit of this position is that the magnetic flux density is inversely proportional to the radius, $B(r)\propto r^{-1}$.  In an azimuthally symmetric geometry, as is discussed here, $B(r)\propto r^{-1}$ leads to an important consequence. The magnetic flux density is divergence-free, $\vec{\nabla}\cdot  \vec{B}=0$, and hence $\partial B_\mathrm{z}/\partial z = -\frac{1}{r}\partial (r B)/\partial r$. The term to the right of the equal sign is identical to zero for $B(r)\propto r^{-1}$, and, therefore, $\partial B_\mathrm{z}/\partial z=0$. So, no magnetic flux is threading through the coil. This condition is true for the entire plane where $B(r)\propto r^{-1}$, in this case, $z=0$. The flux through the coil is zero independent of coil radius and horizontal position.  That means, in weighing, the result is to first-order independent of the precise horizontal position and the coil radius~\cite{ab1,li2016discussion}. The latter can change slightly due to ohmic coil heating. The $Bl$ conservation of a $r^{-1}$ field is further detailed in \ref{sec:AppendixB}. In summary, taking advantage of the symmetry at $z=0$ makes the measurement less susceptible to small deviations from the ideal system.

A magnet system employed for Kibble balance measurement should produce a flat field region along $z$ so that when the coil moves with constant velocity, the induced voltage stays stable. For current-carrying coil systems, the easiest way to obtain a flat $B(z)$ profile is to adjust the separation of the double coil $d$ and the horizontal distance of fixed and movable coils, see e.g. \cite{campbell1907standard,driscoll1942absolute,zhang2011recent}. In figure \ref{fig:coil_system} (e), we take an example to show the magnetic profile distributions with different combinations of $d$ and outer coil radius $r$ (the inner coil radius is fixed at $r_0=100$\,mm). It can be seen by either adjusting $d$ with a fixed $r$ or the opposite (changing $r$ when $d$ is fixed), a flat magnetic profile (in this case, $d=160$\,mm, $r=150$\,mm) can be achieved. 
	
Figure \ref{fig:coil_system}(e) shows that the magnetic field produced is weak, below 1\,mT, even with comparably large  ampere-turns $NI_\mathrm{F}=200$\,A. From the uncertainty relationship shown in figure \ref{fig:uncertainty_Bl}, the measurement error for the induced voltage is considerable at small $Bl$ values. However, choosing a longer wire $l$ to increase the $Bl$ value will also enlarge the wire resistance and the ohmic heating. In summary, the weak field that is produced by conventional coils is a major drawback. And, hence, these systems are no longer in use for Kibble balances.
	
\subsection{Multi-coil magnet system}
	
\begin{figure*}
		\centering
		\includegraphics[width=1.2\textwidth]{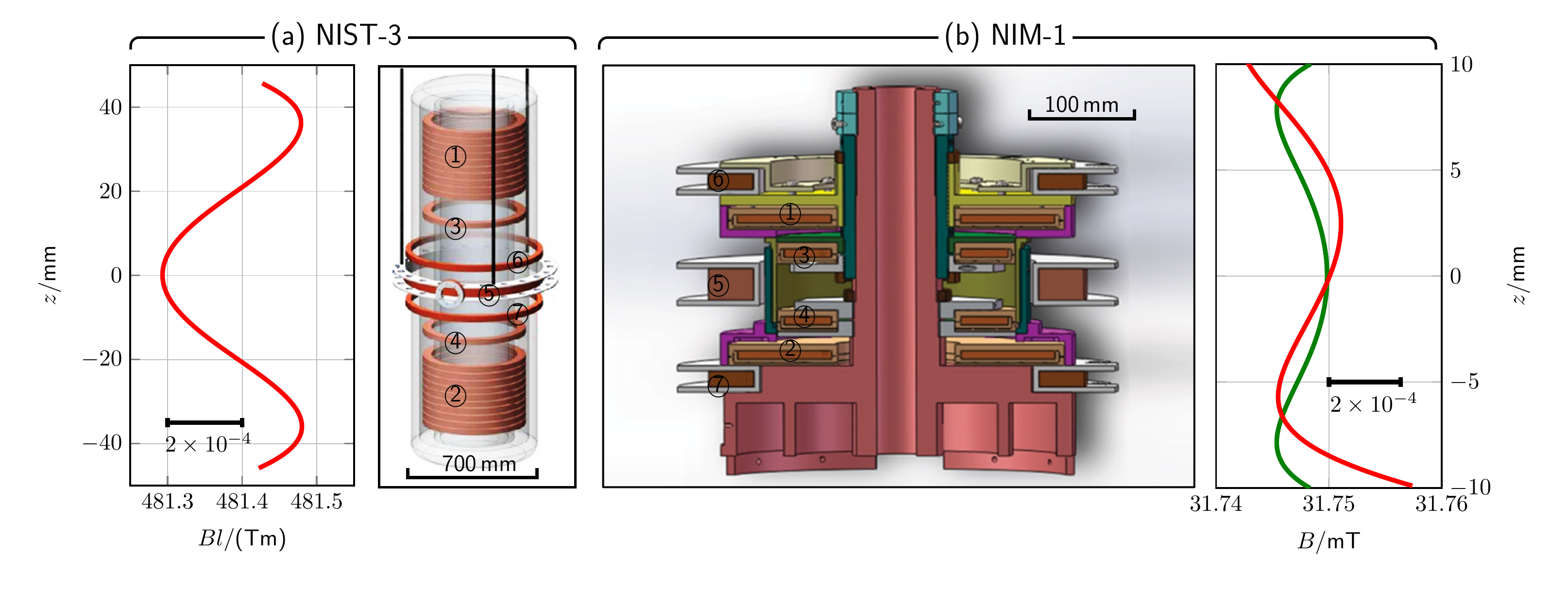}
		\caption{(a) The NIST-3 superconducting magnetic system. The left plot shows a measurement of the magnetic profile in the velocity phase (reproduced from \cite{steiner2005details}), and the right plot is the spatial arrangement of superconducting solenoids. (b) The permanent-magnet-only system was used in the Joule balance experiment. The left shows the construction of the system, and the right presents the designed (green) and achieved (red) magnetic profiles. Reproduced from \cite{xu2016determination}. The labeled components in both systems are: \textcircled{1}--upper main solenoid/magnet ring, \textcircled{2}--lower main solenoid/magnet ring, \textcircled{3}--upper trim solenoid/magnet ring, \textcircled{4}--lower trim solenoid/magnet ring, \textcircled{5}--main movable coil, \textcircled{6}-- upper fixed compensation coil, \textcircled{7}-- lower fixed compensation coil.
		}
		\label{fig:nist-3}
	\end{figure*}

Ohmic heating in the field generating coils and its adverse effect can be eliminated by using superconducting wires.  Researchers at the National Institute of Standards and Technology (NIST, USA) developed a superconducting coil system for the third-generation Kibble balance experiment (NIST-3)  \cite{steiner1997nist,steiner2005details}. The NIST-3 superconducting magnet is shown in figure \ref{fig:nist-3} (a). Two groups of superconducting coils were employed to produce the magnetic field for the measurement. The main solenoids produced a magnetic profile similar to the conventional coil system but with a much larger manetic flux density (sub-Tesla level). Thanks to T.P.~Olsen \cite{ab1}, a pair of trim solenoids were used to compensate for the first order ($z^2$) non-linearities of the main solenoids. Compared to systems shown in figure \ref{fig:coil_system}, this double-layer design allows a quasi-realization of $1/r$ field in a much wider range along $z$. Figure \ref{fig:nist-3} (a) presents a typical NIST-3 velocity measurement result. As is seen, the magnetic profile  changes only by a few parts in 10$^4$ over about 100\,mm $z$ travel. 

Another novel idea implemented in NIST-3 system is that its induction coil consists of two individual coils. One is movable and connected to the balance. The other is fixed in space. In fact, the fixed coil itself consists of two coils that formed together with a virtual coil with the same number of turns as the moving coil. By connecting the moving and the virtually fixed coil in series opposition, the common electromagnetic noise, canceled \cite{NISTmag}, improving significantly the signal-to-noise ratio. The idea is similar to a humbucker in an electric guitar. The double movable coil shown in figure \ref{fig:coil_system} can achieve a similar feature. 
The NIST-3 superconducting magnet was a successful system. It met the magnetic requirements for Kibble balance measurement and produced one of the most precise results for determining the Planck constant at the time \cite{steiner1997nist,nist3,schlamminger2015summary}. One major shortcoming of the superconducting system is the complexity of the operation. On needs a stable current control for the solenoids and liquid helium to reach the transition temperature for the superconductor. For NIST-3 about 250\,L of liquid Helium were necessary for a week of operation. The second problem is the lack of a defined and stable metrological surface. Typically, in velocity mode, the velocity of the coil with respect to the magnet needs to be measured. Very often, that measurement is performed interferometrically with a surface of the magnet providing a mounting surface for the reference arm. A superconducting coil, however, does not offer easy access to a defined surface. The plane of interest, the magnetic center of the superconducting coil, is immersed in liquid helium. A possible surface would be the top of the Dewar, but the stability from that surface to the magnetic center of the coil is not great. For example, vibration, magnetostrictive forces, and thermal expansion due to a change in Helium level in the Dewar can affect the distance between the top of the Dewar and the center of the coil. 
	
A second attempt to improve the field strength and reduce the ohmic heating for the coil system was undertaken by researchers at the  National Institute of Metrology (NIM, China)  for the Joule balance experiment \cite{xu2016determination}. The idea is to replace the field generating coils with permanent magnets yielding two advantages: 1) the ohmic heating of the field generating coils is removed, and 2) a stronger magnetic field is created. The construction of the NIM-1 magnet system and the magnetic profile are shown in figure \ref{fig:nist-3} (b). A flat magnetic profile of about 30\,mT over 1\,cm was obtained. Compared to superconducting coils, the permanent-magnet-only system is simpler and more compact. However, the field strength was several times weaker than that of the NIST-3 system, and to produce a 4.9\,N force (weight of 500\,g mass), the ohmic heating caused by the moving coil of 0.7\,W  was significant. In reality, a large-volume permanent magnet is challenging to manufacture, and hence the rings are usually realized by gluing small pieces together. Typically, the field strength of different parts can vary by as much as 1\%, and the magnetization difference can yield unknown field gradients in open circuits, causing misalignment errors. Besides, the remanence of the permanent magnet has a significant temperature coefficient of $-3\times10^{-4}$/K (SmCo magnet) to $-10^{-3}$/K (NdFeB magnet), without an efficient heat sink, the magnet temperature needs to be well controlled during the measurement.  

The magnet systems described above are open. The magnetic flux is not guided and, therefore, can penetrate the entire room where the Kibble balance is installed. Thus, the following considerations are essential: 1) There will be a vertical field gradient at the mass. As a consequence, a considerable magnetization force occurs when the mass is made from soft magnetic materials, such as stainless steel \cite{davis1995determining}. 2) The magnetic flux density at the coil position can be influenced by iron in its vicinity. Great care has to be taken to avoid iron, and if iron is unavoidable, it has to be mounted such that it does not move with respect to the magnet system. A change of the relative positions may alter the field profile and cause systematic effects. To suppress these effects and, at the same time, further increase the field strength, controlling and aligning the magnetic flux path by introducing soft yokes to the permanent magnet system became inevitable.

\subsection{Flat permanent magnet system}
	
\begin{figure}
		\centering
		\includegraphics[width=0.7\textwidth]{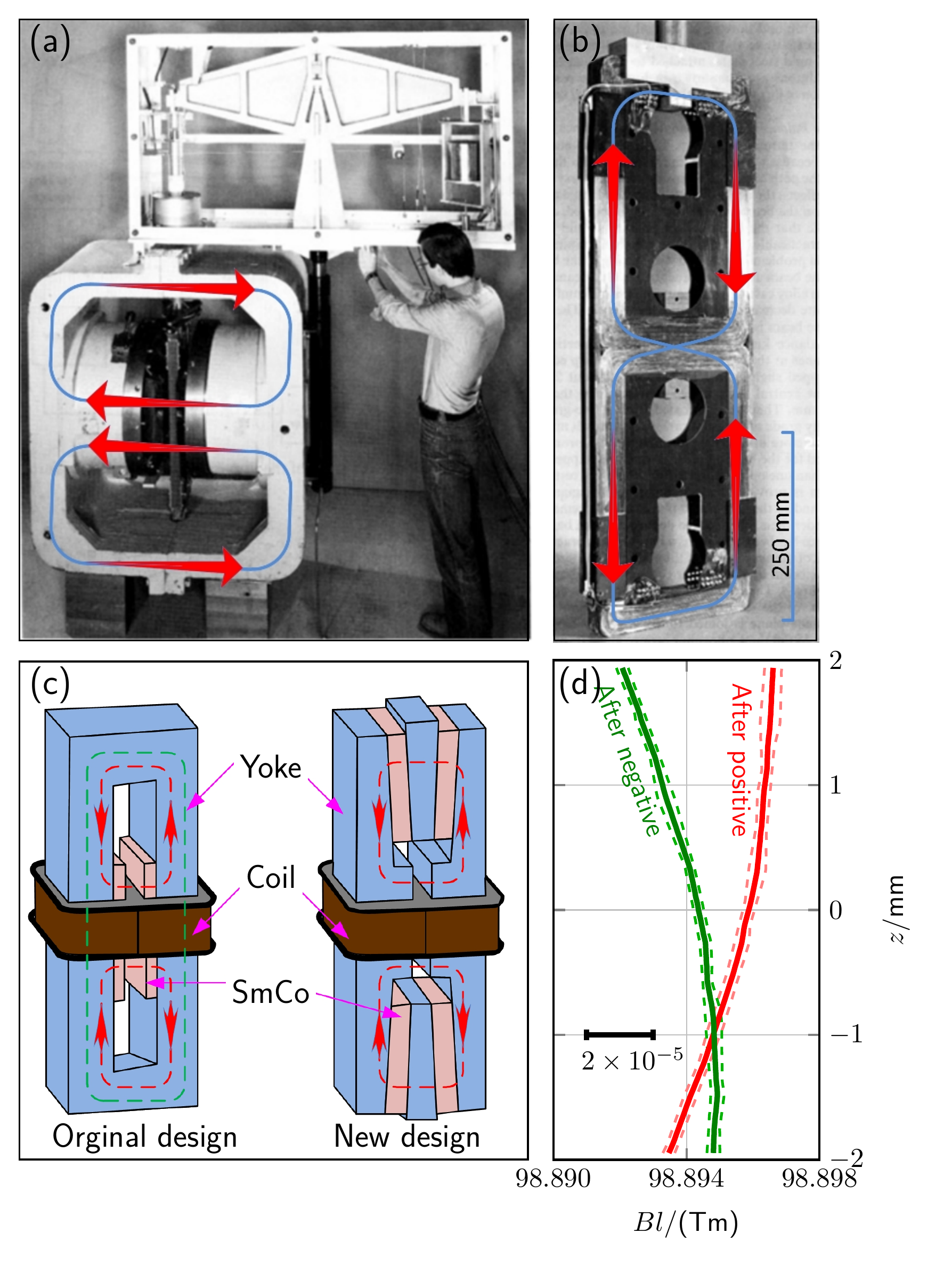}
		\caption{(a) The NPL-Mark I magnet system. The arrows denote the path of magnetic flux. (b) The right plot shows the figure-eight-shaped coil. The arrow is the current flow in the weighing measurement. (a) and (b) are reproduced from \cite{kibble1990realization}. (c) The METAS-Mark I magnet system. The left plot is the original magnetic circuit, and the right is the improved design. (d) shows the hysteresis effect of the original METAS-Mark I magnetic circuit. (c) and (d) are reproduced from \cite{eichenberger2011determination,eichenberger2004new,beer2003status}.}
		\label{fig:NPL-1}
	\end{figure}

The first yoke-based permanent magnet system was employed by the first generation Kibble balance experiment (NPL-Mark I) at the National Physical Laboratory (NPL, UK) \cite{kibble1990realization}. The NPL-Mark I system is shown in figure \ref{fig:NPL-1}(a) and (b). The construction was similar to an air-gapped transformer, but permanent magnet disks created the flux. The magnetic flux was guided horizontally through a 56\,mm width, 0.3\,m$\times$0.3\,m sectional area air gap. The magnetic field in the air gap center was 0.68\,T. A flat magnetic profile was achieved with a figure-eight-shaped coil located vertically in the center of the air gap. The total coil height was larger than the gap height, which ensured that in velocity mode, the magnetic flux through one half of the coil increased while the other half decreased. With symmetry, the difference between upper and lower segments gave a linear change of magnetic flux over $z$.       	
	
The strong magnetic field in the NPL Mark I system was achieved by compressing the flux in a relatively small measurement region. Almost no flux was wasted to the outside of the measurement region. Therefore, the dissipation in the coil during weighing was no longer a limiting factor for the measurement. The magnetic shielding, compared to coil systems, has been improved. The only downside of this design is that only a tiny fraction of the wire length contributes to the force.
The system was massive: the magnet weighs 6000\,kg, and the coil 30\,kg. A large mass can increase the thermal capacity and damp the effects of temperature. However, it is cumbersome to put such an extensive magnet system in a vacuum. Another disadvantage is that the fringe field goes through upper and lower coil segments. Hence, a large part of the fringe field is a common mode in the velocity and the force measurement. 	
The first generation Kibble balance experiment (METAS Mark I) at the Federal Institute of Metrology (METAS, Switzerland) employed a magnetic circuit that is similar to the one in NPL's Mark I. The original design is shown in the left plot of figure \ref{fig:NPL-1}(c). The magnetic circuit principle was the same as NPL Mark I, and a magnetic flux density in the 7\,mm width air gap, of about 0.5\,T, was achieved \cite{beer2003status}. The main difference was that the '8' shape coil was arranged horizontally through the air gap. Note that this setup leaves a closed yoke loop shown as the green dashed line in figure \ref{fig:NPL-1}(c). Ideally, with the same ampere-turns of two segments of the '8' shape movable coil, the total magnetic flux through the closed loop is zero. However, the asymmetry during the weighing measurement, e.g., a non-synchronization of loading or removing the coil current, can considerably shift the yoke $BH$ status and introduce a magnetic hysteresis during the mass-on and mass-off measurement loop. Figure \ref{fig:NPL-1}(d) presents a typical profile measurement after different current polarities \cite{beer2003status}. It shows the hysteresis effect was at the order of $10^{-5}$, which became the major limitation for further improving the overall measurement accuracy. 

Later, the METAS Mark I magnet system was redesigned to address the hysteresis issue. As shown in figure \ref{fig:NPL-1}(b), the permanent magnets (SmCo) were removed from the center and inserted into respectively the upper and lower ends of the circuit \cite{eichenberger2004new}. In this new design, the permanent magnets act also as spacers to cut the previously closed yoke loop. With this increase in the magnetic reluctance for the coil flux path, the hysteresis was significantly reduced \cite{eichenberger2011determination}. The design used for METAS Mark I succeeded in realizing a compact design using a one-dimension horizontal magnetic field. Still, the '8' shape coil suffers from a bad active-to-passive coil ratio. Only $\approx$25\% of the coil contributes to the Kibble principle, but all 100\% contribute to the resistive loss in the weighing mode.
	
\subsection{Radial permanent magnet system}
\begin{figure}
	\centering
	\includegraphics[width=0.475\textwidth]{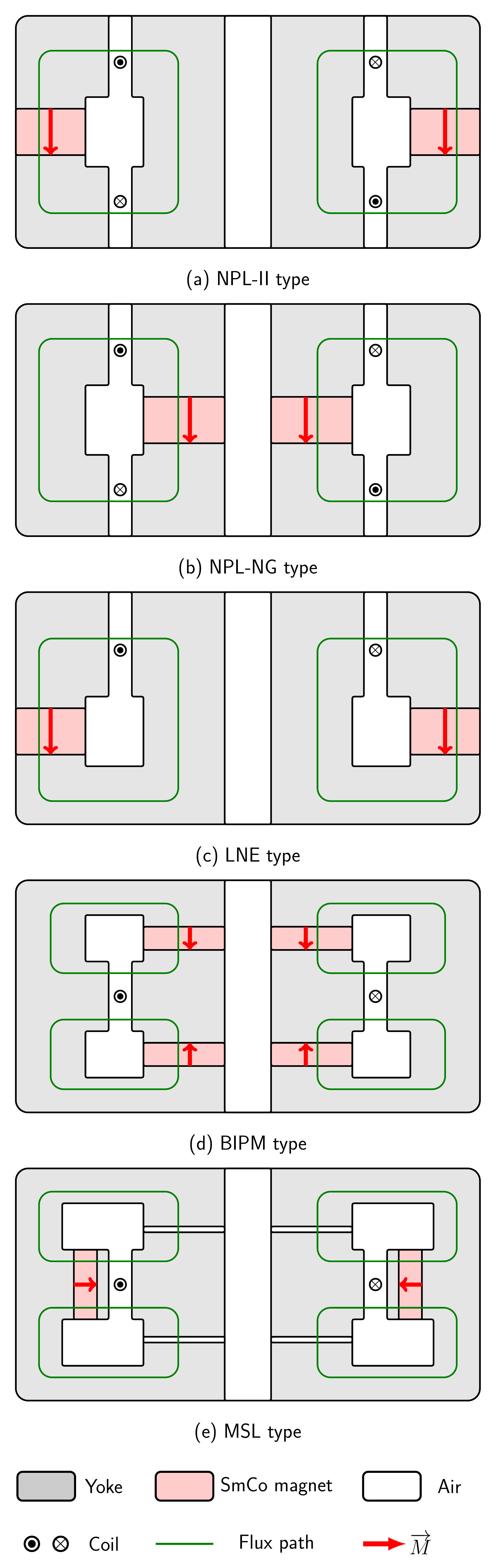}
	\caption{Different radial permanent magnet systems for Kibble balances.}
	\label{fig:NPL-2}
\end{figure}

As shown in Figure \ref{fig:NPL-2}(a), the second generation Kibble balance at the NPL \cite{Robinson2007An,journals/tim/Robinson09}, known as NPL Mark II, used a radial magnetic system and utilize all the wire in the coil for the Kibble principle. In weighing mode, every piece of wire that has dissipation also produces a force. The active-to-passive coil ratio is one.
This design is the first with a cylindrical air gap. The NPL Mark II design has up-down symmetry, and soft yokes guide the magnetic flux of the permanent magnet ring (SmCo) through the upper and lower air gaps. The movable coil, split into two segments in opposite connection similar to the magnet shown in figure \ref{fig:coil_system}, uses the full wire length to produce an electromagnetic force in the weighing and the induction in the velocity phase. The radial field in the center part of each air gap is close to the $1/r$ field distribution, satisfying Olsen's idea. The splitting of the coil significantly suppresses the common noise and produces a very quiet measurement in the velocity phase \cite{NRC}.  
	
The shielding of the NPL Mark II system has been improved compared to the Mark I system. But still, since the SmCo ring is located at the outer yoke and the air gaps contain open ends on the top/bottom surfaces. Flux leaks out at these locations. After the Mark II apparatus was transferred to the National Research Council (NRC, Canada) in 2009, the NPL group started a new generation Kibble balance experiment \cite{NPL3}, referred to here as the NPL-NG system. The NPL-NG experiment still uses the two-coil design with significant improvement on the magnet shielding: As shown in figure \ref{fig:NPL-2}(b), the permanent magnet ring is located inside the inner yoke. Additional shielding has been considered for the NPL-NG design to cut the coupling between the magnet flux and the external flux. 
	
It is easy to imagine the NPL two-gap design with one gap closed. Closing one gap further compresses the magnetic flux, and yields an increased flux density in the remaining gap.
This idea has been implemented at the Laboratoire National de M\'{e}trologie et d'Essais (LNE, France). The LNE magnet is shown in figure \ref{fig:NPL-2}(c). With a  9\,mm width air gap, an average field in the air gap of 0.95\,T was obtained \cite{LNEmag}. This field strength is the strongest magnetic field used in Kibble balance experiments by far. As a result of the broken up-down symmetry, the theoretical magnetic profile over $z$ in the air gap will be sloped (shown in figure \ref{fig:profile_components}(d)) because the inner flux path has a lower reluctance compared to that of the far-end path. To correct it, fine adjustments, detailed in section \ref{sec4}, are required. 
`	
As shown in figure \ref{fig:NPL-2}(d), in 2006, researchers at the Bureau International des Poids et Mesures (BIPM) proposed a novel permanent magnet circuit design that guides the magnetic flux of two permanent magnets (SmCo) rings through one air gap \cite{BIPMmag2006}. Its construction is equivalent to the symmetrical assembly of two LNE-type magnets with SmCo rings inserted in the inner yoke. This design has three advantages: 1) Soft yokes entirely surround the magnet circuit, and therefore the magnetic shielding is nearly perfect \cite{BIPMmagShielding}. Imperfections in the shielding are created by holes that are required to connect to the coil. 2) Similar to the LNE design, since there is only one gap, the flux density in the gap is high and almost no flux is wasted. 3) Since the geometry is symmetric about $z=0$, so is the profile. Hence at that vertical position, the radial field is proportional to  $1/r$. Due to the symmetry, several systematic errors such as nonlinear magnetic effects \cite{linonlinear,linonlinear2} are reduced. 
	
The attractive force at a horizontal plane where the magnet can be opened for coil installation can be very strong (kN level). Therefore in the BIPM magnet system, the coil should be inserted before the circuit is closed. Accessing the coil is difficult after the magnetic circuit is closed, which may be inconvenient for in-situ coil adjustments. By far, the BIPM type magnet design is the most popular magnetic system applied in worldwide Kibble balance experiments, e.g., \cite{NISTmag,METAS,KRISS,UME,NIM}.

The Kibble balance experiment at the Measurement Standards Laboratory (MSL, New Zealand) employs a magnetic circuit as shown in figure \ref{fig:NPL-2}(e) \cite{MSL}. The permanent magnet is a cylinder with a radial magnetization inserted in the outer yoke pole in the one-gap structure. This design can lower the coil flux coupling around the air gap \cite{li17}. However, similar to the original METAS Mark I design, a low reluctance path exists along the yoke. The addition of two spacers in the inner yoke reduces the magnetic hysteresis. The spacers increase the magnetic reluctance for the main flux path and lower the magnetic field in the measurement gap. 
	
\begin{figure}
	\centering
	\includegraphics[width=0.7\textwidth]{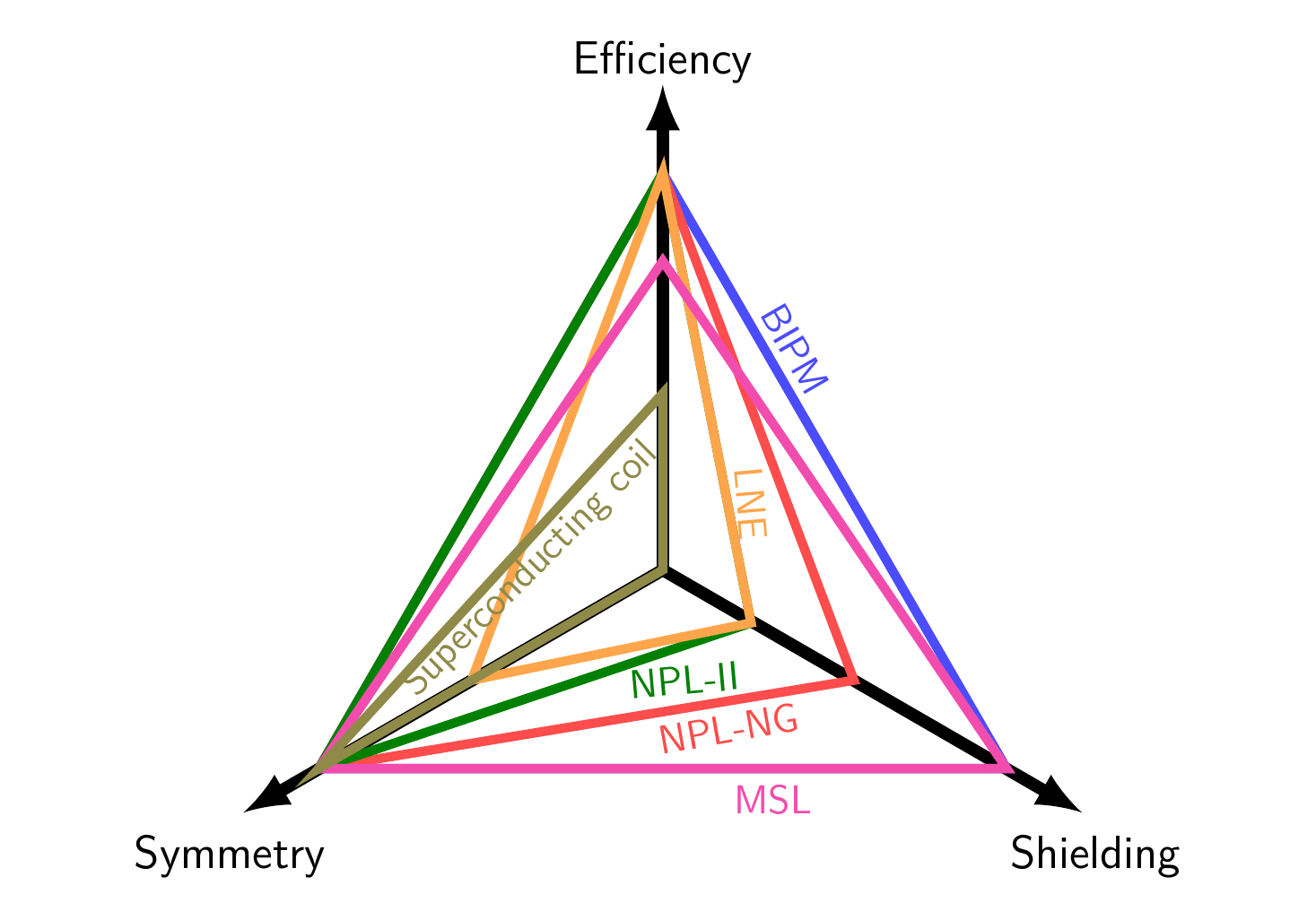}
	\caption{Comparison of different radial magnetic systems along three dimensions: Symmetry, shielding, and efficiency.}
	\label{fig:magcom}
\end{figure}

In summary, yoke-based radial magnetic systems can produce a strong (sub-Tesla), robust and uniform magnetic fields for Kibble balance measurements. In addition to the high field quality, the magnet size is compact, and its operation cost, compared to the superconducting system, is low. Hence, the popularity of these designs in current ongoing Kibble balances. Figure \ref{fig:magcom} compares the performance of different yoke-based radial systems, including the NIST-3 superconducting system. Three features are compared: 1) the efficiency of creating the required magnetic field. 2) the magnetic shielding. 3) the symmetry for Kibble balance measurement. It can be seen that the BIPM-type magnet system has good performances for all three features. We believe the BIPM-type circuit is one of the best Kibble balance magnetic systems, and it will be taken as examples in most cases of the following discussions.  

%% file: 04_DesignMagnets.tex
\section{Design of a permanent magnet}

In this section, we discuss the design of the magnet system in more detail. The equations that were introduced in section~\ref{sec2} are now applied. We start by discussing the material selection. 
Next, we will provide an example calculation of the magnetic flux density in the gap. Then, we show how to calculate the working point of the yoke. After that, we will consider several ways to improve the flatness of the profile. In the subsection that follows that we provide a more detailed analysis of the force that is required to open the magnet and how to reduce this force.  We will end the section with an examination of the thermal properties of the magnet.

\label{sec3}
\subsection{Selecting materials}
\label{sec3:mat}
	
\begin{figure}
\centering
\includegraphics[width=0.7\textwidth]{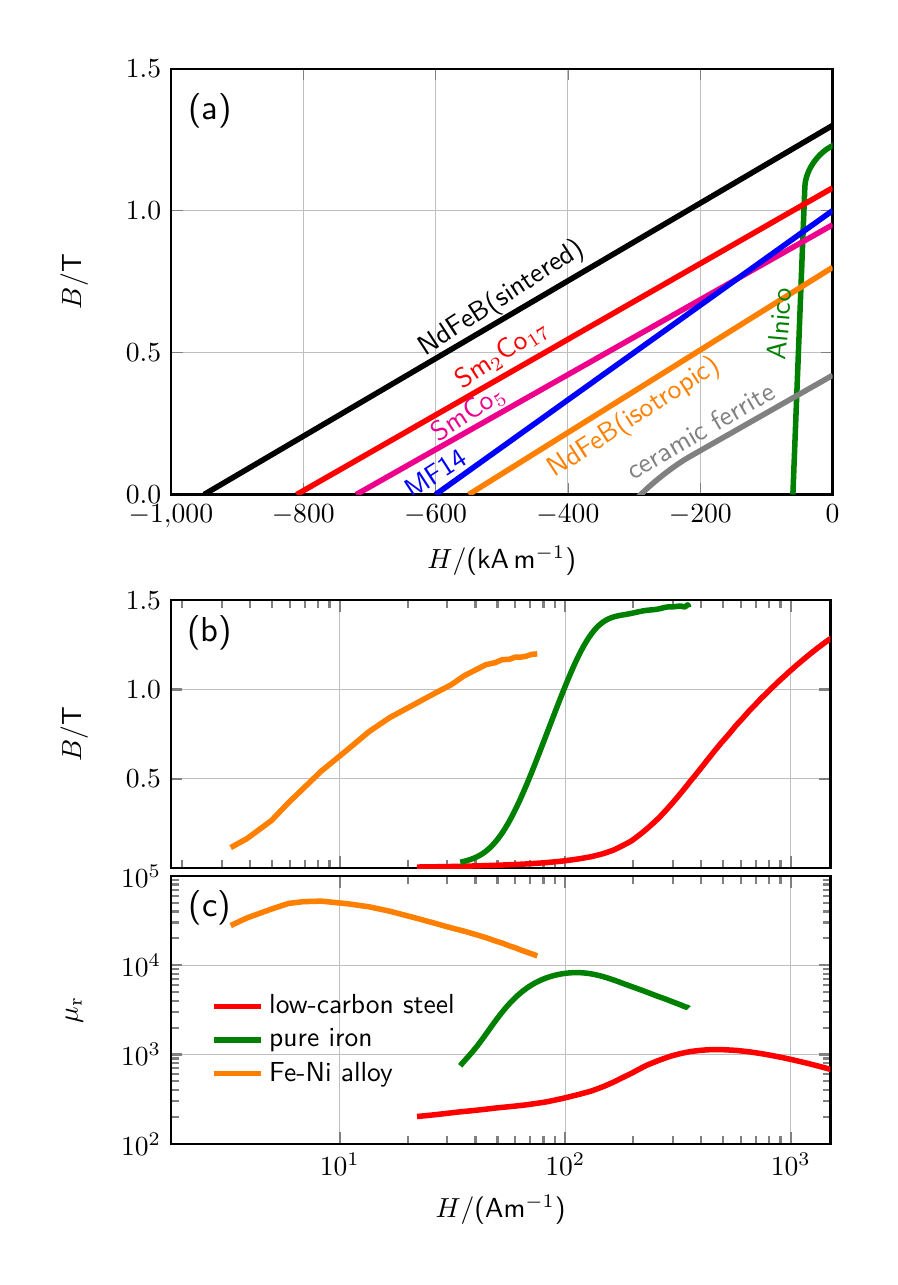}
\caption{(a) Demagnetization curves for some commercially available rare earth magnetic materials along with ferrites and AlNiCo. (b) $B$-$H$ curve and (c) $\mu_\mathrm{r}$-$H$ curve of three typical soft yoke materials. The low-carbon steel, pure iron, and Fe-Ni alloy (50/50) are respectively used for building NIST-4, NIM-2, and BIPM magnets \cite{NISTmag,zhang2015coils,hysteresis}.}
\label{fig:yokeBH}
\end{figure}

The primary components of the air-gap type magnet are the active magnetic material (rare earth) and the yokes.  For the active magnetic materials, the critical graph is the demagnetization curve. That is the part of the $B$-$H$ relationship, also called the hysteresis curve, in the second quadrant (negative $H$, positive $B$.)  
Figure \ref{fig:yokeBH}(a) shows the demagnetization curves of several commercially available magnet materials. The rare-earth magnet materials have two unique features. (1) their demagnetization curves are almost straight lines. (2) they have a large maximum energy product $(BH)_\mathrm{max}$. The maximum energy product is the largest rectangle with sides parallel to $B$ and $H$ that can be found underneath the magnetization curve.

Since, at the percent level, the demagnetization curve for rare earth magnet materials can be considered linear, only two parameters are required to describe it. The magnetic flux produced by the magnet $B_\mathrm{m}$ as a function of $H_\mathrm{m}$ is given by
\begin{equation}
B_\mathrm{m}=\mu_\mathrm{m}\mu_0(H_\mathrm{m}-H_\mathrm{C}),
\label{eq:pm}
\end{equation}
 where $\mu_\mathrm{m}$ is the relative permeability of the material, and $H_\mathrm{C}$ the coercivity, i.e., the magnetic field required to drive the magnetic flux produced by the magnet to 0. For most of today's magnet materials, $\mu_\mathrm{m}\approx1$. We use this approximation for all calculations below. 
	
A larger $H_\mathrm{C}$ value will create a stronger flux density in the air gap in Kibble balance magnetic circuits (details are discussed in \ref{subsec:B}). 
A large magnetic flux $B$ is desired for Kibble balance magnets according to the considerations in section~\ref{sec1}. Hence, high $H_\mathrm{C}$ materials, such as NdFeB and SmCo, are great candidates for the magnetic material for a Kibble balance magnet.
To date, the highest possible $H_\mathrm{C}$ is achieved with sintered NdFeB magnets. Its $H_\mathrm{C}$ is about 10\,\% larger than the $H_\mathrm{C}$ achieved with $\mbox{Sm}_2\mbox{Co}_{17}$, and, hence produces 10\,\% more magnet flux with the same volume of the permanent magnet material. 
So, it seems NdFeB would be the best material to use in a Kibble balance. However, there is a second parameter that should be considered, the temperature coefficient.

In general, the magnetic flux in the Kibble balance must be as stable as possible for environmental influences. One such influence is temperature. The sensitivity of the magnetic materials to temperature changes is expressed in the temperature coefficient of the magnetic material. It denotes the fractional change of the remanence per one-kelvin change of temperature and is often abbreviated by $\alpha$. For NdFeB, $\alpha\approx-1\times10^{-3}$/K and for $\mbox{Sm}_2\mbox{Co}_{17}$, $\alpha\approx-3\times10^{-4}$/K. Hence, the SmCo is about three times more stable to temperature changes. Most designers prefer the smaller (in absolute, irrespective of the sign, terms, $|\alpha|$) temperature coefficient of SmCo and accept a 10\% smaller remanence. This decision was made even harder with the recent discovery of $(\mbox{Gd,Sm})_2\mbox{Co}_{17}$. There, Gadolinium (Gd) is alloyed with Samarium before sintering it with Cobalt, and the result is a magnetic material with an even smaller temperature coefficient,  $\alpha \approx-1\times10^{-5}$/K. However, using $(\mbox{Gd,Sm})_2\mbox{Co}_{17}$, instead of $\mbox{Sm}_2\mbox{Co}_{17}$, will reduce the magnetic flux by another 20\,\%~\cite{benz1975permanent,marangoni2019magnet}. 

A good proxy to quickly evaluate the temperature sensitivity of any magnetic material if the temperature coefficient is not readily available is the Curie temperature, $T_\mathrm{c}$. At the Curie temperature, the magnet loses all its magnetization. The lower the Curie temperature, the higher the temperature coefficient. For SmCo, $T_\mathrm{c}=\SI{825}{\celsius}$, for NdFeB, $T_\mathrm{c}=\SI{310}{\celsius}$.

Another way to decrease the temperature coefficient of the complete magnet system is to use a shunt. This technique is described in the last part of this section. A lower temperature coefficient is achieved, but also the magnetic flux density at the coil is smaller because some of the flux is diverted from the air gap through the shunt.

Having discussed the magnetic material, it is time to say a few things about the second component in the magnet system, the yoke. The yoke aims to guide the magnetic flux from the permanent magnet to the air gap and back. For this purpose, the reluctance of the yoke has to be small according to equation~(\ref{eq:sec2:Flux}). The reluctance is $R_\mathrm{y}=l_\mathrm{y}/(\mu_0\mu_\mathrm{r} S)$. Hence besides a large cross-sectional area $S$ and a short magnetic path $l_\mathrm{y}$, a large relative permeability $\mu_\mathrm{r}$  is desired.  Including the small reluctance, choosing a material with a large relative permeability has the following three advantages:
\begin{enumerate}
\item  It conducts more flux, and it increases the efficiency of the circuit.
\item  It is easier to engineer the profile in the gap, as the side walls made from high $\mu_\mathrm{r}$ are at more uniform potentials~\cite{BIPMmag2017}.
\item It helps to reduce nonlinear magnetic errors \cite{linonlinear,linonlinear2}, because the weighing current influences the magnetic flux in the air gap to a lesser extent.
\end{enumerate}

Note that although some sheet materials can have very high permeability, such as $\mu$-metal, steel sheet, they are typically not used in Kibble balance magnet systems for two reasons. First, the yoke needs to withstand a typical attraction force at the kN level \cite{NISTmag}. Therefore solid material instead of sheet stock is preferred. Second, while the stack of sheets seems feasible, the tiny air gaps in the stack structure increase the reluctance of the yoke and decrease the uniformity of the flux in the air gap. 

In practice, materials with relative permeabilities of 1000 or more are suitable for yokes in Kibble balance magnets. Figure \ref{fig:yokeBH}(b) and (c) reproduce the $B$-$H$ curve and the permeability of three typical yoke materials, i.e., low-carbon steel, pure iron, and Fe-Ni alloy (50/50), which were used respectively in NIST-4, NIM-2, and BIPM systems \cite{NISTmag,zhang2015coils,hysteresis}. It is recommended to heat-treat the parts after machining. The machining process can lower the permeability, and heat treatment can reverse the loss in permeability \cite{NISTmag,hysteresis}. 

\subsection{Magnetic flux density in the air gap}
\label{subsec:B}
	
In the following paragraphs, we calculate $B$ in the gap using the BIPM type magnet as an example. Other magnetic circuits can be analyzed similarly.

\begin{figure*}[tp!]
\centering
\includegraphics[width=\textwidth]{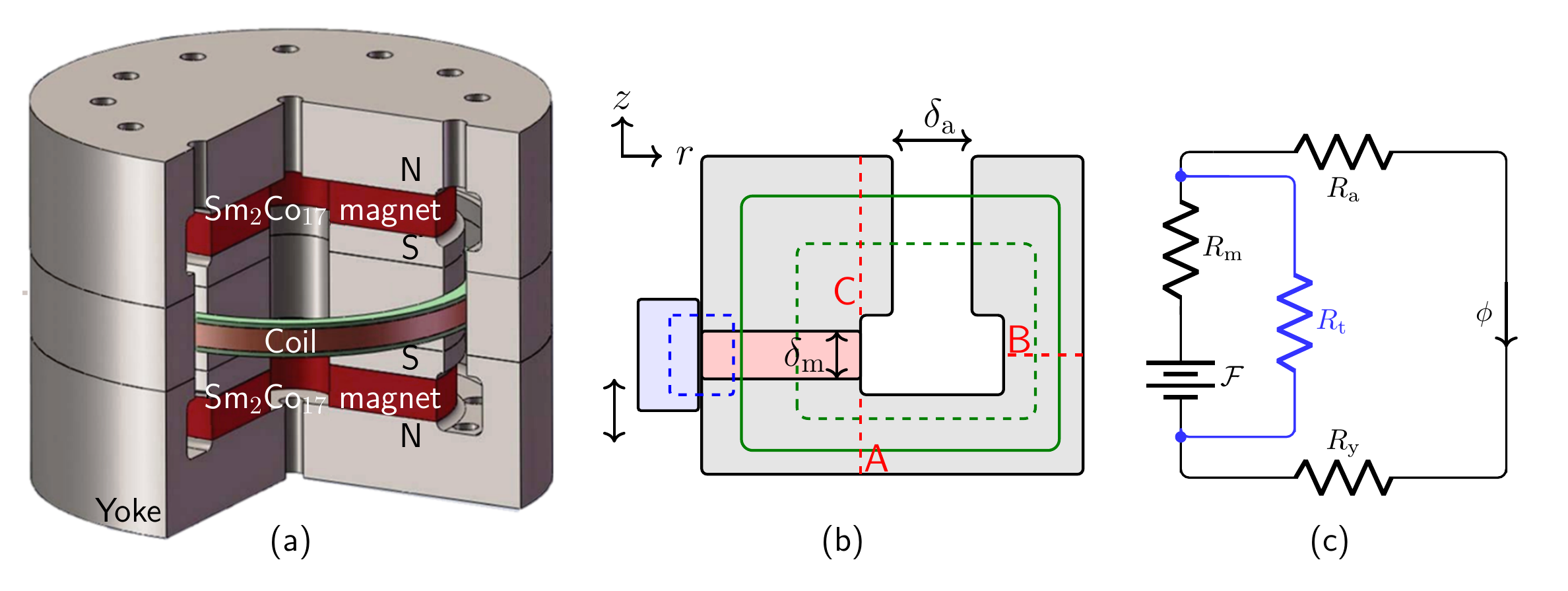}
\caption{(a) shows the 3D construction of a BIPM-type magnetic circuit. (b) The analysis unit for the BIPM-type magnet. (c) The equivalent electrical circuit. The green and dashed green curves denote different main flux paths. The red letters A, B, and C indicate cut-planes at three different locations, see text. }
\label{fig:analysis}
\end{figure*}

The first step in the design process is to find the dependence of the magnetic flux density in the air gap on the dimensions of the system. The symmetry of the magnet system can simplify this process. The number of green flux paths in figure \ref{fig:NPL-2} shows the symmetry of the system. For the BIPM type magnet, only a quarter of the complete circuit needs to be analyzed as indicated in figure \ref{fig:analysis}(a) and (b). As we show in section \ref{magcircuit}, the easiest way to analyze this circuit is to convert it into an equivalent electrical circuit. The permanent ring is providing the  MMF (similar to the voltage source), and the reluctance (resistance in an electrical circuit) of three components, i.e., the permanent magnet, the yoke, and the air gap, is written as
\begin{equation}
R_\mathrm{m}=\frac{\delta_\mathrm{m}}{\mu_\mathrm{m}\mu_0 S_\mathrm{m}},~~R_\mathrm{y}=\frac{\delta_\mathrm{y}}{\mu_\mathrm{y}\mu_0 S_\mathrm{y}},~~R_\mathrm{a}=\frac{\delta_\mathrm{a}}{\mu_0 S_\mathrm{a}},
\label{magR}
\end{equation}
where $\delta_\mathrm{m}$, $\delta_\mathrm{y}$, $\delta_\mathrm{a}$ denote the length; $S_\mathrm{m}$, $S_\mathrm{y}$ and $S_\mathrm{a}$ the cross-sectional areas; and  $\mu_\mathrm{m}$, $\mu_\mathrm{y}$, and $\mu_0$ the relative permeabilities of the permanent magnet, the yoke and the air gap. For now, we assume the flux completely penetrates the air gap, ignoring fringe fields at the upper and lower end of the air gap. It is shown in section \ref{magcircuit} that a permanent ring can be seen as a battery with MMF $\mathcal{F}=-H_\mathrm{C}\delta_\mathrm{m}$ while leaving the space as vacuum (air), i.e. $\mu_\mathrm{m}\approx1$. The three reluctances form a series connection, hence by Ohm’s law, 
\begin{equation}
(R_\mathrm{m}+R_\mathrm{y}+R_\mathrm{a})\phi=-H_\mathrm{C}\delta_\mathrm{m}.
\label{magOhm}
\end{equation}
For a high permeability yoke, $R_\mathrm{y}<<R_\mathrm{a},R_\mathrm{m}$. Without loss in generality, we set $R_\mathrm{y}=0$. The total flux through the air gap and the magnet is the same and can be written as the product of the cross-sectional area and the flux density, i.e.,
\begin{equation}
\phi=S_\mathrm{m}B_\mathrm{m}=B_\mathrm{a}S_\mathrm{a}.
\label{magflux}
\end{equation}
	
Substituting equations (\ref{magR}) and (\ref{magflux}) into (\ref{magOhm}), the magnetic flux density in the air gap can be solved \cite{schlamminger2013design}. It is
\begin{eqnarray}
B_\mathrm{a}&=&\frac{-H_\mathrm{C}}{\displaystyle \frac{S_\mathrm{a}}{S_\mathrm{m}}\frac{1}{\mu_\mathrm{m}\mu_0}+\frac{\delta_\mathrm{a}}{\delta_\mathrm{m}}\frac{1}{\mu_0}}\nonumber\\
&\approx&\frac{-\mu_0H_\mathrm{C}}{\displaystyle \frac{S_\mathrm{a}}{S_\mathrm{m}}+\frac{\delta_\mathrm{a}}{\delta_\mathrm{m}}}\approx\frac{\SI{1}{\tesla}}{\displaystyle \frac{S_\mathrm{a}}{S_\mathrm{m}}+\frac{\delta_\mathrm{a}}{\delta_\mathrm{m}}}.
\end{eqnarray}

Note that the last result is for Sm$_2$Co$_{17}$ magnets, for which $\mu_0 H_\mathrm{C}\approx-\SI{1}{\tesla}$ \cite{NISTmag}. It can be seen the air gap magnetic field strength $B_\mathrm{a}$ is determined mainly by two ratios, $S_\mathrm{a}/S_\mathrm{m}$ and $\delta_\mathrm{a}/\delta_\mathrm{m}$. 
As mentioned above, the fringe fields at the edge of the air gap have been neglected. It can be taken into account by multiplying $S_\mathrm{a}$ with a geometrical factor $\gamma>1$. This mathematical trick pretends that the air gap is taller than it actually is. More on this topic and how to reduce the fringe field can be found in section~\ref{sec:sec4:fringe}.

\subsection{Magnetic working point of the yoke}
	
Two conditions are desired for the yoke. First, the yoke should not be saturated at any point. Second, the average yoke permeability should be high.
	
One can investigate the first condition by examining the cross-sectional area of the yoke along the flux path.  In figure \ref{fig:analysis}(b), three sectional planes are indicated by the letters A, B, and C. The cross-sectional areas are $S_\mathrm{A}$, $S_\mathrm{B}$, and $S_\mathrm{C}$, respectively.
Since flux is conserved the flux density in one area, here for example in region B, is given by
\begin{equation}
B_\mathrm{B} = B_\mathrm{a} \frac{S_\mathrm{a}}{S_\mathrm{B}}.
\end{equation}
The area $S_\mathrm{B}$ needs to be large enough to keep the $B_\mathrm{B}$ below saturation in the yoke's $B$-$H$ curve. 
The size and the weight of the magnet can be kept small by setting  $S_\mathrm{A}=S_\mathrm{B}$ \cite{you2016designing}. $S_\mathrm{C}$ is determined by dimensions of the permanent magnet and the air gap, and for most cases, $S_\mathrm{C}>S_\mathrm{A}, S_\mathrm{B}$ to obtain enough coil movement range with a uniform field distribution.  

Maintaining a large average permeability in the yoke is important for three reasons. (1) To keep the MMF drop over the yoke small, delivering more flux to the air gap, (2) to minimize nonlinear errors that occur when the coil carries current \cite{linonlinear,linonlinear2,hysteresis} and (3) to achieve a flat field profile. A high yoke permeability makes the two sides of the air gap equipotential surfaces, and the flux transverses uniformly through the gap. Yoke materials, such as the Fe-Ni whose $B$-$H$ curve is shown in figure \ref{fig:yokeBH}, have very high permeabilities so that all three points can be achieved.
	
\subsection{Profile flatness}

The phrase ``flatness of the field'' or ''flatness of the magnetic profile'' includes two related goals. (1) the radial component of the magnetic flux density $B_\mathrm{r}$ should be constant with the traveling range of the coil along $z$. (2) the radial component of the flux density multiplied by the radius, $rB_\mathrm{r}$ should be constant along $r$. The first property ensures that the $Bl$ is independent of the exact weighing position along $z$, see~\ref{sec:ConvCoil}. With the second property, $Bl$ becomes independent of the coil radius and, hence, of the coil's thermal expansion during weighing. 

\begin{figure*}[tp!]
		\centering
		\includegraphics[width=1.1\textwidth]{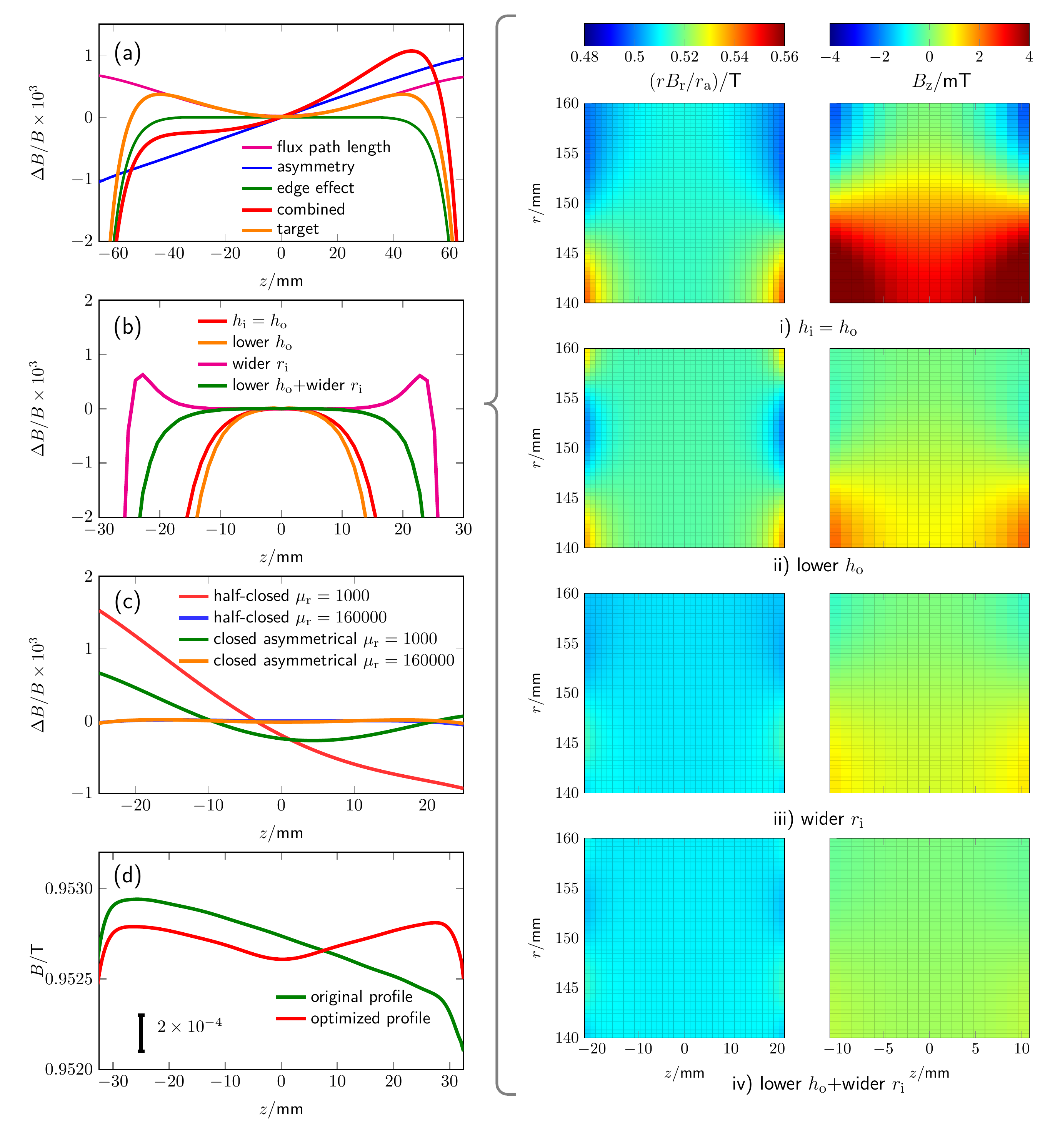}
		\caption{(a) general factors that affect the magnetic field flatness along the vertical direction. The scales are obtained by FEM simulations based on the NIST-4 system. (b) compares the magnetic profiles with several different designs of air gaps. The right subplots show the flatness of $rB_\mathrm{r}$ and $B_\mathrm{z}$ components, respectively. The plots are reproduced from \cite{li2020simple}. (c) presents how the magnetic profile is related to the magnet asymmetry with different yoke permeability. Reproduced from \cite{BIPMmag2017}. (d) shows the magnetic profile of the LNE Kibble balance magnet. Reproduced from \cite{LNEmag}.}
		\label{fig:profile_components}
	\end{figure*}

Maxwell's equation link the the two components of the flux density together via,
\begin{eqnarray}
	\displaystyle\frac{1}{r}\frac{\partial (rB_\mathrm{r})}{\partial r}+\frac{\partial B_\mathrm{z}}{\partial z}&=&0,\label{maxwell1}\\
\displaystyle\frac{\partial B_\mathrm{r}}{\partial z}-\frac{\partial B_\mathrm{z}}{\partial r}&=&0\label{maxwell2}.
\end{eqnarray}
If $B_\mathrm{z}$ were constant, the two properties for a flat field would be met. However, this perfection cannot be achieved over the entirety of the gap. Thinking about this property reveals three sources of deviation from field flatness.  (1) the fringe field at the end of the gap (edge effect), (2) magnet asymmetry, and (3) the flux path length.  All three effects are summarized  in figure \ref{fig:profile_components} (a).

\subsubsection{Reducing the fringe field}
\label{sec:sec4:fringe}

The field inside the gap depends on the aspect ratio of the gap. A narrow and tall air gap has a much more uniform field than a wide and short air gap.  This effect is analog to a similar problem in electrostatics, the field in a parallel plate capacitor. At the end of the air gaps, the flux lines bulge outward. Per unit area, the yoke near the end of the air gaps carries less flux than in the center of the gap. In other words, the reluctance near the air gap end is larger, and hence the magnetic flux density is smaller. The edge effect can easily lead to a magnetic field reduction at the percentage level. With the edge effect, the region where the field is uniform, e.g., $\Delta B/B<5\times10^{-4}$, can be significantly smaller. As a result, a flat profile can only be obtained near the center of the magnet, and only about 50\% of the gap may be usable\cite{NISTmag,BIPMmag2017}. 
For a gap with parallel sides of equal height, $h_\mathrm{a}$, \cite{li2020simple} gives an equation for the relative deviation of the radial magnetic flux as a function of vertical position. It is,
	\begin{eqnarray}
    \frac{B_\mathrm{r}(z)}{B_\mathrm{r}(0)}-1=&-&\exp\left[-\left(1+\frac{\pi(h_\mathrm{a}-2|z|)}{\delta_\mathrm{a}}\right)\right]\nonumber \\
    &+&\exp\left[-\left(1+\frac{\pi h_\mathrm{a}}{\delta_\mathrm{a}}\right)\right],
    \end{eqnarray}
where $h_\mathrm{a}$ and $\delta_\mathrm{a}$ denote the air gap height and width, respectively.
	
In a BIPM-type magnet, the yoke-air boundary at the end of the air gap, however, is not symmetric even when the heights of inner and outer yokes are equal ($h_\mathrm{i}=h_\mathrm{o}$) \cite{li2017two}.
The difference is noticeable: the inner boundary contains the permanent ring and the yoke, while the outer yoke has only yoke material.
As a result, the magnetic flux lines at the gap end will slope further towards the outer yoke. Because of this, magnetically, the outer yoke is higher than the inner yoke, even when they are geometrically the same.  The red line in figure \ref{fig:profile_components}(b) and the right top plot i) present the $B_\mathrm{r}$ and $B_\mathrm{z}$ distribution in the central region of the air gap with $h_\mathrm{i}=h_\mathrm{o}$. Large $B_\mathrm{z}$ gradients are seen in both $r$ and $z$ directions, and therefore the profile quality in this case ($h_\mathrm{i}=h_\mathrm{o}$) is not high. 

So far, we tacitly assumed that the air gap is bounded by vertically aligned iron pieces of the same height. In other words, the air gap has perfect symmetry. However, as described above, the MMF is not symmetrically placed to the air gap. The source of the magnetic field is closer to the inner yoke. Hence, the path to the outer yoke has more reluctance, and the symmetry is broken. As a consequence, the flux lines bend towards the outer yoke. It appears that the magnetic height of the outer yoke is higher than the physical height. This phenomenon can be remedied in two ways. First, the symmetry can be restored by adding additional magnetic material on the outside yoke~ \cite{li2017two}.  Second, the magnetic symmetry can be restored by lowering the outer yoke such that magnetically the two sides of the air gap have identical heights. This idea is described in  \cite{bielsa2015alignment}.

The effect of changing the outer yoke height is illustrated in figure~\ref{fig:profile_components}.  On the right side of the figure, the magnetic flux in the air gap is shown. The top row shows it for the case where both heights are identical. The second row shows it for $h_\mathrm{o}<h_\mathrm{i}$. Lowering the outer yoke improves $B_\mathrm{z}$.  The top right surface plot in the figure shows all shades from dark red to blue, where the plot one row below only shows the colors in the middle of the range. Disappointingly, the radial field is not much improved. This point is also illustrated in panel (b) of figure~\ref{fig:profile_components}. The orange line shows $ B_\mathrm{r}(z)-B_\mathrm{r}(0)$ as a function of $z$. The orange line is calculated for $h_\mathrm{o}<h_\mathrm{i}$ and the red line $h_\mathrm{o}=h_\mathrm{i}$. There is a small but not significant gain in uniformity for $h_\mathrm{o}<h_\mathrm{i}$ compared to $h_\mathrm{o}=h_\mathrm{i}$. A similar (small) effect can be achieved by adding a pair of SmCo magnets to the outer yoke. However, doing so will compromise the shielding property of the yoke. Fluctuating external fields will be able to reach the coil.

Reference \cite{li2020simple} proposes another technical solution to improve field flatness: Adding a piece of iron rings with a rectangular cross-section at the upper and lower edges of the gap. These features decrease the gap size at the end of the gap, effectively reducing the reluctance and increasing the flux. The flatness of $B_\mathrm{r}(z)$ can be optimized by adjusting the two parameters of the rectangle, the height, and the width of the rectangle. An example (parameters were shown in \cite{li2020simple}) is shown in the third right subplot iii) and magenta curve in figure \ref{fig:profile_components}(b). It can be seen in this case the $1/r$ field distribution for $B_\mathrm{r}(r)$ has better quality than is achieved by lowering $h_\mathrm{o}$. More important, the usable measurement range for $B_\mathrm{r}(z)$ has been greatly increased compared to the original design ($h_\mathrm{i}=h_\mathrm{o}$). As shown in subplot iv) and the green curve in figure \ref{fig:profile_components}(b), a flatter field distribution of $rB_\mathrm{r}$ and $B_\mathrm{z}$ can be obtained if both techniques of lowering $h_\mathrm{o}$ and widening $r_\mathrm{i}$ are applied.

\subsubsection{Improving magnet symmetry}	

The second factor that can significantly improve field flatness is, in general, the overall symmetry, and more specifically, the up-down symmetry of the magnet system. 
By design, the BIPM type magnet system exhibits perfect mirror symmetry around $z=0$. However, this symmetry can be broken due to machining tolerances, assembly, and material inhomogeneities. Concrete examples that break the symmetries are
\begin{itemize}
    \item The gap could be slightly tapered due to machining tolerances. 
    \item Dowel pins or bolts used to align and fasten components of the magnet system could introduce magnetic asymmetries. 
    \item The magnetization of the upper SmCo ring could be different from the lower ring. 
    \item The permeability of the iron could be inhomogeneous.
\end{itemize}    
The latter is especially troublesome because the permeability depends not only on the stresses induced during fabrication but also on the magnetic history. For example, during the construction of NIST-4, it was discovered that the procedure used to close the magnet had an effect on the permeability of the yoke and changed the profile flatness~\cite{NISTmag}.

Materials with high permeability ease some of these problems. Ideally, the two sides of the gap are equipotential surfaces. So the MMF-drop is the same between any points on each side of the gap. Materials with a high $\mu_\mathrm{r}$, such as  Fe-Ni (50/50) alloy, can be used to achieve the equipotential surface. An impressive illustration of the power of high $\mu_\mathrm{r}$ materials is the BIPM magnet~\cite{BIPMmag2017}. The top cover of the BIPM magnet is missing, but the field is reasonably flat. This fact is demonstrated in panel  (c) of figure~\ref{fig:profile_components}. These four curves are compared. The curves are obtained with an FEA calculation. The magnet is either half-open or complete. When the magnet is complete, the bottom SmCo disk has 10\% more magnetization than the top disk. For each case, the field was calculated for $\mu_\mathrm{r}=1000$ (soft iron)  or $\mu_\mathrm{r}=160000$ (50\%Fe-50\%Ni).  For the latter case, there is no difference if the magnet is open or closed. This graph impressively demonstrates how a lack of symmetry can be overcome with a high permeable material. It recovers a perfect field even with half the flux path missing or, in the other case, with a 10\% difference in magnetization.  

In summary, we would advise the designer to start with a symmetric plan and build the yoke with high permeable materials if the construction budget allows these materials. The use of these more expensive materials can compensate for unwanted deviations in the production process.

\subsubsection{Equalizing the flux paths}	

The third factor that has an influence on the shape of the magnetic profile is the length of the flux path. The length of the flux path changes the profile, independent of the presence of a fringe field at the end of the gap. As shown in figure \ref{fig:analysis},  the reluctance along the solid green line is greater than that of the dashed green line, and hence, the magnetic field in the gap center, $B_\mathrm{r}(z)$, distributes as an 'M' shape. The $B_\mathrm{r}$ is lower at the center of the gap and then increases before it rolls off to the end of the gap.

The length of the flux path is a powerful but yet simple argument, and it can guide our intuition for the pot magnet system employed by LNE~\cite{LNEmag}. 
The field at the top of the gap has to be smaller because of the larger reluctance in the flux path necessary to reach the top. The reluctance increase can be counteracted by making the gap smaller by introducing a taper. The $B_\mathrm{r}$ of a parallel and tapered gap of the LNE magnet is shown in panel (d) in figure~\ref{fig:profile_components}. The taper runs from the center of the gap to the top. The gap with a nominal width of \SI{9}{\milli\meter} is \SI{8}{\micro \meter} smaller at the top. The yoke material used here does not have an exceptionally high $\mu_\mathrm{r}$. 

In summary, visualizing the length of the flux path in the magnet is a valuable tool to get a qualitative understanding of the profile in the magnet. Differences in flux paths can be compensated by adjusting the gap size. The effect that different length flux paths have on the profile is more pronounced if the permeability of the yoke is low. So, these differences can be evened out by using high permeable materials.

\subsection{Force required to  open the magnet}
\label{sec5:force}
	
Magnet systems that entirely enclose the coil apart from a few holes to attach the coil are called closed yokes.
These designs have superior shielding performance compared to the open-yoke designs. 
However, the yoke needs to be opened and closed at least once to install the coil. 
The force required to open the magnet can be large, on the order of several kN.
Therefore, a dedicated device called a magnet splitter is required to open and close the magnet system in a controlled way.
The magnet splitter, such as the one used for NIST-4\cite{NISTmag} can only be used when the magnet is not installed in the balance.
It is difficult to integrate such a device into the balance for in-situ adjustments. 

\begin{figure}
	\centering
	\includegraphics[width=0.5\textwidth]{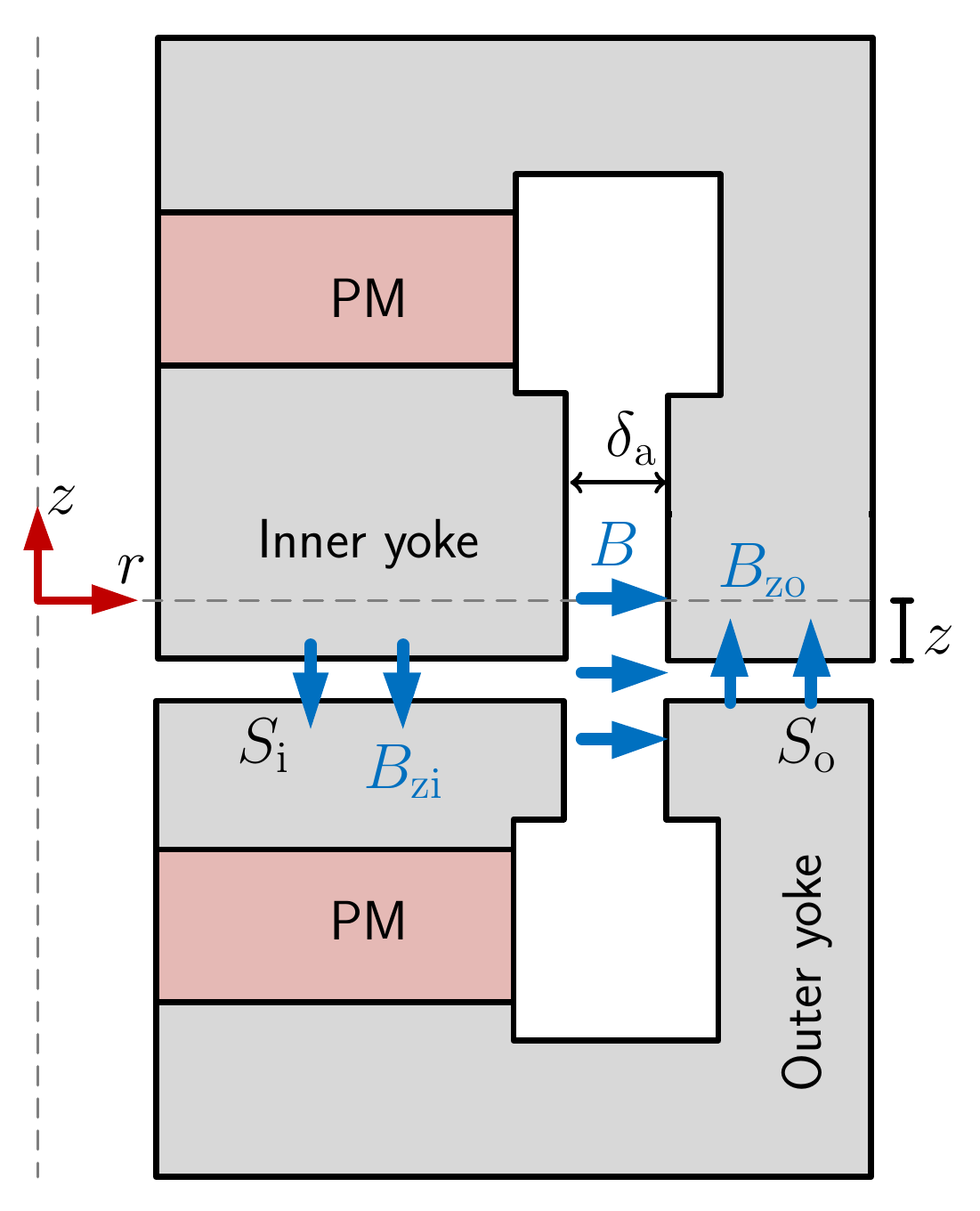}
	\caption{Schematic integration of the Maxwell stress tensor to determine the force required to split the magnet. The symmetry plane is given by the horizontal dashed line. The split plane is a distance $z$ away from the split plane.}
	\label{fig:sec2:tensorplot}
\end{figure}

Here we follow the Maxwell stress tensor method and the derivation given in the appendix of~\cite{marangoni2019magnet} and show how to reduce the splitting force as much as possible. 
It is assumed that the split plane is horizontal, and, hence, $\mathrm{d}\mathbf{S}$ is vertical.
In this case, only the last row of $\mathbf{T}$  is relevant. 
As it is shown in figure~\ref{fig:sec2:tensorplot}, the vertical force simplifies to,
\begin{equation}
F_\mathrm{z}(z)=\frac{\pi r_\mathrm{a}B^2}{\mu_0}\left(\delta_\mathrm{a}-\frac{2\pi r_\mathrm{a}(S_\mathrm{i}+S_\mathrm{o})}{S_\mathrm{i}S_\mathrm{o}}z^2\right),
\label{eq:fz}
\end{equation}
where $r_\mathrm{a}=(r_\mathrm{i}+r_\mathrm{o})/2$ is the mean radius of the air gap, $B$ the average magnetic flux density in the air gap, $S_\mathrm{o}$, $S_\mathrm{i}$ the cross-sectional area of the outer and inner yokes at the splitting plane, and  $B_\mathrm{o}$, $B_\mathrm{i}$ the magnetic flux density at these surfaces.  The distance of the split plane to the symmetry plane of the magnet is denoted by $z$.

 The force calculated with  equation (\ref{eq:fz}) occurs at the initial separation, at the moment when the contact between the metal surfaces breaks, i.e, $d=0$. This force can be made zero by choosing
\begin{equation}
|z|=z_0:=\sqrt{\frac{\delta_\mathrm{a}S_\mathrm{i}S_\mathrm{o}}{2\pi r_\mathrm{a}(S_\mathrm{i}+S_\mathrm{o})}}.
\end{equation}
With increasing $|z|$, the direction of the force changes. For
 $|z|<z_0$, $F_\mathrm{z}(d=0,z)$ is repulsive and for  $|z|>z_0$, it is attractive.

\begin{figure}[tp!]
\centering
\includegraphics[width=0.8\textwidth]{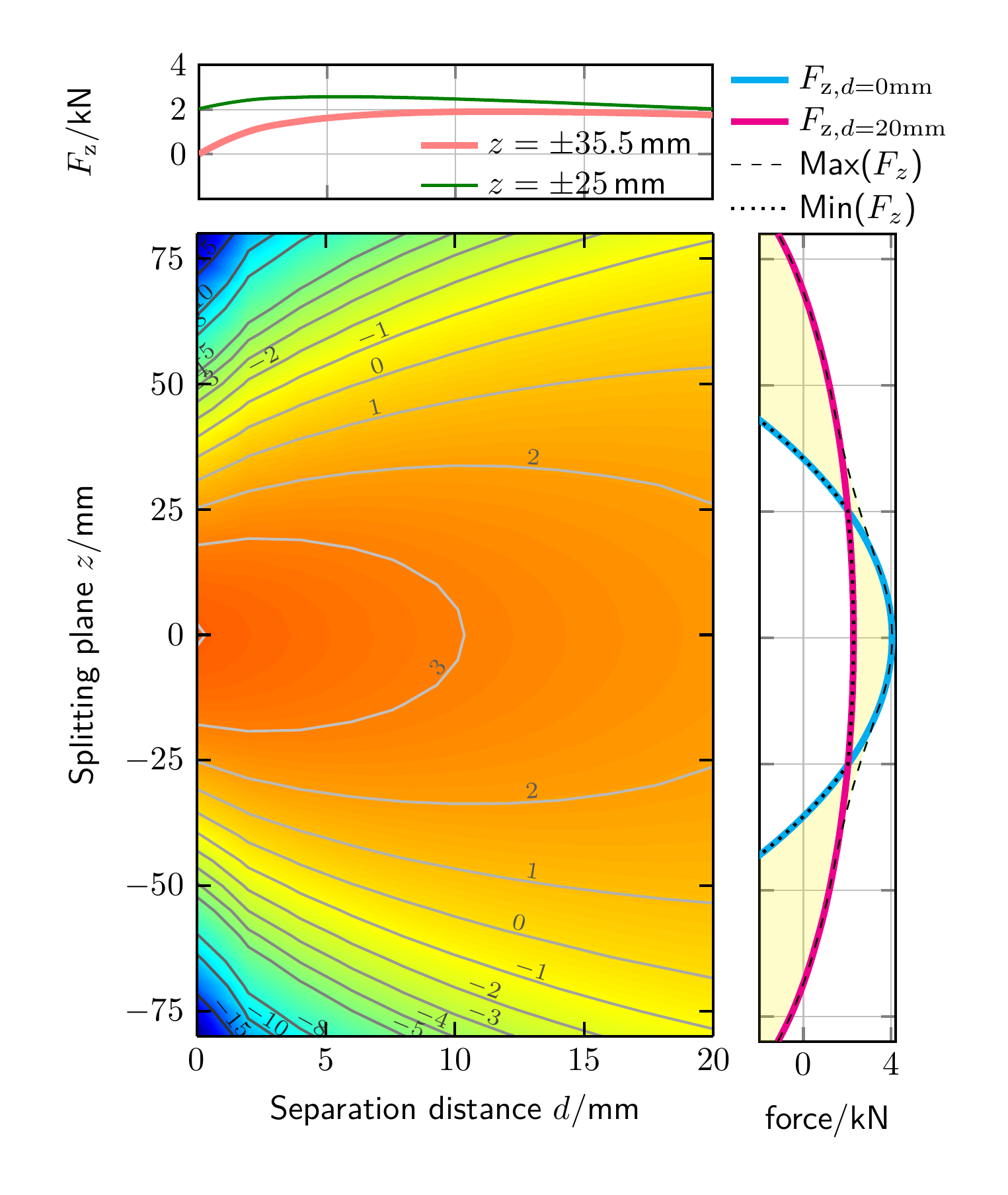}
\caption{The main plot shows the splitting force $F_\mathrm{z}$ as a  function of two parameters, the position of the splitting surface $z$ and the separation between the yoked $d$. The unit of the force is kN.  A positive (negative) force means the two parts of the magnet repel (attract) each other. For this magnet, the precision gap is between \SI{-50}{\milli\meter}  and \SI{50}{\milli\meter}. The right plot presents the magnetic force $F_\mathrm{z}$ along two vertical lines at $d=0$\,mm and $d=20$\,mm, as well as the force boundaries (maximum and minimum values of $F_z$) required to open the magnet within 20\,mm at different splitting planes. The top subplot shows the magnetic force as a function of the separation distance $d$ at the two locations,  $z=\pm25$\,mm and $z=\pm35.5$\,mm. 
}
	\label{fig:openforce}
\end{figure}

Equation~(\ref{eq:fz}) gives and analytical expressions for the force required to open the magnet \cite{marangoni2019magnet}. In reality, however, the magnetic force changes as the gap opens, because $B_\mathrm{zi}$, $B_\mathrm{zo}$ and $B$ are functions of the yoke separation $d$.	
In order to insert the coil into the magnet, a separation greater than the coil height, i.e. $d>h_\mathrm{c}$ is required. Therefore, the magnetic force $F_\mathrm{z}$ should also contain the dependence on $d$. Here we take the NIST-4 magnet system as an example. By finite element analysis (FEA), the distribution magnetic force $F_\mathrm{z}$ as a function of the two parameters $z\in(-80,80)$\,mm and $d\in(0,20)$\,mm is shown in figure \ref{fig:openforce}. At the start of the separation, $d=0$, the force calculated by FEA agrees with (\ref{eq:fz}), see the cyan curve in the right subplot. 
In red, this subplot also contains a calculation of $F_z$ for $d=20\mathrm{mm}$.  It can be seen that the latter curve has a similar behavior to the one at $d=0$, but is much flatter.

For the construction of the splitter, it is necessary to know how much the force changes during the splitting process. The yellow shaded area in the right subplot of figure \ref{fig:openforce} indicates the dynamic range of the force. For a given $z$ the yellow shading extends from the $F_\mathrm{z,min}$ to $F_\mathrm{z,max}$. The reader can identify three regions. For $|z|>\SI{68.5}{\milli \meter}$, both extreme forces are negative. Hence, during the splitting process, there will always be an attractive force between the two parts of the magnet. For $|z|<\SI{35.5}{\milli \meter}$, the splitting force is repulsive for all $d$. For all other $z$ values, the splitting force changes sign. It is attractive at first when the magnet opens ($d=0$) and becomes repulsive with increasing $d$. Such a load change is important to take into account when designing the splitter. Interestingly, with the split plane at $z\approx\SI{25}{\milli \meter}$, the force stays fairly constant for the entire opening process. 

The forces required to split the magnet (several kN) are about the same order of magnitude as the weight of the whole or at least part of the magnet ($g$ times several 100 kg). The weight of one of the two parts that the magnet is split into, can be used to reduce the force that the splitter must generate. For example, the total mass of the NIST-4 magnet is \SI{850}{\kilo \gram}. If the splitting is performed at $z=25$\,mm, the weight of the upper piece is $\approx$ 3.3\,kN and that of the lower $\approx$ 5.2\,kN. Conceivably, one could use the 3.3\,kN to work against the repulsive force and reduce the maximum $F_z(d=\SI{5}{\milli\meter},z=\SI{25}{\milli\meter})=\SI{2.57}{\kilo\newton}$ and the minimum $F_z(d=\SI{0}{\milli\meter},z=\SI{25}{\milli\meter})=\SI{2.03}{\kilo\newton}$ to combined downward forces of 0.73\,kN and 1.27\,kN, respectively. Note, this is not what researchers at NIST are doing. The split plane was chosen at $z=\SI{-50}{\milli\meter}$ to be outside of the precision gap, and the heavier two-thirds of the magnet is lifted off. The theoretical $F_z$ at $z=-50$\,mm has a minimum value of 4\,kN. Adding the weight of the top piece, the maximum splitting force required is 9.2\,kN. The example should show, however, that by clever selection of the location of the split plane and use of the weight, the split force can be well minimized. 
A force at \SI{1}{\kilo\newton} level is achievable with lead screws. Therefore, it seems possible that such a system can be integrated into the Kibble balance.  Then, the splitting and maintenance of the coil could be made in situ. One significant advantage would be that the Kibble balance can be used to measure the profile, and one does not need to have a dedicated profile measurement system for fine adjustments of the profile.

Note, the smallest splitting force, including the weight, occurs at $z=\pm\SI{25}{\milli\meter}$, but the precision gap extends from $\SI{-50}{\milli\meter}$ to $\SI{50}{\milli\meter}$. If the split plane is located at one of these locations, the magnetic field profile near the break is disturbed. Hence, the magnetic flux will only be smooth in a length of $\SI{75}{\milli\meter}$. Placing the break in the magnet is a trade-off. As can be seen in figure~\ref{fig:openforce}, the smallest splitting force occurs in or near the region of the precision gap. However, that is the location where the location of the split plane is least desirable.

\subsection{Thermal considerations}
\label{sec:sec4:thermal}
	
The magnetization of the SmCo material has a temperature coefficient of about $-3\times10^{-4}$/K. Although the magnetic field drift is very smooth due to a large thermal capacity and can be removed by ABA~\cite{swanson_2010} measurement in Kibble balance, it is preferred to reduce the temperature coefficient to a smaller level. While this step is optional for SmCo, it is mandatory for NdFeB because its temperature coefficient is much higher. 

Besides potentially adding a systematic bias to the measurement, a large temperature coefficient has another significant downside. At pump down, most surfaces cool down due to the evaporation of a thin water film. Since the magnet's thermal mass is large and is well insulated in a vacuum, it takes weeks for the magnet to equilibrate fully thermally. Thus, if the temperature coefficient of the magnetic material is significant, the measurement will drift for a long time. The drift adds uncertainty to the measurement, makes the investigation of systematic effects difficult, and is commonly not desired.  

In general, there are two avenues to reduce the temperature coefficient of the magnet. First, one can choose an active magnetic material with a very low temperature coefficient, for example, (Sm,Gd)Co~\cite{marangoni2019magnet,METAS}, see section~\ref{sec3:mat}. Second, the magnetic circuit can be designed to be less sensitive to temperature. The latter idea is illustrated by the blue rectangle in figure~\ref{fig:analysis}. Part of the magnetic flux is routed through a shunt whose reluctance varies with temperature. Given the temperature dependence of its reluctance, the geometry of the shunt can be finely tuned such that the flux in the air gap is relatively independent of temperature at the design temperature. 

As is shown in the equivalent circuit in ~\ref{fig:analysis}, an additional reluctance, the shunt, is parallel to the permanent magnet. A  small amount of magnetic flux $\phi'$ goes through the shunt with reluctance $R_\mathrm{t}$. The circuit, now,  has two loops. One is carrying the main flux $\phi$, the other the shunted flux $\phi'$. Using Kirchhoff's laws, one obtains,
\begin{eqnarray}
(\phi+\phi') R_\mathrm{m}+\phi (R_\mathrm{a}+R_\mathrm{y})&=&\mathcal{F}\\
(\phi+\phi') R_\mathrm{m}+\phi' R_\mathrm{t}&=&\mathcal{F}.
\label{eq:T}
\end{eqnarray}
Eliminating $\phi'$ in equation (\ref{eq:T}) and ignoring $R_\mathrm{y}$ yields
\begin{equation}
\phi\approx\frac{\mathcal{F}}{R_\mathrm{m}+R_\mathrm{a}+\displaystyle\frac{R_\mathrm{m}R_\mathrm{a}}{R_\mathrm{t}}}.
\label{eq:phiT}
\end{equation}
To keep $\phi$ insensitive to temperature $T$, i.e. $\partial \phi/\partial T=0$, equation (\ref{eq:phiT}) is written as
\begin{eqnarray}
\frac{1}{\mathcal{F}}\frac{\partial \mathcal{F}}{\partial T}&=&-\frac{1}{R_\mathrm{m}+R_\mathrm{a}}\frac{R_\mathrm{m}R_\mathrm{a}}{R_\mathrm{t}^2}\frac{\partial R_\mathrm{t}}{\partial T},\nonumber\\
\left[\frac{1}{\mathcal{F}}\frac{\partial \mathcal{F}}{\partial T}\right]&=&-\frac{R_\mathrm{m}R_\mathrm{a}}{(R_\mathrm{m}+R_\mathrm{a})R_\mathrm{t}}\left[
\frac{1}{R_\mathrm{t}}\frac{\partial R_\mathrm{t}}{\partial T}\right]
.\label{eq:comp}
\end{eqnarray}
The expressions in square brackets denote the relative temperature coefficient of the MMF and the shunt reluctance. Temperature compensation with a shunt is possible because the relative temperature coefficient of the MMF is negative, but that of the shunt is positive, typically a few percent per kelvin. Hence, $R_\mathrm{t}$ can be chosen such that equation~(\ref{eq:comp}) is valid. In that case, the temperature dependence of the flux in the gap vanishes.

%% file: 05_TestMeasurement.tex
\section{Delivering design to reality}
\label{sec4}

Once the magnet has been designed, it's time to build it. Once it's made, it must be verified. The engineers and scientists have to determine the field and the flatness of the profile. Perhaps the magnet must be split open to insert the coil. The shielding properties of the magnet system must be measured, and finally, the temperature coefficient of the complete system must be determined. This section explains all these tasks in detail. So far, we have dealt with an ideal magnet. Here, reality sets in.

\subsection{Mechanical assembly and alignment}

Ideally, on the horizontal plane $xy$ at $z=0$, the magnetic flux density should be uniform in the azimuthal direction and be proportional to $1/r$  in the radial direction. 
A deviation from these two desired goals could be caused by a nonuniformity of the magnetic materials, machining defects, misalignment during assembly, and other problems that break the symmetry.
Although a slightly different coil placing can minimize these effects (see  below), the best is to use good design to avoid these problems from the start with the following three tips:
\begin{enumerate}
\item 	Use symmetric magnet rings. If two rings are employed, they should be as identical as possible in size and magnetization. Very often, these rings are much larger than the size that can be reasonably magnetized. In this case, each ring will be composed of smaller segments that can be magnetized. It's best to measure each segment and assemble each ring such that the average magnetization is identical. Also, scramble the segments in each ring so that azimuthal uniformity is achieved as best as possible.
\item Use high permeability yokes.
High yoke permeability helps create equal potential boundaries and, therefore, can largely average out the asymmetry. As the machining process could significantly lower the yoke permeability, heat treatment before assembly is necessary.
\item Keep the assembly as symmetrical as possible. The magnetic working point on any material depends on its magnetic history. During the assembly, two yoke pieces touch at one point instead of evenly around the circumference. The flux that flows through the point of contact can be very high, altering the magnetic working point at that spot. With the altered magnetic working point, the reluctance of the section has been changed, and the azimuthal symmetry of the magnet system is broken. Hence, try to assemble the pieces that carry magnetic flux in an even, symmetric, and controlled fashion.
\end{enumerate}
	
\begin{figure}[tp!]
\centering
\includegraphics[width=0.7\textwidth]{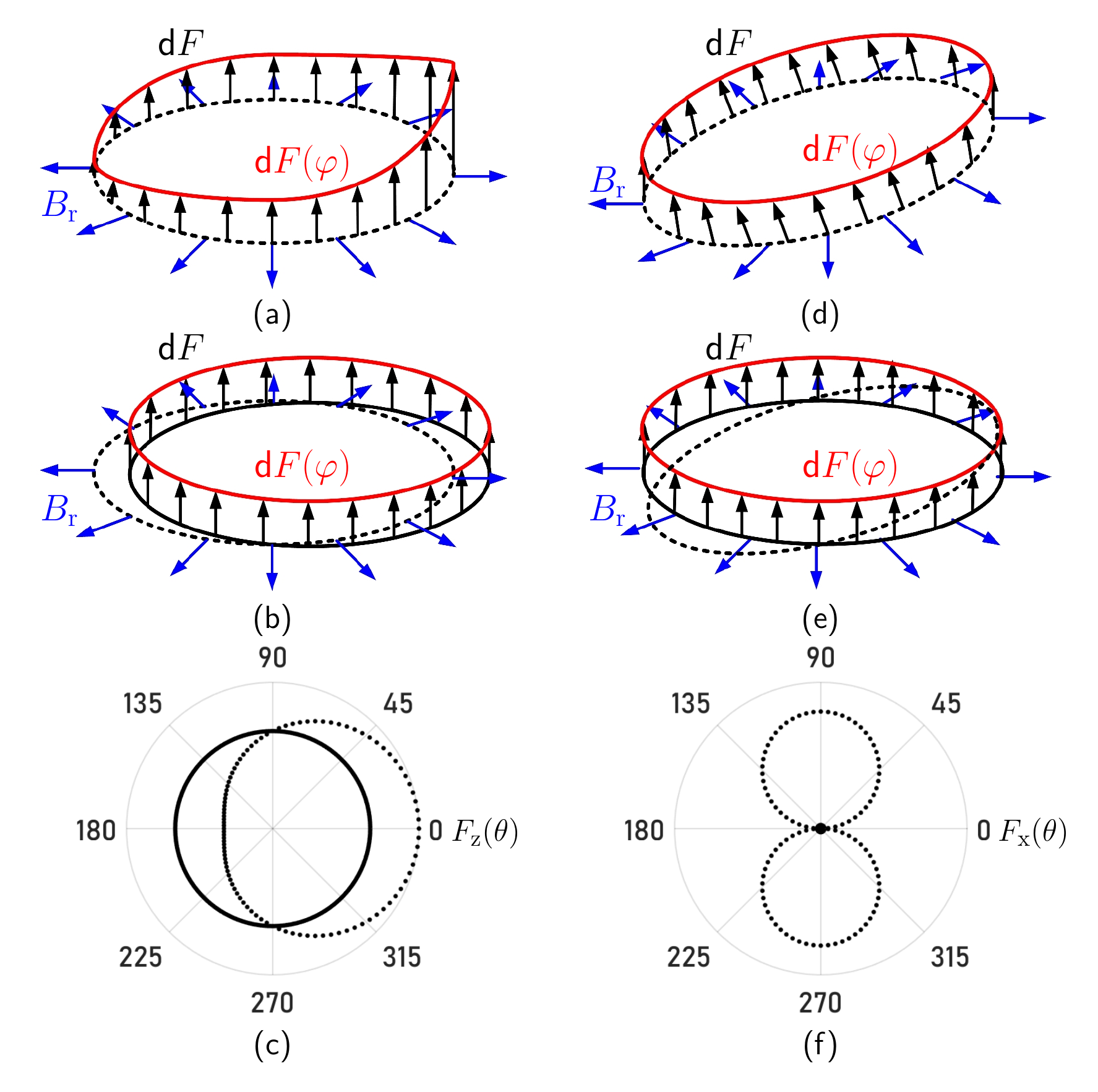}
\caption{The schematic of the coil alignment. For the left column, the coil is horizontal, but the inner yoke of the magnet is eccentric, such that the flux is higher on the right side. For the right column, the magnet is perfect but the coil is initially tilted. The first row shows the coil and the forces on the coil in the original alignment. In all cases, the flux is purely horizontal. The situation on the left can be ameliorated by displacing the coil and the situation on the right by tilting the coil. The improved situation is shown in the middle row. The last row shows the vertical and horizontal force for the left and right scenarios, respectively. The dashed line shows the force in the original and the solid line in the improved situation. The forces are shown as polar plots. 
}
\label{fig:aligment}
\end{figure}

Next, we discuss how the geometric factor depends on the mechanical assembly and how it is affected by the coil alignment. The magnetic flux density in the air gap, $B_\mathrm{r}$, is determined by the width of the gap. 
As we will see below, asymmetries can be compensated by placing the coil eccentrically in the gap. However, the amount of eccentricity for the coil placement is limited since the coil should not touch the yoke. Therefore, this argument gives an upper bound on how much asymmetry can be allowed.
For now, the inner and outer yokes are assumed to be perfect cylinders. In this case, misalignment can occur when (a) the cylinder axes are not parallel with one another, or (b) if the cylinder axes are not coincident at $z=0$ and (c) a combination of (a) and (b).
All three cases will cause an azimuthal variation of $B_\mathrm{r}$. In a perfect symmetrical magnet, $B_\mathrm{r}$ is independent of the azimuth, i.e., $\partial B_\mathrm{r}/\partial \varphi=0$. Let's assume that this assumption is no longer true.
An example where the inner yoke is displaced along the negative $x$ axis is shown in Fig.~\ref{fig:aligment}(a). Since the gap is smallest along the positive $x$ axis, the force is largest. The force on a coil carrying current is indicated by the black vectors and the red curve connecting the tips of the vectors.  The force is no longer isotropic but is larger at $\varphi=0^\circ$ and smaller at $\varphi=180^\circ$.
Interestingly, the total $Bl$ along the whole coil is conserved, as is shown in \cite{li2016discussion}. 
This is because, to first order, the reluctance of the gap does not change by displacing the inner yoke and hence the flux through the coil and with that the flux gradient or $Bl$ (see \ref{sec:AppendixA})  remains the same.

If one were to plot the vertical force $F_\mathrm{z}$ as a function of azimuthal angle $\varphi$, one would obtain a cosine shifted by an offset. The maximum would occur at $0^{\circ}$ and the minimum at $180^{\circ}$. This relationship can be visualized by a polar plot, as shown in Fig.~\ref{fig:aligment} (c). The force on the right side of the coil is larger than on the left. Hence, a torque $\tau_\mathrm{y}$ about the $y$ axis occurs. Correspondingly, in velocity mode, an induced EMF can arise if the coil rotates around $y$ while sweeping vertically~\cite{Stephan16}. This additional EMF can lead to a bias in the measurement. 
	
There is an easy way to avoid the bias in velocity mode and eliminate $\tau_\mathrm{y}$:  Place the coil eccentric to the coordinate center.  The amount the coil needs to be moved is
\begin{equation}
\Delta r=\frac{\Delta B_\mathrm{r}}{2{\overline{B}_\mathrm{r}}}r_\mathrm{c},
\end{equation}
where $\overline{B}_\mathrm{r}$ is the average radial magnetic flux density at the coil and $\Delta B_\mathrm{r}$ is the difference between the maximum and the minimum of $B_\mathrm{r}$.  The coil has to be moved toward the maximum field. So, in the above example, in the direction of $+x$. Note an analytic equation for $\Delta B_\mathrm{r}$ based on the eccentricity of the inner yoke can be found in~\cite{li2016discussion}.

The gap width is given by the difference in radius of the outer and inner yoke, $\delta_\mathrm{a}=r_\mathrm{o}-r_\mathrm{i}$. By subtracting the width of the coil from $\delta$ the air space around the coil is obtained. If the coil is centered, which is usually the case, half of the air space is inside and the other half outside of the coil. The maximum distance the coil can be moved is given by
\begin{equation}
\Delta r< \frac{\delta_\mathrm{a}-w_\mathrm{c}}{2}\label{eq:boundary}
\end{equation}
Hence, the maximum relative asymmetry that can be cancelled with this technique is given by
\begin{equation}
\frac{\Delta B_\mathrm{r}}{\overline{B}_\mathrm{r}}< \frac{\delta_\mathrm{a}-w_\mathrm{c}}{r_\mathrm{c}}
\end{equation}

The second coil misalignment discussed here is a tilt. One can tilt the coil to reduce the angle between the coil and the magnetic field plane so that the horizontal motion of the coil is minimum. As shown in figure \ref{fig:aligment} (d), when the coil is tilted with respect to the $B_\mathrm{r}$ field, a horizontal force is generated due to the vertical current. Here, we assume the $B_\mathrm{r}$ to be horizontal. The distribution of the horizontal force component along the wire circular is shown in figure \ref{fig:aligment}(f).  To fix this, it requires to tilt the coil to where the coil displacement (proportional to horizontal force) is zero during mass-on and mass-off. Note, the same is true if the magnetic field is inclined.  Then one can find a coil tilt, where the horizontal force is zero. But, ideally, of course, both coil and magnetic field are horizontal.

For a perfectly machined magnetic circuit, all reference surfaces are either parallel or perpendicular to each other. Especially, the top surface is parallel to the magnetic flux density at the center of the magnet. It can be used to align the field horizontal which is important to produce only a vertical force in weighing mode. In some experiments, the weighing is performed at multiple vertical  positions~\cite{robinson2011simultaneous,fang2014watt,qian2020preliminary}. In such cases, the top surface of the magnet is not good enough to be used as the field reference.  Reference \cite{bielsa2015alignment} gives a practical and elegant way to measure the field inclination. A rotating magnetometer that is instrumented with capacitive probes is lowered into the gap at different positions. At each position, the probe is centered in the gap using the signal of the capacitive probes. From the reading of the magnetometer, the tilt of the magnetic field can be obtained.
Finally, the experimenter has to be aware that changing the tilt of the magnet will also require changing the position of the coil if one wants to generate a purely vertical force in the weighing mode, see  (\ref{eq:boundary}). Hence, one has to be aware of the available parameter space. Is it possible to tilt the magnet by the desired angle without the coil touching the yoke? Only if the answer is in the affirmative, does it make sense to carry on with the experiment.

\subsection{Profile measurements}
	
After the magnet is assembled, it is advisable to measure the flatness of the profile before integrating the system into the experiment. In this way, it is much easier to tweak the magnetic profile, i.e., shim the magnet, should it become necessary.  
	
There are two principal ways one can measure the profile of the magnetic flux density. The measurement can be performed at selected points with a probe, or an integrated flux ($Bl$) can be measured with a coil. The information provided by the latter measurement is more applicable to the Kibble balance experiment. The measurement at discrete points is often easier to carry out and does not require dedicated hardware.
	
Using a probe, one must be aware that the field gradient along $r$ direction is large, and thus the probe measurement requires a perfect vertical motion relative to the yoke surface. For example, the field gradient of the NIST-4 system is $\Delta B/B=-\Delta r/r_\mathrm{c}$, and for a $\Delta B/B$ resolution of $1\times10^{-4}$, the probe variation along the $r$ direction, $\Delta r$, should be less than $10^{-4}\cdot r_\mathrm{c}=\SI{21.5}{\micro \meter}$.  

\begin{figure}[tp!]
	\centering
	\includegraphics[width=0.7\textwidth]{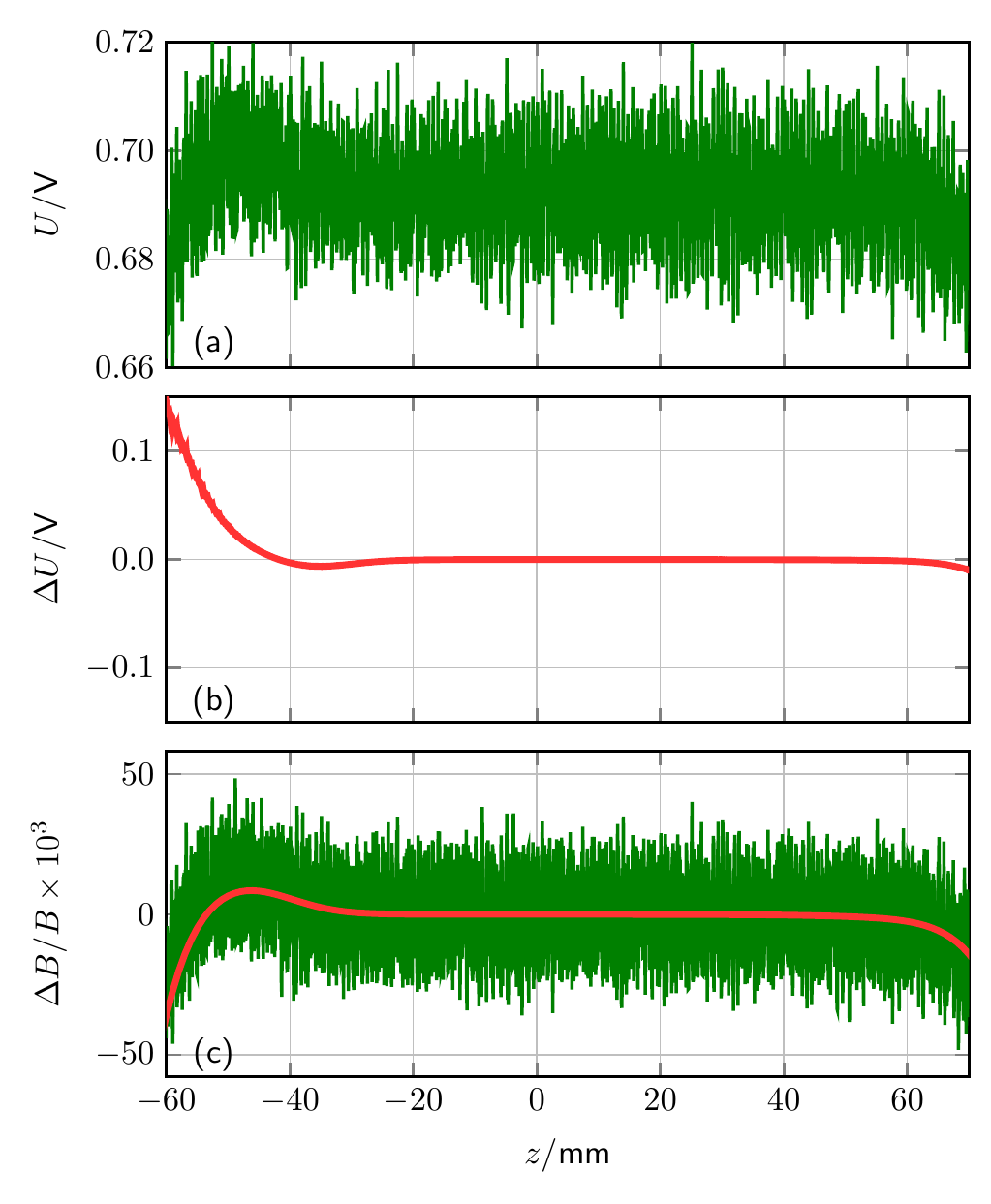}
	\caption{An example of measurement result obtained with the gradiometer coil. (a) shows the absolute induced voltage and (b) is the differential signal of upper and lower coils. (c) compares the magnetic profiles represented by the single coil measurement and the gradiometer coil measurement.}
	\label{fig:gradient}
\end{figure}

Unlike a probe that can be used to spot-check the profile, a coil can be used to get an integrated result along a path. Similar to the Kibble balance experiment, voltages $U$ and velocities $v$ are measured along the trajectory. However, since the measurement is performed before the magnet is integrated into the balance, an auxiliary device must be used to move the coil. As a result, the velocity cannot be measured precisely and may not be constant along the trajectory. Furthermore, the measurement is often noisy due to vibrations.  The noise is further exacerbated by the fact that the velocity and voltage measurements are typically and, unlike in a Kibble balance measurement, not perfectly synchronized. Hence, vibrations do not cancel when the quotient $U/v$ is calculated. This is illustrated by the green curves shown in figure \ref{fig:gradient}~(a) and (c).  The data in this figure are from actual data measured at NIST~\cite{NISTmag}.  The peak-peak value of $\Delta B/B$ is over $5\times10^{-2}$, and even with careful averaging, it is difficult to measure a change in the magnetic flux density with relative uncertainties below $1\times10^{-3}$. 

Reference~\cite{NISTmag} provides an excellent solution to overcome these measurement challenges.
The idea is to use a gradiometer coil to detect the field variation in small ranges. Then, a detailed profile can be obtained by merging the field gradient information with an averaged absolute profile determination (deduced from $U/v$ measurement). 
A gradiometer coil contains coils with the same parameters (number of turns, average radius) separated by a vertical distance $\Delta z$. In the measurement, the induced voltages in one of the two coils, $U$, and their differential output $\Delta U$ are measured simultaneously. Note that it is easier to measure voltages simultaneously than to measure voltage and velocity at the same time. The ratio of two signals can be written as
\begin{equation}
\frac{\Delta U}{U}=\frac{B_\mathrm{r}(z)-B_\mathrm{r}(z-\Delta z)}{B_\mathrm{r}(z)}\approx\frac{\Delta z}{B_\mathrm{r}(z)}\frac{\partial B_\mathrm{r}}{\partial z}.
\label{eq:dU}
\end{equation}   
 Using $B_\mathrm{r}(z)=U/(2\pi r_\mathrm{c}Nv)$, the magnetic profile $B_\mathrm{r}(z)$ can be solved  by numerically integrating equation (\ref{eq:dU}) along the measurement interval, i.e.,
\begin{equation}
B_\mathrm{r}(z)=\frac{1}{2\pi r_\mathrm{c}Nv\Delta z}\int_{z_\mathrm{start}}^z \Delta U(z')dz'+O,
\label{eq:Br:gr}
\end{equation}
where the constant $O$ is a chosen such that $B_\mathrm{r}(0)=U/(2\pi r_\mathrm{c}Nv)$. Figure \ref{fig:gradient} (b) shows the differential voltage $\Delta U$ in the same example, and (c) compares the magnetic profiles obtained by $U/v$ measurement and the gradiometer coil determination. It can be seen that the signal-to-noise ratio has been improved by three orders of magnitude by the simple gradiometer coil. 
	
It is also possible to build a radial gradiometer coil. Here two coils are placed radially separated. The researchers may use a radial gradiometer coil to check the  $1/r$ dependence of the magnetic flux density \cite{NISTmag}.

We want to add a few words to the uncertainty consideration. For the gradiometer to work, the researcher has to know the spacing $\Delta z$, and the two coils have to be nearly identical. For example, if $\Delta z$ is misstated by 10\,\%, then according to equation (\ref{eq:Br:gr}), the calculated $B_r$ is off by 10\,\% also. Hence, an uncertainty in $\Delta z$ does not change the shape of the obtained curves but the absolute calibration.

If one coil, has more turns or produces relatively more voltage by the factor $\delta$, then
\begin{equation}
\frac{\Delta U}{U}=\frac{B_\mathrm{r}(z)-(1+\delta)B_\mathrm{r}(z-\Delta z)}{B_\mathrm{r}(z)}\approx\frac{\Delta z}{B_\mathrm{r}(z)}\frac{\partial B_\mathrm{r}}{\partial z}-\delta.
\end{equation}
Using this result in equation (\ref{eq:Br:gr}) will produce a linear slope. So the higher polynomial terms of the profile are correct, but a linear term could be produced by a difference in the technical data of the coils. Such a difference can be found out by installing the coil upside down. In this case, the linear term would flip sign. By precise machining, the coils can be made very close to being the same, and in practice, these concerns are small. The profile obtained with a gradiometer coil is superior to that obtained by other methods. 

\subsection{Flattening the  profile}
	
The magnetic profile obtained after assembly may not be flat over the desired measurement range. If this is the case, experimenters need to tweak the profile slope to achieve two requirements. (1) There exists at least one point in the sweep range of the coil where the derivative of $B_\mathrm{r}$ with respect to $z$ is zero (flat spot). This location will be used as the weighing position. (2) The $B_\mathrm{r}(z)$ profile is uniform with relative variations of a few parts in $10^4$ over the sweep range of the velocity measurements.
	
The magnetic field is inversely proportional to the air gap width, and hence, the profile slope can be reduced by slightly enlarging the gap width at regions with a stronger magnetic field. As shown in figure \ref{fig:profile_components}(d), this technique has been successfully applied in flattening the $B_\mathrm{r}(z)$ profile in the LNE Kibble balance magnet system \cite{LNEmag}. For example, to remove a relative slope of  $1\times 10^{-3}$, the gap width $a$ must be changed by  $10^{-3}\delta_\mathrm{a}$. For the LNE case, the maximum yoke radius difference at the top end of the air gap is only 4$\mu$m. Therefore, this approach requires precise control of machining, and even with great care, several iterations (measurement, grinding, measurement) may be necessary. Note, there is no guarantee that this method will ultimately converge to a flat profile because the magnetic working point of the yoke can change during assembly after grinding. The profile changes caused by this parasitic magnetization can be similar in order of magnitude to the original problem.\cite{NISTmag}.  

\begin{figure}[tp!]
	\centering
	\includegraphics[width=0.68\textwidth]{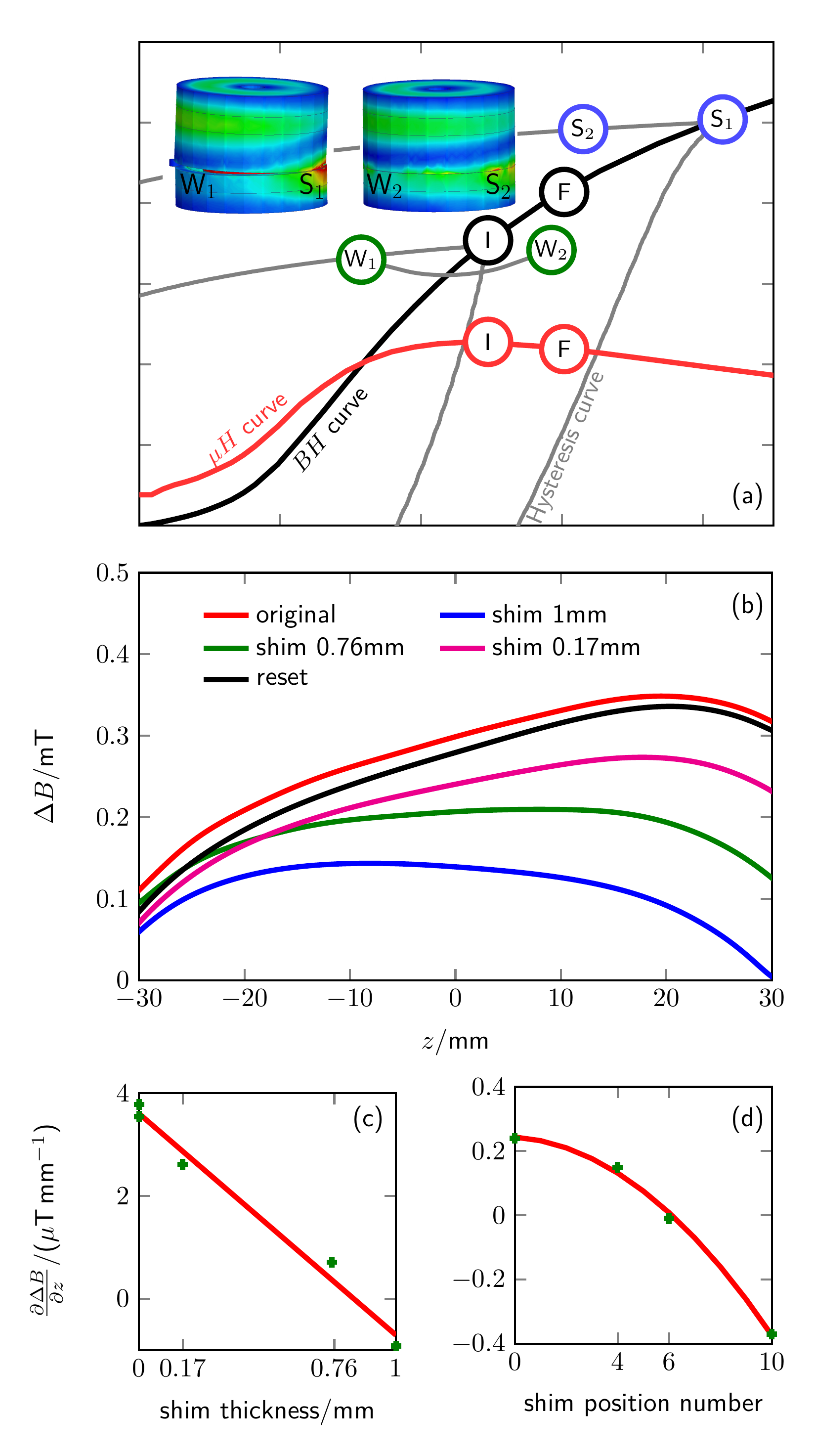}
	\caption{(a) shows the schematic procedure of one typical shimming operation. The black and red curves represent the main magnetization $BH$ curve and the $\mu H$ curve. The gray curves are hysteresis $BH$ curves. The two 3D renderings show the magnetic status during shimming (left) and after shimming is removed (right). Note \textcircled{W} is the shimming point while \textcircled{S} is the magnetic saturation point.  
	The points \textcircled{I} and \textcircled{F} denote the initial and final states of the shimming procedure. The two states are shown on the  $\mu H$ curve and $BH$ curve. (b) presents the profile adjustment using different shim thicknesses with the NIST-4 magnet system. (c) and (d) are the magnetic profile slope change (at $z=0$\,mm) as a function of shim thickness and shim position number.}
	\label{fig:shimming}
\end{figure}
	
Researchers of the NIST-4 Kibble balance team proposed a novel method, so-called 'shimming,' to flatten the magnetic profile \cite{NISTmag}. This method was invented after the measuring and grinding iteration did not converge, as is discussed above.  
The method takes advantage of the yoke hysteresis. 
By creating at least one point with a high flux density in the yoke, the equivalent $BH$ working point of the yoke can then walk towards a higher $H$ direction and lower the average permeability and, hence, more reluctance. The higher reluctance causes a lower field at the gap in the vicinity of the point. This process is repeatedly applied around the gap. The process is explained in the magnetization curve of the yoke in figure~\ref{fig:shimming}(a). 
The shimming reduces the field strength near the point under consideration. 
At this point, the initial magnetic status of the yoke, with the yoke fully closed, is given by \textcircled{I}. The magnet is then opened and a   small non-magnetic piece is inserted at the opposite side of the point under consideration. With the piece in place, the yoke closes on the opposite side first. All the flux from the lower permanent magnet has to go from one part of the yoke to the other. The reluctance through the contact point is lower, and hence a majority of the flux is conducted there. 
The point under consideration is in that area. The yoke $BH$ working point at the contact surface goes from \textcircled{I} to \textcircled{S$_1$} while the other non-contact out yoke has less $H$, which will go from \textcircled{I} to \textcircled{W$_1$} on the hysteresis curve. After we remove the shim and close the magnet, the working point of the yoke will have shifted from \textcircled{S$_1$} to \textcircled{S$_2$} at the contact surface. Opposite the contact surface, the working point will have shifted from \textcircled{W$_1$} to \textcircled{W$_2$}. Due to the yoke hysteresis, the final yoke states at the two opposite sides are different. They are shown as \textcircled{S$_2$} and \textcircled{W$_2$} in the figure. Note both started out at \textcircled{I}, but the hysteresis prevents the states to come back to the initial point. At the end, the average state of the yoke is in the middle between \textcircled{S$_2$} and \textcircled{W$_2$} , denotes as  \textcircled{F}. The initial and final states are also drawn on the red $\mu$-$H$ curve. The magnet field of the final state is higher than the initial state. The change of the permeability depends on where the maximum of the permeability is. In figure~\ref{fig:shimming}(a) the maximum is to the left of the initial state, hence the yoke permeability will decrease with the shimming procedure.

When the magnet is fully opened and left there for a while, the procedure is reset. Its final state will depend on the exact closing. Where will the two yokes make contact first? If the magnet splitter is not perfect and brings to yoke to touch at one point, then, unbeknownst to the user, a shimming step has been executed. Hence, we recommend not to fully open the magnet during the shimming procedure but instead keep the gap small and be aware of the contact points between the yokes.

As shown in figure \ref{fig:shimming}(b)-(d), the magnetic profile change is a function of shim thickness and shim position number. It can be seen that the change in profile is approximately proportional to the shim thickness.
The shimming is also more effective the more points are used, see (d). Therefore, we recommend evenly distributing the shim points along the azimuthal direction. We also suggest using more points (at least four) and a thin shim instead of a thick shim that is only inserted at a few locations.  The described shimming procedure is a customizable and valuable tool to flatten the $B_\mathrm{r}(z)$ curve. 

Other methods to balance the magnetic flux between the upper and lower air gap exist. For example, leaving a small air gap in the splitting plane also results in a flat profile \cite{NISTmag}. The gap functions as an added reluctance to the lower part of the magnet circuit. The disadvantage, however, is that the shielding of the magnet is compromised. The shimming and the introduction of the gap add reluctance to the stronger, here, the lower part of the magnet. The stronger magnet can also be weakened by adding a magnetic shunt in parallel to the permanent magnet. As shown in figure \ref{fig:analysis} and equation (\ref{eq:phiT}), the magnetic shunt returns part of the magnetic flux, which is then no longer available for the gap. The NIM-2 magnet system used both shimming and a magnetic shunt to remove the $B_\mathrm{r}(z)$ slope \cite{you2018shimming}. Since it is difficult to demagnetize the shunt once it is saturated, we recommended to use high-saturation material for the shunt, such as HiperCo50 (saturation magnetic flux density is about 2.4\,T). 
	
\subsection{Shielding performance} 
	
A magnet with an enclosed air gap has better shielding performance than a magnet whose gap is open to the environment. Here, we briefly discuss the two aspects of the term ``shielding performance''. First, the magnetic flux originating from the permanent magnets should stay inside the magnet system and not leak outside. Second, any external flux produced, for example, by electrical currents in the vicinity, should not flow through the coil. Both aspects are related. A yoke that contains the internal flux does not admit outside flux to the precision gap.  

Flux leaking outside the magnet is more problematic for the Kibble balance because it can produce a magnetic force on the test mass. The vertical force on the mass with a volume susceptibility $\chi$ and a permanent magnetization $M$ is, according to  \cite{davis1995determining},
\begin{equation}
F_\mathrm{\chi}=-\frac{\chi}{2}\frac{\partial}{\partial z}\int B\cdot H\mathrm{d}V-\mu_0\frac{\partial}{\partial z}\int M\cdot H\mathrm{d}V.
\end{equation} 
To achieve relative uncertainties below $10^{-8}$, the researchers must evaluate $F_\mathrm{\chi}$ in their Kibble balance. First, the magnetic flux density and its derivative at the mass position must be measured. Then, a combination of these measurements with the magnetic properties of the test mass  ($\chi$, $M$) allows the determination of $F_\mathrm{\chi}$.
The magnetic field on the symmetry axis central ($x=0,y=0$) is close to vertical and decays rapidly with increasing distance to the surface of the magnet.  A typical save distance for the test mass from the top of the magnet is approximately 10\, cm. For a BIPM-type magnet, the effect is at the order of $10^{-9}$ for a  $E_\mathrm{1}$ class steel mass \cite{NISTmag}. For an open gap circuit, the relative contribution of the parasitic magnetic force to $F_\mathrm{z}$ can reach  $\approx1\times 10^{-6}$. In this case, a mass with low magnetic susceptibility, such as one made from PtIr, should be used for the measurement \cite{BIPMmag2017,NRC}.
	
Regarding shielding flux from the outside, reference~ \cite{BIPMmagShielding} shows that a BIPM-type magnet rejects flux from sources that are far away, e.g. the earth magnetic flux, very well. Magnetic flux from close sources can, however, penetrate the magnet system \cite{xu2018research}. Not always does such parasitic flux lead to systematic effects. The effect cancels, for example, if the parasitic flux stays constant between velocity and weighing mode. Nevertheless, the time-changing parasitic flux will increase the noise, especially in the velocity mode.
We recommend placing sources of varying magnetic flux, such as power supplies, away from the Kibble balance to avoid interference. For experiments where the magnet, as opposed to the coil, is moving, e.g., \cite{NIM, UME}, sources of external flux should be handled even more carefully. For these systems, not only does the external flux interfere with the velocity mode, but it will also affect the weighing measurement because the force, $F=I Bl$, does not distinguish between a  $B$ produced by the magnet system or an external source.  The relative bias caused by an external field can reach the $\sim 10^{-7}$. By adding additional shielding, the relative bias can be reduced to $\sim 10^{-8}$\cite{xu2018research,xu2019elimination}. 

One way to test the quality of the shielding is to measure the power spectrum of the induced voltage in the coil when the coil position is fixed \cite{NISTmag}.  As an added benefit from the obtained spectrum, quiet regions may be found. These quiet regions should be exploited in the measurement by choosing an integration time in the voltage measurement that corresponds to these quiet regions. This practice will improve the signal-to-noise ratio, especially in velocity mode.

\subsection{Determining the actual temperature coefficient}
	
Without compensation, the temperature coefficient of a typical SmCo magnet is  $\approx-3\times10^{-4}$/K. 
It is good practice to measure the temperature coefficient on the actual magnet to verify that the true coefficient is not significantly larger than that.
An easy way to measure the temperature coefficient is to introduce a temperature change and continuously measure the magnet field and the temperature. 
If a Hall effect sensor is used to measure the magnetic field, its temperature coefficient must be known and calibrated out.
For the temperature measurement, the coupling to the magnet is also a concern. 
Since the magnet system has a large thermal mass, its temperature is delayed from the room's temperature.  
Hence it is recommended to clamp the temperature sensor to the metal of the yoke and shield it from air currents.
If the measurement is carried out in a vacuum, physical contact with the temperature sensor is not optional but essential.
Once a reliable temperature sensor is installed in a vacuum, the Kibble balance measurement can be used to determine the temperature coefficient \cite{NIST}.	
	
In that context, we would like to consider the typical temperature drift that the experimenter can expect.
Typically, after the system is evacuated, the temperature of the magnet drops from room temperature. 
The latent heat required to evaporate a water film on the surface of the magnet is the cause of the temperature drop.
After that, the magnet will slowly drift back to ambient temperature. 
Since the magnet is very often insulated in a vacuum, it takes several days for the magnet to become thermalized. 
During the thermal recovery, the drift is large, and one must employ techniques to suppress the drift, such as ABA-type measurements~\cite{swanson_2010} for the Kibble balance. 
Another technique that is, for example, employed at the NRC Kibble balance is to pre-heat the magnet before pumping.
This technique can considerably shorten the time required to achieve thermal equilibrium.

%% file: 06_MagneticEffects.tex
\section{Effects of the magnet on the result}
\label{sec5}
	 
As we have discussed, the core idea of the Kibble balance is that the geometric factors in the weighing and velocity mode are identical, and hence, cancel in the final result. 
In this section, we examine how well this idea holds up.
The geometric factor in velocity measurement, $(Bl)_\mathrm{v}$, is usually taken as a reference, and the geometric factor in weighing is compared to it. 
This choice is usually made because the geometric factor in weighing mode, $(Bl)_\mathrm{w}$,  depends on more factors, such as, the current in the coil, the coil expansion due to ohmic heating, and so forth.
In the weighing mode, usually two measurements, one with the mass on the balance pan (mass-on) and one without (mass-off) are made.  The average currents in the coil required to balance the system are  $I_\mathrm{off}$ and $I_\mathrm{on}$, respectively. 
Usually, the tare mass on the balance is chosen such that the currents are equal and  $I_\mathrm{off}=-I_\mathrm{on}$ which yields
 \begin{eqnarray}
 F_\mathrm{on} &=& (Bl)_w I_\mathrm{on}, \\
F_\mathrm{off} &=& (Bl)_w I_\mathrm{off}\\
\mbox{with} &&I=I_\mathrm{off}=-I_\mathrm{on}\; \mbox{it is,}\nonumber\\
F_\mathrm{off}-F_\mathrm{on}  &=& 	 (Bl)_\mathrm{w} I + (Bl)_\mathrm{w} I.
\label{eq:force:diff:bl}
 \end{eqnarray}
If $(Bl)\mathrm{w}$ depends linearly on the current, i.e., $BL_\mathrm{w}=Bl_0(1+\alpha_1 I)$, the force difference is independent of $\alpha_1$.
In general, all odd powers of $I$ in $(Bl)_\mathrm{w}$ will cancel in the $F_\mathrm{off}-F_\mathrm{on}$. 
In the following section, only one state in the weighing is discussed, and we use $I$ to describe the current in the coil.
The other important variable is the vertical position of the coil, abbreviated with $z$. Similar to \cite{Robinson2007An}, we expand $Bl$ to second order in $I$ and $z$. We obtain
\begin{equation}
\frac{(Bl)_\mathrm{w}}{(Bl)_\mathrm{v}}-1\approx\alpha_1 I+\alpha_2 z+\alpha_3(Iz)+\beta_1 I^2+\beta_2z^2+\beta_3(Iz)^2,
\label{eq:terms}
\end{equation}
where $\alpha_1$, $\alpha_2$, $\alpha_3$ are respectively the linear coefficients of coil current $I$, coil position $z$ and the mixed term $Iz$; $\beta_1$, $\beta_2$, $\beta_3$ are the quadratic coefficients for the same terms.
Note, we assume these coefficients to be constant, neither dependent on $I$ nor $z$. The following subsections discuss the effects associated with these terms.
	
\subsection{Coil self-inductance}
\begin{figure}[tp!]
\centering
\includegraphics[width=0.7\textwidth]{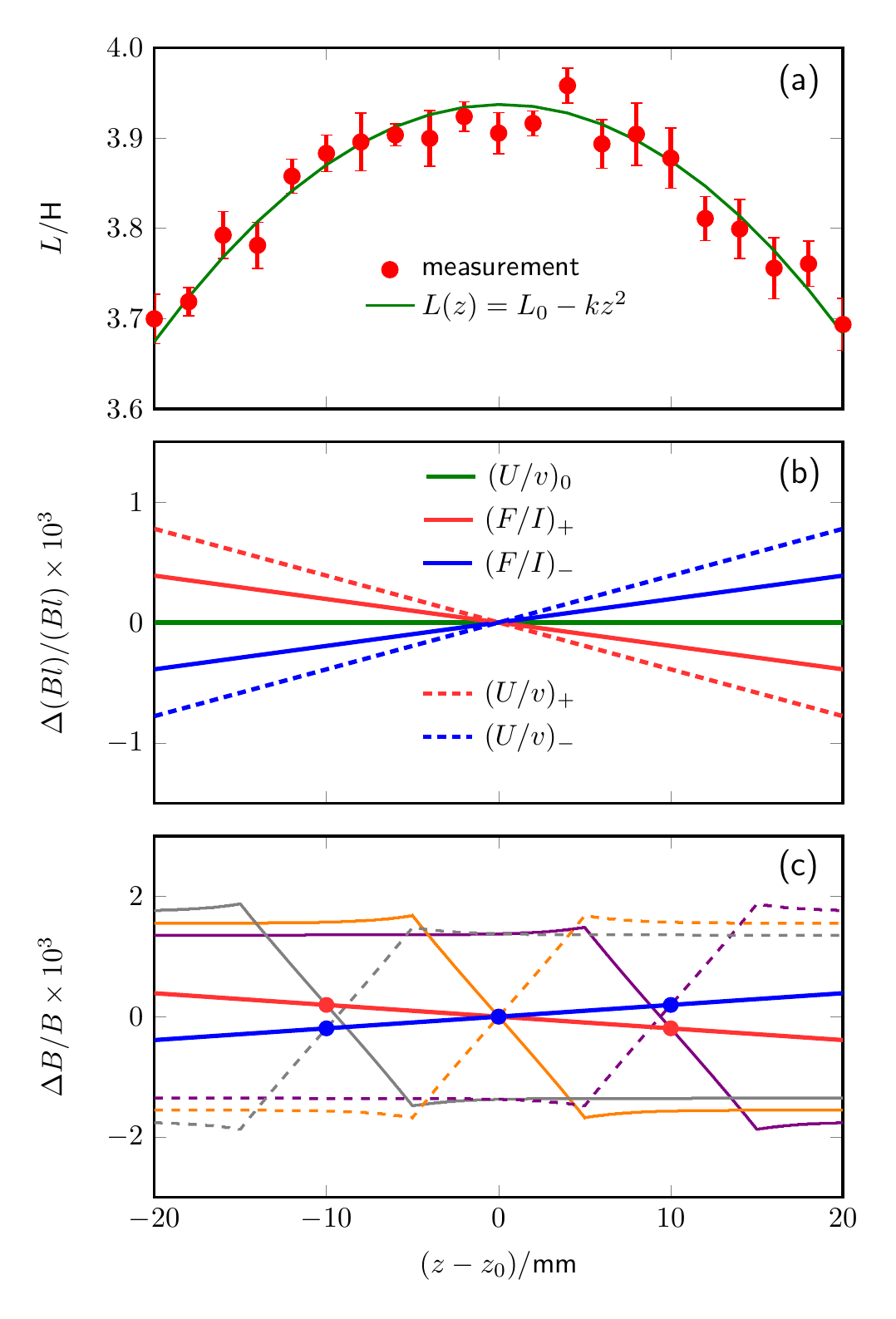}
\caption{(a) is an experimental termination of the dc value coil inductance by the frequency extrapolation method for the BIPM Kibble balance. (b) shows the magnetic profile change related to the coil current, where $(U/v)_0$, the velocity measurement profile without current, is used as a reference. Conventional Kibble balances employ $(U/v)_0$, $(F/I)_+$ and $(F/I)_-$ profiles, while one-mode measurement scheme uses $(U/v)_+$, $(U/v)_-$, $(F/I)_+$ and $(F/I)_-$. (c) shows the actual magnetic distribution along $z$ in the weighing measurement at different coil positions. Note the yoke center $z_0=-3.7$\,mm due to asymmetrical construction,  i.e., the top cover of the magnet was not installed. Reproduced from \cite{li17}.}
\label{fig:currenteffect}
\end{figure}

Any current-carrying coil has energy due to its self-inductance. 
The energy is given by $E=\frac{1}{2}I^2L(z)$, where $L(z)$ is the self-inductance of the coil (dc value) at the weighing position $z$.
Since the energy depends on $z$, a force appears in the direction where $L$ has a maximum.
The negative derivative of the energy gives this force on the coil
\begin{equation}
F_\mathrm{L}=-\frac{1}{2}I^2\frac{\partial L}{\partial z}-IL\frac{\partial I}{\partial z}.
\label{eq:inductanceforce}
\end{equation}
Here, we assume constant current, and the second term vanishes. Still, the force depends on $z$.
If the measurements in force mode were used to calculate a $(Bl)_\mathrm{w}$, one would obtain a relative change of
\begin{equation}
\frac{\Delta (Bl)}{Bl}	:=\frac{(Bl)_\mathrm{w}}{(Bl)_\mathrm{v}}-1=-\frac{1}{2}I\frac{\partial L}{\partial z}.
\end{equation}
The additional parasitic force is proportional to the current squared and would drop out if the currents are perfectly balanced.  
Recall, how odd powers of $I$ in $Bl$ cancel in equation~(\ref{eq:force:diff:bl}).
In an up-down symmetrical magnet, $L$ can be assumed to be a quadratic function of coil position $z$ with $L$ being maximal with the coil in the middle, i.e., $L(z)=L_0-kz^2$. More specifically, $L(z)$ can be written as
\begin{equation}
L(z)=\frac{\mu_0\pi r_\mathrm{a}N^2}{\gamma h_\mathrm{a}}[(\gamma h_\mathrm{a})^2-z^2],
\label{eq:Lz}
\end{equation}
where $r_\mathrm{a}$ is the mean radius of the air gap, $h_\mathrm{a}$ half-height of the air gap, and $\gamma$ is a factor that corrects the height of the air gap for the fringe fields. Typical values of $L(z)$ can be seen in figure~\ref{fig:currenteffect}.
	
The $L(z)$ function can be measured with different methods.
For example, the ac parameters of the coil (complex impedance) can be measured at different vertical positions at different low frequencies.
The measurements can then be extrapolated to $f=0$~\cite{NISTmag,li17}. Alternatively, the force-current ratio, $mg/(I_{+}-I_{-})$, can be measured at different weighing positions $z$. 
The $\partial L/\partial z$ and finally $L(z)$ can be obtained by comparing the profile change to the voltage-velocity ratio over the range.  
Figure \ref{fig:currenteffect} (a) shows such a determination of  $L(z)$ carried out with the BIPM Kibble balance \cite{li17}. Note, measurements in \cite{NISTmag} show that $L$ has a considerable frequency dependence, related to the skin effect. Therefore, it is important to keep the coil current stable during the weighing measurement to suppress unwanted electrical noise or systematic bias.

The position, $z_0$, where $L$ has the maximum value (typically at the Henry level depending on the coil parameter and the gap dimensions), is named the yoke center.
At this location, the inductance force $F_\mathrm{L}=0$ and therefore, the $Bl$ value is independent of the coil current $I$. 
Note that the yoke center may neither coincide with the maximum in $Bl$ nor with $z=0$.
On either side of $z_0$, the inductance force acts in the opposite direction.
The coil is always pulled toward the center of the yoke.
The situation is analog to a solenoid actuator, where an iron slug is drawn into an energized coil.  
The only difference is that the iron piece is on the outside and the coil on the inside in a Kibble balance.

The sum of the forces, $ Bl I +F_\mathrm{L}$, is different from $BL I$ by $F_\mathrm{L}$ that is primarily linear in $z$. Hence, the measured maxima of $(Bl)_\mathrm{w}$ and $(Bl)_\mathrm{v}$ differ by the effect that this additional slope has. 
That relative difference is illustrated in Figure \ref{fig:currenteffect}(b), where the baseline shown in green is $Bl$ without current, i.e., $(U/v)_0$.
The $(F/I)_+$ (solid red) and $(F/I)_-$ (solid blue) curves are obtained with the weighing measurement.
Different from most Kibble balances, the BIPM balance uses a so-called 'one-mode, two-phase' measurement scheme \cite{BIPM}.
There, the weighing current is flowing through the coil also in velocity measurements.
Hence, a bifilar coil is required. 
One coil is employed for induction while the other coil provides the current to counterbalance the test mass.
For an ideal bifilar coil, the inductance of each coil $L$ equals the mutual inductance of the two coils $M$.
Hence, an additional term appears in the induced voltage of the velocity mode.
It is
\begin{equation}
\Delta U=I\frac{\partial L}{\partial z}v+L\frac{\partial I}{\partial z}v.
\label{eq:additionalU}
\end{equation}
Again, assuming constant current, only the first term remains. 
Hence,  $\Delta(Bl)_\mathrm{v}=\partial L/\partial z$.  Comparing this term to equation (\ref{eq:inductanceforce}) yields $\Delta(Bl)_\mathrm{v}=2\Delta(Bl)_\mathrm{w}$. 
This effect is demonstrated by the dashed lines in figure \ref{fig:currenteffect}(b). 
It can be seen that the lines representing $(U/v)_+$ and $(U/v)_-$ have twice the slope of the lines $(F/I)_+$, $(F/I)_-$. 
Even though the corrections are different in weighing and velocity, the true magnetic field changes follow the weighing profile, see~\cite{li18}.
As shown in Figure~\ref{fig:currenteffect}(c), the weighing profile change is resulted by an average of the magnetic field over the coil wire region. 
We discuss below two systematic effects that appear in weighing mode due to self-inductance. These effects are present in both the traditional, two-mode, and newer, one-mode measurements. We write the relative change in $(Bl)_\mathrm{w}$ as
\begin{eqnarray}
\frac{\Delta (Bl)}{(Bl)_\mathrm{w}}&=&\frac{\displaystyle I_+^2\left(\frac{\partial L}{\partial z}\right)_{z_+}-I_-^2\left(\frac{\partial L}{\partial z}\right)_{z_-}}{2(Bl)_v(I_+-I_-)}\nonumber\\
&\approx& \frac{\displaystyle (I_++I_-)\left(\frac{\partial L}{\partial z}\right)_{z_\mathrm{a}}}{2(Bl)_v}\nonumber\\
&&+\frac{\displaystyle\left(\frac{I_+^2+I_-^2}{2}\right)\left(\frac{\partial^2L}{\partial z^2}\right)_{z_\mathrm{a}}(z_+-z_-)}{2(Bl)_v(I_+-I_-)},
\label{eq:error1}
\end{eqnarray}
where $z_\mathrm{a}=({z_++z_-})/{2}$ is the average weighing position.
The first term in the sum can be reduced, potentially even to zero, by symmetrizing the current,  $I_+=-I_-$. 
Even if the currents can not be made perfectly equal, the term can vanish if the average weighing position is close to the magnetic center, where $\partial L/\partial z=0$. 
Unfortunately, the second term in the sum depends on the second derivative of the inductance with respect to the vertical position.
Of the three factors of the second term, only $z_+-z_-$ can be made small. 
The reason $z_+\ne z_-$ is because of the finite stiffness of the coil suspension. Suppose the balance is controlled to the same position. 
In that case, the coil can be at slightly different positions because the forces acting on the coil suspension differ by $mg$ and can lead, depending on the suspension stiffness, to a change in position of a few up to a few tens micrometers. 
It was shown in a typical Kibble balance configuration, 1$\mu$m coil position change would cause a bias of $1\times10^{-8}$ \cite{li17,BIPMmag2017}.
A possible solution would be to slightly change the target in the feedback mechanism for the balance control to maintain a constant coil position instead of a constant balance position.
If that cannot be achieved, a correction of the size of the second term in equation (\ref{eq:error1}) including a reasonable uncertainty must be applied. 
Another concern that applies for the one-mode measurement is increased noise. The term $(\partial I/\partial z) v$ in equation (\ref{eq:additionalU}) is identical to $\partial I/\partial t$.
Hence noise in the current flowing through one coil will give rise to noise in the voltage measurement. 
Researchers at the BIPM noticed a significant increase in noise of the $U/v$ measurements when the current source was powered by the mains and not by batteries.
We recommend using a current supply with low internal noise and adding a low pass filter between the current source and the coil to minimize the additional noise.
	
To summarize this section, the bias caused by the coil inductance is proportional to the coil current $I$ and the weighing position $z$. It is minimized by keeping these two variables as close as possible for the two weighing measurements, mass-on and mass-off. Its effect is summarized by $\alpha_3(Iz)$ in in equation (\ref{eq:terms}).

\subsection{Diamagnetic force}
	
Diamagnetic material suspended from the balance inside the magnet is inevitable for any Kibble balance. For example, the coil wire is made of copper, whose volume magnetic susceptibility is about $-1\times10^{-5}$. 
In addition, glass and ceramic pieces are typically mounted on the coil.
Ceramic is often used as a coil former.
On the former, flat or corner cube mirrors are mounted to allow the researchers to determine the coil's position and velocity. 
These elements are made from glass. 
The diamagnetic force acting on a part in the air gap can be written as \cite{frog}
\begin{equation}
F_\mathrm{d}=\frac{\int \chi \mathrm{d}V}{\mu_0} \frac{1}{2}\frac{\partial}{\partial z} \left(B+b\right)^2,
\label{eq:diaF}
\end{equation}
where $\int\chi dV$ denotes the magnetic susceptibility integrated over the volume of the part, $B$ and $b$ ($\propto I$) the magnetic flux density created by the magnetic circuit and the coil itself. 
Typically $b$ is much smaller than $B$, with the ratio  $b/B$ being usually $\approx 10^{-3}$.

\begin{figure}[tp!]
\centering
\includegraphics[width=0.7\textwidth]{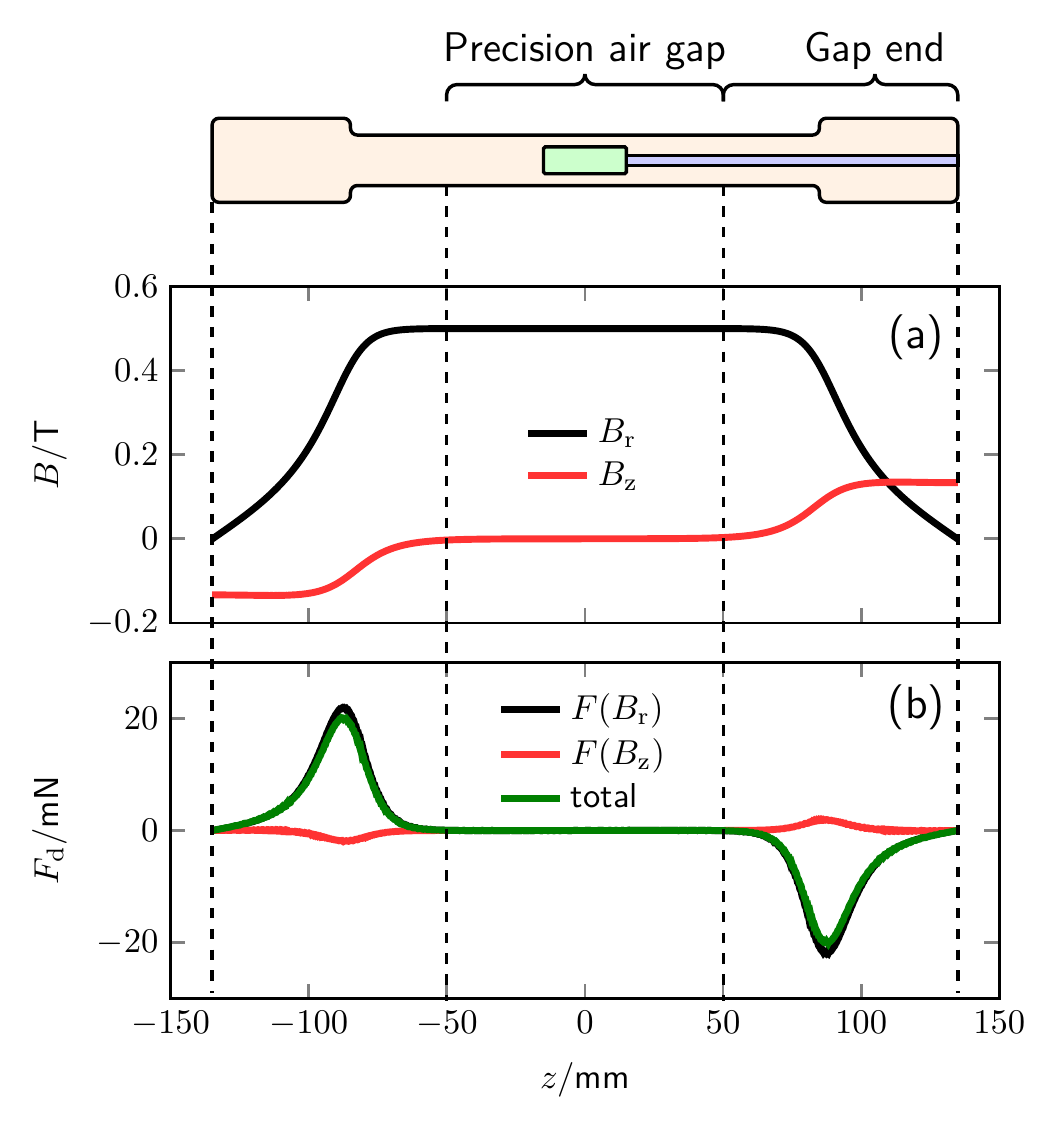}
\caption{(a) presents the magnetic flux density distribution along $z$ in the middle of the air gap. $B_\mathrm{r}$ and $B_\mathrm{z}$ denote the radial and vertical components, respectively. These results are obtained by FEA calculation using the NIST-4 parameters \cite{NISTmag}. (b) shows the vertical diamagnetic force based on $B_\mathrm{r}$ and $B_\mathrm{z}$. For the calculation a coil wit a 20\,mm$\times$20\,mm cross-section entirely made from copper ($\chi=-1\times10^{-5}$) was assumed. The drawing above both plots shows schematically the cross-section of the gap. The coil is presented as the light green rectangle. The with is scaled differently than the length to save space. The shaded blue area indicates the support rods for the coil. Note, the diamagnetic force on them is not included for the calculation of $F_\mathrm{d}$.}
\label{fig:diaforce}
\end{figure}

As shown in figure \ref{fig:diaforce}, the effect of $F_\mathrm{d}$ ought to be analyzed in two different regions:
Inside the precision gap and near the end of the gap. 
The latter is the region that is not part of the precision air gap but is still inside the yoke.  
Figure \ref{fig:diaforce} (a) shows the magnetic field distribution along with the whole gap space in the NIST-4 system. 
At the end of the gap end, the magnetic gradient $|\partial B/\partial z|$ is large and dominates the diamagnetic force.

In figure \ref{fig:diaforce}(b), the diamagnetic force calculated to act on a copper ( $\chi=-1\times10^{-5}$) coil with a cross-section of 20\,mm$\times$20\,mm is shown.
At the very end of the gap, the diamagnetic force is considerable, 20\,mN, or relatively $2\times10^{-3}$.
Of course, in the actual experiment, the coil would not move into this region. However, other mechanical elements, most importantly the suspension rods, have to transverse this region.
Hence, a significant diamagnetic force on the coil stirrup is possible.
In 2017, researchers at METAS measured this diamagnetic force on their experiment and found the diamagnetic force to change by about 100\,mg per centimeter of travel. 
This number corresponds to a relative change of $(Bl)_\mathrm{w}$ per length of about $1\times10^{-8}/\SI{}{\mu m}$.
To keep the effect at or below the target uncertainty of the balance, the suspension position change during mass-on and mass-off should be less than 1\,$\mu$m. 
	
The effect discussed here is captured by the term $\alpha_2z$  in equation (\ref{eq:terms}). 
In reality, the relationship between systematic force and the diamagnetic effect is more complicated. 
As mentioned above, the force acting on the coil suspension changes by $mg$ between the mass-on and mass-off measurement. 
Since the coil suspension has a finite stiffness, movement occurs between the two measurement states.
Typically, the force feedback is designed to maintain a constant balance position.
Hence, the stretching of the coil support causes the coil to move by a small amount, a few $\SI{}{\mu m}$. 
With that, the amount of material that is in the end region of the gap changes.
To our knowledge, this effect has not been described in the literature.
Therefore, we believe that the change of the diamagnetic force due to coil support stretching should be investigated in the near future.

Inside the precision air gap, the profile is flat, and $\frac{\partial B}{\partial z}$ has a much lower value than at the end of the gap. 
The weighing position is usually chosen where $\frac{\partial B}{\partial z}\approx0$. 
The only term that remains is the flux gradient that is produced by the coil, $\frac{\partial b}{\partial z}$. 
In contrast to the $\frac{\partial B}{\partial z}$, the term $\frac{\partial b}{\partial z}$  is proportional to the coil current $I$, and therefore, does not drop out in the difference between mass-on and mass-off measurement. 
A relative bias of 
	\begin{equation}
	\frac{\Delta (Bl)}{(Bl)\mathrm{w}}=-\frac{r_\mathrm{a}}{\gamma S_\mathrm{a}}\int \frac{\chi }{r}dS
	\label{eq:diamagbias}
	\end{equation}
is introduced to the weighing measurement. Here,  $S$ denotes the cross-sectional area of the segments in the air gap, and $\gamma$ the factor to correct the height of the gap for the fringe fields, as defined in equation (\ref{eq:Lz}). 

The bias given in equation~(\ref{eq:diamagbias}) is by far the most dominant bias discussed in this section. 
Using the BIPM magnet system as an example, the term $\frac{\Delta (Bl)}{(Bl)_\mathrm{w}}$ can be as large as  $1\times10^{-6}$ for a copper coil.
Details on how to calculate this effect can be found in \cite{diamagnetic2020}. Because the term is 100 times larger than the smallest uncertainties reported Kibble balances, there was considerable debate in the community if such a diamagnetic force exists in the real world.
This dilemma has recently been solved \cite{diamagnetic2020}: 
The diamagnetic force does exist, but the analysis found the same bias in the velocity measurement when weak magnetization materials are used. 
It turns out that two effects perfectly cancel each other. 
At most, a relative bias of order $10^{-9}$ may result due to small asymmetries.

In the end, the last two effects discussed here are minimal. 
Referring back to equation~(\ref{eq:diaF}), they are captured by  $b\cdot\partial b/\partial z$ and $b\cdot\partial B/\partial z$.
Since $b$ is so much smaller than $B$ they can be considered second-order effects.

\subsection{Nonlinear effects}
Going back to equation (\ref{eq:terms}), there are three terms that have quadratic, i.e., nonlinear behaviors, $\beta_1 I^2$, $\beta_2 z^2$, and $\beta_3(zI)^2$.  
These effects are, however, small, as can be seen from the following simple consideration.
The differences in the linear counterparts of these same terms between the mass-on and mass-off measurements contribute an effect that is of order $1\times10^{-5}$. 
Hence the quadratic effects must be well below $1\times 10^{-9}$. 
As shown in figure~\ref{fig:currenteffect}, the flux contributed by the coil is between $10^{-4}$ and $10^{-3}$ of the magnetic flux contributed by the magnet.
As a result,  the nonlinear term, $\beta_1 I^2$, should be checked. 
We know of three nonlinear current effects that could contribute to $\beta_1 I^2$, and they are:
	
	\begin{itemize}
		\item Yoke magnetic reluctance change. 
		The flux produced by the coil traverses the air gap twice. 
		A small amount of the coil-produced flux flows through the whole magnetic circuit. 
		The ratio of these two magnetic fluxes depends on the reluctance ratio $R_\mathrm{a}/R_\mathrm{m}$. Here $R_\mathrm{a}$ is the reluctance of the air gap  $R_\mathrm{m}$ that of the magnetic circuit for the coil produced flux.
		Inside the iron, the former/latter flux is parallel/perpendicular to the flux produced by the permanent magnet. 
		The additional flux in the yoke changes its reluctance due to the $\mu_r$ dependence on $H$ and, as a consequence, the magnetic flux density inside the air gap.
		A detailed description of these effects and how they can introduce a systematic error is given in \cite{linonlinear, linonlinear2}.
		For here, it is sufficient to know that the relative effects are well below the typical uncertainty of Kibble balances.
		
		\item Yoke hysteresis effect. 
		The effect described in the previous paragraph assumes that the magnetic state of the yoke changes along the primary hysteresis loop of the material. 
		However, this is not the case. 
		The magnetic state is only slightly disturbed by the weighing current—the magnetic and the state changes along a minor hysteresis loop. 
		Reference \cite{hysteresis} investigated the $Bl$ shift caused by a change in yoke hysteresis. 
		It shows that a possible relative bias of order  $10^{-8}$ can occur in a BIPM-style magnet system in the conventional operation scheme.
		For large dimensional air gaps, e.g., \cite{NISTmag,NRC}, this effect is negligible. 
		But for magnets with smaller air gaps, a careful investigation of this effect should be conducted.
		
		\item  Coil heating effect. 
		The coil heating contributes to the $\beta_1 I^2$ term in equation (\ref{eq:terms}).
		In contrast to the aforementioned effects, the coupling of the current to the magnet is not magnetic but thermal.
		In conventional two-mode measurement, there is no current flowing through the coil in velocity mode and the coil cools.
		Once the balance is switched to force mode, current flows in the coil and the ohmic power dissipation causes heating.
		This effect can be readily measured by observing the coil resistance, i.e., the voltage drop over the coil divided by the current in the coil.
		The coil resistance goes up as the coil heats up.
		Typically the heat transfer between the coil wire and the magnet is very weak because it is radiation only. 
		Still, the magnet temperature can rise, and due to the temperature coefficient of the permanent magnet material, the magnetic flux density drops.
		For the wire, the temperature change is below 1~K, and the time constant for this effect is several minutes. 
		The effect is much smaller for the permanent magnet system, and the time constant is much longer.
		Yet, this effect should not be forgotten in comprehensive uncertainty analysis of a Kibble balance measurement.
	\end{itemize}

In addition to a theoretical study, the size of the term $\beta_1 I^2$ in equation~(\ref{eq:terms})  can also be determined experimentally. 
For that, masses with different nominal values need to be measured and the results compared to results obtained by the classical sub-division scheme using standard balances.
Before the redefinition, one would simply measure the Planck constant with different nominal values. 
The researchers at NIST and NRC have done that~\cite{NIST,NRC}. 
Both systems employ wide air gaps, and as predicted by the theoretical analysis, the $\beta_1 I^2$ is very small, only a few parts in $10^9$, as is predicted by theoretical analysis \cite{linonlinear,linonlinear2,hysteresis}.
For magnet systems with narrower air gaps, careful theoretical evaluations and experimental checks of the nonlinear terms should be carried out.
	
The heating effect can be checked similarly but has an additional parameter that can be used to our advantage.
While the magnetic effect is nearly instantaneous, the heating effect is delayed. 
Hence, time should be a variable in the investigation in one of the following two ways.
(1) a delay of varying length can be added between the weighing and velocity modes. (2) the duration of each of both modes can be changed.
Balances utilizing the one-mode scheme are not subject to any of the current related effects discussed above.
In that case, the current is present while measuring the $U/v$. 
All current effects on the magnet system are already included in the measurement,  and the researcher does not have to worry about it — a significant advantage of the single-mode scheme.

%% file: 07_conclusion.tex
\section{Summary}
\label{sec7}
	
The magnet system supplies the magnetic flux density $B$ that is part of the geometric factor, $Bl$ for the Kibble balance.
Kibble's theory relies on the fact that the $Bl$ factors in force mode and velocity mode are identical. 
Hence, it will cancel out in the final equation that equates electrical power to mechanical power.
One would think that because $Bl$ cancels out, not much thought should be given to designing the magnet system. H
However, the opposite is true. Because the $Bl$ in weighing mode, where current is present, cannot differ by more than is tolerable in the final uncertainty budget from the $Bl$ in velocity mode, where there is no current,  a good design is crucial. 
The requirements ease somewhat for Kibble balances that use the one-mode measurement, where the measurements with and without current are conducted simultaneously using a bifilar coil.
	
In this article, we have attempted to collect the physics and engineering principles that are important for the magnet designer.
About ten years ago, the literature lacked articles on magnets for Kibble balances. 
Since then, many articles have been written, and the bibliography gives a comprehensive overview of this body of work.

This article is structured in six chapters.
The introduction gives the basic equations of the Kibble balance and makes clear what role the magnet system plays.
In the second chapter, we introduce the basic quantities that are needed to analyze a magnetic circuit.
The third chapter shows how the magnet system of the Kibble balance has evolved over the decades.
It classifies the known predominant yoke-based permanent magnets with a radial field. 
The fourth chapter shows the choices that must be made designing a magnet system. These include material choices but also choices of geometry.
The fifth chapter titled "Delivering design to reality" gives insight on how to deal with imperfections of the final magnet system. 
Designing something on paper is one thing, but the reality is another due to machining tolerances and material flaws. 
The chapter shows how to overcome some of these problems. 

Last but not least, the effects of the magnet system on the result are discussed. 
In the sixth chapter, several systematic effects are described. 
In addition, it contains valuable tips and procedures to determine the size of these effects.

After working through the seven chapters, it may appear intimidating to design a magnet system. 
However, we would like to remind the reader: 
The difficulty comes because of the exquisite small uncertainty that the Kibble balance is aiming for. 
The best ones achieve relative uncertainties of \SI{1e-8}{}.
At this level, all metrology is hard.  
If it weren't for the ambitious measurement goal, the design of the magnet system would be easy, because of the inherent symmetry of Maxwells' equation.
It's not only that the $Bl$ cancels out in the final Kibble equation but also most parasitic effects, at least at the \SI{1e-6}{} level. Below that, the hard work begins.

%% file: 08_Appendix.tex
\appendix

\section{$Bl$ integral and flux derivative}
\label{sec:AppendixA}
Assuming a coil with a single turn, the force on  a small segment of wire is given by
\begin{equation}
    \ud  \vec{F} = I \ud \vec{l} \times \vec{B}.
\end{equation}
The total force on the wire loop is the integral over the closed contour $C$,
\begin{equation}
     \vec{F} = -I \underset{C}{\mathlarger{\oint}} \vec{B} \times \ud  \vec{l} .
\end{equation}
We are interested in the $z$ component  of the force and we perform the integral in Cartesian coordinates. Hence,
\begin{equation}
     F_\mathrm{z} = -I \underset{C}{\mathlarger{\oint}} (B_\mathrm{x}\ud y - B_\mathrm{y}\ud x ).
\end{equation}
According to Green's theorem the line integral over the closed contour can be written as the 2d integral over the enclosed area A,
\begin{equation}
     \underset{C}{\mathlarger{\oint}} (B_\mathrm{x}\ud y - B_\mathrm{y}\ud x) = -\underset{A}{\mathlarger{\iint}} \left( \frac{\partial B_\mathrm{x}}{\partial x} +\frac{\partial B_\mathrm{y}}{\partial y} \right)\ud x\,\ud y.
\end{equation}
Since, $B$ is divergence free,
\begin{equation}
\frac{\partial B_\mathrm{x}}{\partial x} +\frac{\partial B_\mathrm{y}}{\partial y}  = - \frac{\partial B_\mathrm{z}}{\partial z},
\end{equation}
the following equivalency can be obtained
\begin{equation}
     \underset{C}{\mathlarger{\oint}} (B_\mathrm{x}\ud y - B_\mathrm{y}\ud x) = \underset{A}{\mathlarger{\iint}}  \frac{\partial B_\mathrm{z}}{\partial z}\ud x\,\ud y = \frac{\partial}{\partial z}  \underset{A}{\mathlarger{\iint}} B_z \ud x\,\ud y = 
     \frac{\partial \Phi}{\partial z}
\end{equation}
Hence, for a closed contour,
\begin{equation}
     Bl:=\bigg(\underset{C}{\mathlarger{\oint}} \vec{B} \times \ud  \vec{l} \,\bigg)_\mathrm{z}= \frac{\partial \Phi}{\partial z}.
\end{equation}

\section{$Bl$ conservation in a $1/r$ field}
\label{sec:AppendixB}
Without loss in generality, we consider a coil with a single turn.
The magnetic flux density is given in the range of $r_i<r<r_o$, where $r_i$ and $r_o$ denote the inner and outer radii of the air gap, respectively. 
The mean radius of the coil is $r_c$, where the magnetic flux density is $B_c$. Let $B(r)=\frac{G}{r}$, where $G$ is a constant.
At the coil position, $G=B_cr_c$ and $Bl_0=2\pi r_cB_c=2\pi G$. Using this notation, we investigate four scenarios.

\begin{figure}[tp!]
	\centering
	\includegraphics[width=0.45\textwidth]{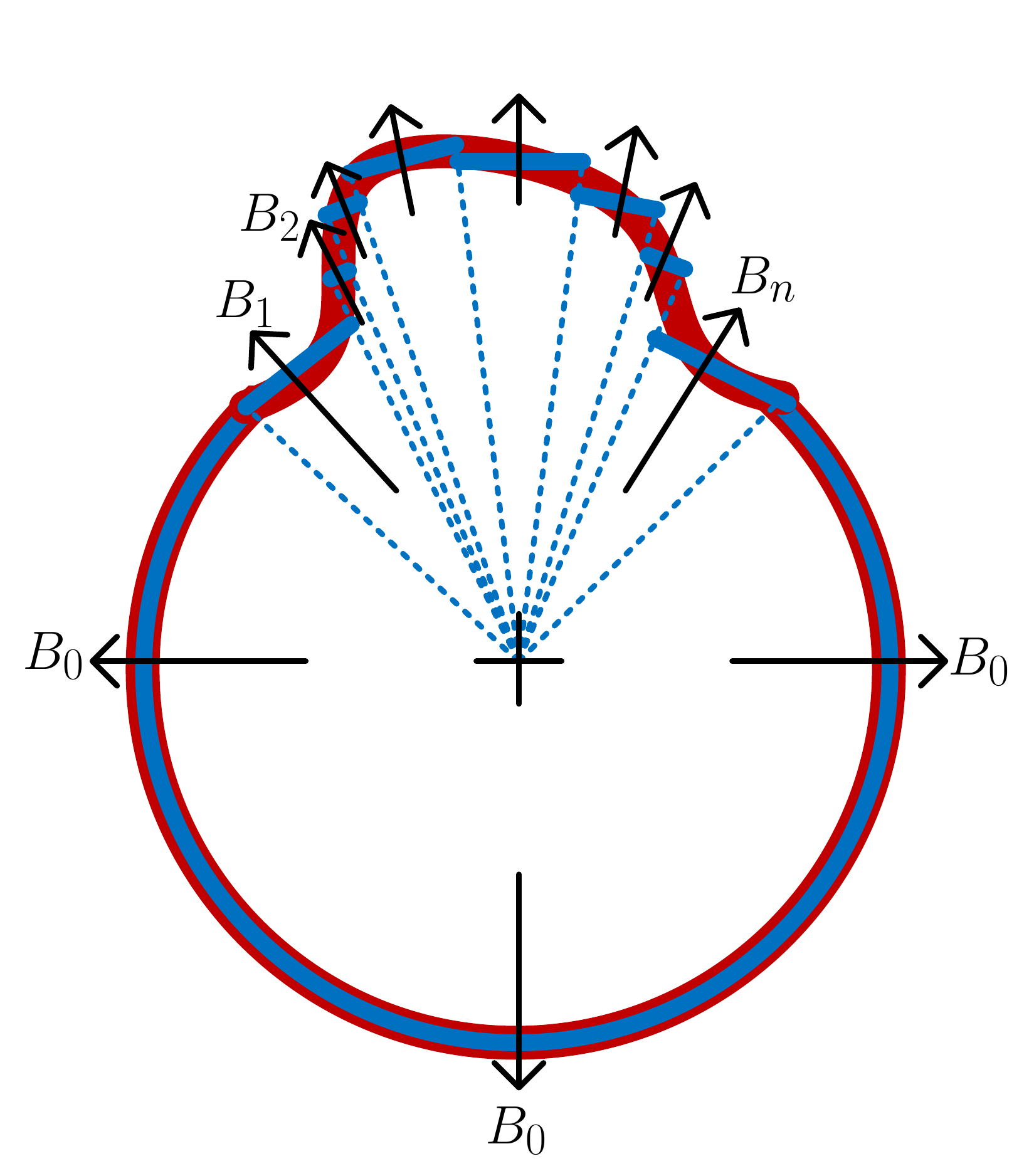}
	\caption{A circular approximation of a non-circular shape coil. }
	\label{fig:coilshape}
\end{figure}

\begin{enumerate}
    
    \item Coil thermal expansion under ideal alignment. Assume that the coil is aligned to the air gap center, and now consider the coil radius is changed by thermal expansion,
    $r_\mathrm{c}'=r_\mathrm{c}+\Delta r$.
    Then the $Bl=B'l$ of the coil with radius $r_\mathrm{c}'$ is,  
    \begin{equation}
        Bl= B' \cdot 2\pi (r+\Delta r)  = \frac{G} {r+\Delta r} \cdot 2 \pi(r+\Delta r) = 2\pi G =Bl_0,
    \end{equation}
    is independent to $\Delta r$. 

    \item Horizontal displacement of the coil. Here we  assume that the coil is no longer centered, but  horizontally displaced by $\delta r=\sqrt{\Delta x^2+\Delta y^2}$. This case was discussed in \cite{li2016coil}. For an eccentric coil, the flux integral is
    \begin{eqnarray}
        Bl
        &=& \frac{Bl_0}{2\pi}\int_{0}^{2\pi}\displaystyle \frac{r_\mathrm{c}^2+r_\mathrm{c}(\Delta x \cos\theta+\Delta y \sin\theta )}{(r_\mathrm{c}\cos\theta+\Delta x)^2+(r_\mathrm{c}\sin\theta+\Delta y)^2}\mbox{d}\theta,\nonumber\\
        &=&Bl_0.
    \end{eqnarray}
    The integral evaluates to $2\pi$. This relationship has been proven in appendix A of \cite{li2016coil}.
    The $Bl$ is independent of horizontal displacement in a $1/r$ flux density. 
    
\item A coil of non-circular shape. Without losing generality, we consider a coil with a shape as the red curve in figure \ref{fig:coilshape}. The shape is approximated by $n$ arcs that are aligned to the center. The magnetic flux density at these arc radii is $B_0$, ..., $B_n$. The $Bl$ of such an approximation can be written as
\begin{equation}
    Bl=\sum_{i=0}^nB_ir_i\theta_i,
\end{equation}
where $\theta_i$ and $r_i$ denote the angle and radius of the $i$th arc.
As shown above, $B_ir_i\theta_i=B_cr_c\theta_i$. Hence, $Bl$ can be written as
\begin{equation}
    Bl=B_cr_c\sum_{i=0}^n\theta_i=2\pi r_cB_c=Bl_0.
\end{equation}
The exact expression $Bl=Bl_0$ is obtained with $n\rightarrow\infty$. In summary, the $Bl$ is independent of the coil shape in a $1/r$ magnetic field.  Note, the logic used here can also be applied to prove (ii).

\item A tilted coil of non-circular shape. It can be deduced from equations (\ref{maxwell1}) and (\ref{maxwell2}) that in an $1/r$  $B$-field, $\frac{\partial B}{\partial z}=0$ and hence $B$ is uniform along $z$. Hence, only the $xy$ projection of the coil contributed to the $Bl$. As we have shown in (iii), the $Bl$ is conservation and equal to $Bl_0$. 

\end{enumerate}

Strictly speaking, the yoke-based radial magnet has the $1/r$ field distribution only at the $z$ planes where $\frac{\partial B}{\partial z}=0$. For other $z$ values, the field has a vertical gradient component and hence is no longer proportional to $r^{-1}$. The related effects on $Bl$ for different parasitic motions are described in \cite{li2016coil}.

%% file: 09_References.tex
\section*{References}

\normalem